\documentclass[a4paper,11pt]{article}
\pdfoutput=1 
\usepackage{jheppub} 
\usepackage[T1]{fontenc}
\usepackage[italian,english]{babel}
\usepackage{hyperref}
\hypersetup{
	colorlinks=true,
	linkcolor=[rgb]{0.0,0.42,0.89},
	filecolor=magenta,      
	urlcolor=[rgb]{0.0,0.42,0.89},
	citecolor=[rgb]{0.64,0.0,0.0},
}
\usepackage{ifpdf}
\usepackage{subfigure}
\usepackage{gensymb}
\usepackage{amssymb}
\usepackage{amsfonts}
\usepackage{epsf}
\usepackage{rotating}
\usepackage{graphicx}
\usepackage{amsmath}
\usepackage{fancyhdr}
\usepackage{lineno}
\usepackage{babel, blindtext}
\usepackage{graphics}
\usepackage{color}
\usepackage{multirow}
\usepackage{pstricks}
\usepackage{xcolor}
\usepackage{multirow}
\usepackage{hhline}
\usepackage{cancel}
\usepackage{framed}
\usepackage{mathtools}
\usepackage{diagbox}
\usepackage[lmargin=1.7in,bmargin=0.2in,footskip=0.5in,total={6.9in,9.5in}]{geometry}

\newcommand{\nn}{\nonumber}
\newcommand{\lsim}{\mathrel{\mathop{\kern 0pt \rlap
  {\raise.2ex\hbox{$<$}}}
  \lower.9ex\hbox{\kern-.190em $\sim$}}}
\newcommand{\gsim}{\mathrel{\mathop{\kern 0pt \rlap
  {\raise.2ex\hbox{$>$}}}
  \lower.9ex\hbox{\kern-.190em $\sim$}}}

\newcommand{\be}{\begin{equation}}
\newcommand{\ee}{\end{equation}}
\newcommand{\bea}{\begin{eqnarray}}
\newcommand{\eea}{\end{eqnarray}}

\title{\boldmath Discerning Singlet and Triplet scalars at the electroweak phase transition and Gravitational Wave}

\author[a]{Priyotosh Bandyopadhyay,}
\author[a,b]{Shilpa Jangid}

\affiliation[a]{
Indian Institute of Technology Hyderabad, Kandi,  Sangareddy-502284, Telangana, India}
\affiliation[b]{
		Asia Pacific Center for Theoretical Physics, Pohang, 37673, Korea}
\emailAdd{bpriyo@phy.iith.ac.in, ph19resch02006@iith.ac.in, shilpa.jangid@apctp.org }

\preprint{IITH-PH- 0003-21}

\abstract{In this article we examine the prospect of  first order phase transition with a Y=0 real $SU(2)$  triplet extension of the Standard Model, which remains odd under $Z_2$, considering  the observed Higgs boson  mass, perturbative unitarity, dark matter constraints, etc. Especially we investigate the role of Higgs-triplet quartic coupling considering one- and two-loop  beta functions and compare the results with the complex   singlet extension case.  It is observed that at one-loop level, no solution can be found for both, demanding the Planck scale perturbativity. However, for a much lower scale of $10^4$ GeV, the singlet case predicts  first order phase transition consistent with the observed Higgs boson mass. On the contrary, for the two-loop beta functions with one-loop potential, both  the scenarios foresee strongly first order phase transition consistent with the observed Higgs mass with upper bounds of 310, 909 GeV on the triplet and singlet masses, respectively. This mass bound shifts to 259 GeV in case of triplet with the inclusion of two-loop contributions to the effective potential and the thermal masses with two-loop beta functions, consistent with the Planck scale perturbativity and the observed Higgs boson mass value. This puts the triplet in apparent contradiction with the observed dark matter relic bound and thus requires additional field for that. The preferred regions of the parameter space in both the cases are identified  by benchmark points, that predict  the Gravitational Waves with detectable frequencies in present and future experiments. }

\keywords{\footnotesize Higgs bosons, Beyond Standard Model, Dark matter, Electroweak symmetry breaking, Gravitational Wave }

\begin{document}

\maketitle
\flushbottom

\section{Introduction} 
The discovery of the Higgs boson around 125.5 GeV was the last stepping stone in the Standard Model (SM) \cite{ATLAS:2012yve,CMS:2012qbp} and a proof of spontaneous symmetry breaking (SSB) in generating the masses of  some of the SM particles. However, the nature of symmetry breaking is far from understood i.e., the role of another scalar, the order of phase transition (PT), etc, are still to be comprehended. It is intriguing to notice that with one Higgs doublet and the Higgs boson mass around 125.5 GeV, one only finds a smooth cross-over \cite{KAJANTIE1996189,Laine1996,Schiller1997,Csikor1999,Michela2014} but not the first order phase transition.  The requirement of the first order phase transition is vastly  related to the observed baryon number and lepton number in today's universe \cite{Kuzmin:1985mm,Riotto:1999yt,Morrissey:2012db}.  This pushes for additional scalar(s) along with the SM Higgs doublet. The requirement of additional scalar can also be justified as in the SM, there is no cold dark matter(DM) candidate and it can  also provide the much needed stability of the electroweak vacuum of the SM, and its various seesaw extensions \cite{Jangid:2020dqh,Jangid:2020qgo,Bandyopadhyay:2020djh,sjak}.  It is also interesting to see, if  these  additional scalars are consistent with various constraints coming from the  collider experiments, dark matter relic abundance along with the requirement of the strongly first order phase transition. The inspections regarding the first order phase transition exist in various possible extensions of the SM  viz. in the supersymmetric scenarios \cite{Carena1996,Quiros:1999jp,Delepine1996,Laine1998,Grojean2005,Huber2001,Huber2006,Kanemura:2012hr,Cheung:2012pg,Kanemura:2011fy,Chiang:2009fs,Carena:2012np,Giudice,Stanley}, inert doublet model\cite{Chowdhury2012,Borah2012,Gil2012,AbdusSalam2014,Cline2013}, scalar singlet \cite{Vaskonen:2016yiu,Profumo:2007wc,Ahriche:2007jp,Espinosa:2011ax,Cline:2012hg,Cline:2013gha,Barger:2008jx,Gonderinger:2012rd,Ahriche:2012ei,Brauner:2016fla,Carena:2019une,Carena:2018vpt,Parsa,ghorbani2019strongly,Tenkanen,Schicho,Fuyuto:2014yia}, 2HDM \cite{Huber2013,Kanemura:2005cj,Andersen:2017ika,Barman:2019oda}, triplet\cite{Patel:2012pi,Kazemi:2021bzj,niemi,Cho:2021itv,Chiang:2018gsn} and multiple fields \cite{Paul:2019pgt,tripletsinglet2021,Shajiee,Hitoshi,Garcia-Pepin:2016hvs}. Some of these extensions need a revisit considering various recent experimental constraints along with the theoretical perturbative unitarity. 

The first order phase transition originates from the bubble nucleation of the true vacuum at the nucleation temperature $T_n$. These bubbles expand due to the pressure difference between the true and the false vacua and the broken phase extends to the unbroken phase outside \cite{LINDE1983421}. During such bubble expansion in the first order phase transition, bubbles collide and create Gravitational Wave (GW) \cite{LINDE1983421,Witten:1984rs,Hogan:1986qda,Caprini:2019egz,Gould:2019qek,Weir:2017wfa,Hindmarsh:2017gnf,Guo:2020grp,Hindmarsh:2015qjv}. Different scenarios foretelling the first order phase transition, generate different GW frequencies that can be detected by the various present and future experiments.  

It would be interesting to see, if such different extensions can be distinguished either theoretically or experimentally. We are particularly  interested in the study of  SM  extension with a  $Y=0$ $SU(2)$ triplet, which stabilize the electroweak  vacuum till the  Planck scale \cite{Jangid:2020qgo} and  compare the results with with a singlet extension. Such triplet, odd under $Z_2$, provides the much needed dark matter in terms of its neutral component, which should be $\lsim 1.2$ TeV to satisfy the DM relic abundance  \cite{Jangid:2020qgo}. The scenario is especially interesting, as it provides a charged Higgs  boson with displaced  decays, which can be detected in the LHC and the MATHUSLA \cite{Jangid:2020qgo, Bandyopadhyay:2020otm, sneha}.  In the context of supersymmetry $Y=0$ triplet is also motivated for the reappearance of TeV scale supersymmetry consistent with $125.5$ GeV Higgs boson \cite{Bandyopadhyay:2013lca,Bandyopadhyay:2015tva,Bandyopadhyay:2015oga}, predicting correct $B \to X_s \gamma$ \cite{Bandyopadhyay:2013lca, Bandyopadhyay:2014tha}, triplet charged Higgs boson\cite{Bandyopadhyay:2014vma,Bandyopadhyay:2017klv, Bandyopadhyay:2015ifm} and displaced decays of triplinos \cite{SabanciKeceli:2018fsd}. 

In this article, we explore the  possibility of  such triplet providing the much needed strongly first order phase transition and the corresponding bound on the triplet mass parameter. The compatibility with perturbative unitary at one- and two-loop level are also studied along with  the bounds from the Higgs data and DM relic. The scenario is also compared with the complex singlet extension of the SM, which is also odd under $Z_2$ \cite{quiros}. Finally, by measuring the bubble nucleation temperature along with other parameters, we estimate the signal frequencies of the GW created by the bubble collisions. Along with this, the  sound wave of the plasma and the turbulence  contribute substantially, are also considered here. Such frequencies can be detected by various future space interferometer  experiments like Big Bang Observer (BBO) \cite{Yagi:2011wg}, Laser Interferometer Space Antenna (LISA)\cite{LISA:2017pwj} and earth-based detector LIGO (LIGO) \cite{KAGRA:2013rdx,PhysRevLett.116.241103,PhysRevLett.118.221101}, and such regions are identified.

The article is organised as follows. In \autoref{ISM} and \autoref{ITM} we describe the inert singlet and inert triplet model along with the calculation of  thermal corrected potential and masses with broadly defining the regions responsible for first order phase transition. The critical temperature and the effect of the quartic couplings are discussed in \autoref{cricT}. The bounds from perturbative unitarity at one- and two- loops, DM relics are discussed in \autoref{RGb}. The frequencies for the Gravitational Waves and their detectability in various experiments for different benchmark points  are discussed in \autoref{GW}. Finally we conclude in \autoref{concl}.
 
\section{Calculation of finite temperature potential for Inert Singlet scenario } \label{ISM}
The minimal SM is extended with a complex singlet which is considered to be odd under the $Z_2$ symmetry. The SM Higgs doublet $H$ is even under the $Z_2$ symmetry and transforms as $H \rightarrow H$, whereas the singlet $S$ goes to $-S$. Being odd under $Z_2$, the neutral component of the singlet becomes the dark matter candidate. The detailed calculation of the tree-level mass spectrum and the vacuum stability analysis at zero temperature are given in \cite{Jangid:2020qgo}. The corresponding tree-level scalar potential is given by \footnote{This is our notation to use Higgs singlet interaction coupling as $\lambda_{hs}$ and this quartic coupling is defined as $\lambda_{hs}=2 \zeta^2$ in \cite{quiros}.}
\be
 V=-\mu^2  H^\dagger H+m_S^2  S^*S+\lambda_1|H^\dagger H|^2+\lambda_s |S^*S|^2+\lambda_{hs}(H^\dagger H) (S^*S), \label{Eq:2.1}
\ee

\begin{center}
	$	H
	= \left(\begin{array}{c}
	G^+   \\
	\frac{1}{\sqrt{2}}(\phi+h)+i G^0  \end{array}\right) $,
\end{center}
where neutral component $\phi$, of the SM Higgs doublet $H$, acts as the background field. However in case of SM, the field dependent masses for Higgs field $h$, Goldstone bosons $G^0$, the gauge bosons ($W^{\pm}$ and Z boson) and the dominant top quark contributes to the effective potential. The expressions for the field dependent mass contributing to effective potential from SM are given as follows;
\bea
\label{eq:2.2}
\begin{split}
	&M_h^2(\phi)  =  3\lambda_1 \phi^2 - \mu^2,   \quad M_{G^0}^2  =   \lambda_1 \phi^2 -\mu^2  \\
	& M_W^2(\phi) =  {g^2\over 4} \phi^2,  \quad 
	M_Z^2(\phi)  =  {(g^2+g'^2)\over 4}\phi^2  \\
	& M_{t}^2(\phi)  =  {y_t^2 \over 2}\phi^2,
\end{split}
\eea
where $M_t$ is defined as the top-quark mass. As the singlet does not acquire the vacuum expectation value (vev), the field dependent masses for the singlet will be given in terms of SM background field $\phi$ only. The field dependent mass of singlet contributing to the effective potential is calculated as:
\bea
\label{eq:2.3}
\begin{split}
	& M_{S}^2 (\phi)  =  m_S^2 + \frac{\lambda_{hs}}{2}\phi^2.
\end{split}
\eea

The one-loop daisy improved finite temperature effective potential can be written as \cite{Quiros:1999jp,quiros}
\be\label{vpoteq}
\rm V_{eff}(\phi,T)= V_{0}(\phi) +V_1(\phi,0)+ \Delta V_1(\phi, T) +\Delta V_{daisy/ring}(\phi,T),
\ee
where $ V_{0}(\phi) $ corresponds to the tree-level potential $ V_{\rm tree}(\phi)$,
\bea
V_{0}(\phi) =V_{\rm tree}(\phi)=\frac{-\mu^2}{2}\phi^2 + \frac{\lambda_1}{4}\phi^4.
\eea
Here, $V_1(\phi,0)$ is evaluated at one-loop at zero temperature via Coleman-Weinberg prescription \cite{Coleman}. $\Delta V_1(\phi, T) $ presents the one-loop temperature corrected potential. The potential without daisy resummation can be written as 

\be\label{eq:vtot}
\rm V_{tot}=V_{0}(\phi) +V_1(\phi,0)+ \Delta V_1(\phi, T). 
\ee

The total one-loop result ($V_{\rm eff}(\phi,T)$) includes the resummation over a subclass of thermal loops which are defined as ring diagrams or daisy diagrams and the plasma effects are explained by these ring improved one-loop effective potential \cite{Weinberg,Dolan,Kirzhnits:1976ts,Gross,Fendley:1987ef,Kapusta}. These ring diagrams mainly amounts to adding thermal corrections to bosons using $ \Delta V_B$. But this method of adding thermal corrections or resummation is not uniquely defined and  there are two different methods for adding such thermal corrections, one is Parwani  method and second one is Arnold-Espinosa method \cite{Arnold:1992rz}. In Arnold-Espinosa method,  $M^2(\phi)\rightarrow M^2(\phi,T)=\mathcal{M}^2(\phi)$ is done only for the cubic term as in Eq: \autoref {eq:vtot} and not for every term of the effective potential  to obtain the ring-improved effective potential

\bea
\rm V_{daisy/ring}  =   V_{tot}[M^2(\phi)]+\frac{T}{12 \pi}\sum_{ bosons}(M^3(\phi)-M^3(\phi,T)), \quad \rm Arnold-Espinosa \quad method.
\eea

In case of Parwani method $M^2(\phi)\rightarrow M^2(\phi,T)=\mathcal{M}^2(\phi)$ is done for each term in the effective potential as shown below,
\bea
\hspace*{2cm} \rm V_{ring}  = V_{tot}[M^2(\phi,T)] \qquad  \qquad  \rm Parwani \quad method.
\eea

Therefore, there is a difference of two-loop order terms in these two prescriptions and can give us idea about the uncertainties in our calculations if we neglect the higher-order terms in the perturbation theory. However, for this analysis we consider Arnold-Espinosa prescription via considering thermal replacement of mass for the cubic mass terms only. Since, fermions do not contribute in the cubic term, so such replacement are ignored here.

The effective potential in high-temperature limit includes $\phi$ depending mass contributions from bosons and fermions of SM and singlet can be written as;
\bea
\label{eq:2.4}
\rm V_{eff}(\phi,T)= V_{tree}(\phi, 0)+ \Delta V_B (\phi, T) + \Delta V_F (\phi, T),
\eea
where $V_{\rm tree}(\phi, 0)$ is the tree-level potential and $\Delta V_B(\phi, T)$ is the one-loop contribution including thermal corrections from bosons. These one-loop contributions from bosons are defined as;
\bea\label{eqvs}
\begin{split}
	&\Delta V_B  =  \sum_{i=h, G, W_L, Z_L, \gamma_L, W_T, Z_T, \gamma_T, S} n_i \Delta V_i,
\end{split}
\eea
where $G \in \{G^0, G^{\pm}\}$ and $W_L, Z_L, \gamma_L, W_T, Z_T, \gamma_T$ are the longitudinal and transverse components of gauge bosons $W^{\pm}, Z$ and photon $\gamma$ with $\Delta V_i$ as detailed below
\bea
\label{eq:2.11}
\Delta V_i & = & {\frac{m_i^2(\phi)T^2}{24}-\frac{\mathcal{M}_i^3(\phi)T}{12\pi}-\frac{m_i^4(\phi)}{64\pi^2}\Big[ \rm log \frac{m_i^2(v)}{c_B T^2}-2\frac{m_i^2(v)}{m_i^2(\phi)}+\delta_{i G}log\frac{m_h^2(v)}{m_i^2(v)}\Big]}.
\eea 
As mentioned earlier, for fermions only the dominant contribution from top quark is considered in $\Delta V_F(\phi, T)$ and it does not have any cubic term, so no thermal corrections to masses are considered here as shown below, 
\bea
\label{eq:2.7}
\Delta V_F & = & n_t\Big[{\frac{m_{t}^2(\phi)T^2}{48}+\frac{m_{t}^4(\phi)}{64 \pi^2}\Big[log\frac{m_{t}^2(v)}{c_F T^2}-2\frac{m_{t}^2(v)}{m_{t}^2(\phi)}\Big]}\Big]. 
\eea  
In \autoref{eqvs} the number of degrees of freedom for SM fields and triplet bosons are given as;
\bea
\label{eq:2.9}
n_h & = & 1, n_G=3, n_S=2, n_{t}=12  \nn \\
n_{W_L} & = & n_{Z_L}=n_{\gamma_L}=1, n_{W_T}=n_{Z_T}=n_{\gamma_T}=2,
\eea 
while the coefficients $c_B$ and $c_F$ used in above \autoref {eq:2.11} and \autoref {eq:2.7} are defined by: log$c_B$=3.9076, log$c_F$=1.1350. The Debye masses used in Eq: \eqref{eq:2.11},  $\mathcal{M}_i^2(\phi)$ for $i=h, G, T, W_L, W_T, Z_T, \gamma_T$ are as follows;
\bea
\mathcal{M}_i^2 & = & m_i^2 (\phi) + \Pi_i(\phi, T),
\eea
where  $m_i^2 (\phi)$ are the field-dependent masses and $\Pi_i(\phi,T)$ are the self-energy contributions given by;
\bea
\label{eq:2.15}
\begin{split}
	& \Pi_h(\phi,T)  =  \Big(\frac{3g_2^2+g_1^{2}}{16}+\frac{\lambda_1}{2}+\frac{y_t^2}{4}+\frac{\lambda_{hs}}{12}\Big)T^2, \\
	& \Pi_{G}(\phi,T)  =  \Big(\frac{3g_2^2+g_1^{2}}{16}+\frac{\lambda_1}{2}+\frac{y_t^2}{4}+\frac{\lambda_{hs}}{12}\Big)T^2, \\
	& \Pi_T(\phi,T) =  \frac{2\lambda_s+\lambda_{hs}}{6}T^2, \\
	& \Pi_{W_L}(\phi,T)  =  \frac{11}{6}g_2^2T^2, \\
	& \Pi_{W_T}(\phi,T)  =  \Pi_{Z_T}(\Phi,T)=\Pi_{\gamma_T}=0.
\end{split}
\eea
Here, the self energy contribution to the transverse component of gauge bosons $W_T, Z_T$ and $\gamma_T$ is zero and only the longitudinal components get the self energy contribution. The Debye mass expressions for $Z_L$ and $\gamma_L$ are written as follows;
\bea
\begin{split}
	\mathcal M_{Z_L}^2 = \frac{1}{2}[m_Z^2(\phi)+ \frac{11}{6}\frac{g^2}{\cos^2\theta_W}T^2 + \Delta(\phi,T)], \\
	\mathcal M_{\gamma_L}^2 = \frac{1}{2}[m_Z^2(\phi)+ \frac{11}{6}\frac{g^2}{\cos^2\theta_W}T^2 - \Delta(\phi,T)],
\end{split}
\eea
where $\Delta$ is given as
\bea
\Delta^2(\phi,T)= m_Z^4(\phi) + \frac{11}{3} \frac{g^2 \cos^2 2\theta_W}{\cos^2\theta_W}\Big[M_Z^2(\phi)+\frac{11}{12}\frac{g^2}{\cos^2\theta_W}T^2\Big]T^2.
\eea

Now, after getting the full one-loop effective potential including thermal corrections we can do the complete numerical analysis. To see the effectiveness of plasma screening, we can first include the dominant contribution from the singlet field only by neglecting the contributions from other bosons in SM. Considering the contribution from singlet only in \autoref{eqvs}-\autoref{eq:2.11} and substituting in \autoref {eq:2.4}, we get the $\phi$ dependent part of one-loop effective potential as follows;
\bea
\label{eq:2.18}
V(\phi)& = & A(T)\phi^2 + B(T)\phi^4 +C(T)(\phi^2 + K^2(T))^\frac{3}{2}.
\eea
Here the temperature dependent coefficients are given as;
\bea
\begin{split}
	\label{eq:2.19}
	& A(T) =  -\frac{1}{2}\mu_T^2 +\frac{1}{4}\Big(\frac{\lambda_{hs}}{6}+\frac{y_t^2}{2}\Big)T^2,\\
	& B(T)  =  \frac{1}{4}\lambda_S ,\\
	& C(T)  =  -\Big({\frac{\lambda_{hs}}{2}}\Big)^\frac{3}{2}\frac{T}{6\pi},\\
	& K^2(T)  =  \frac{(\lambda_{hs}+2\lambda_s)T^2+ 6 m_S^2}{3 \lambda_{hs}},
\end{split}
\eea
where,
\bea
\mu_T^2 & = & \mu^2-\frac{\lambda_{hs}}{16\pi^2}\Big\{M_{S}^2(v)+m_{S}^2  \log\frac{c_BT^2}{m_S^2(v)}\Big\} + \frac{3}{8 \pi^2}y_t^2 m_{top}^2(v)\log\frac{m_{top}^2(v)}{c_FT^2},\\
\lambda_T & = & \lambda_1+\frac{\lambda_{hs}^2}{32 \pi^2}\log\frac{c_BT^2}{m_S^2(v)}+\frac{3}{16\pi^2}y_t^4\log\frac{m_{top}^2(v)}{c_FT^2}.
\eea
It is clear from \autoref{eq:2.18} that $\phi=0$ is the local minima at very earlier epoch,  if $A(T)>0$, which leads to following constraint;
\bea
-\frac{1}{2}\mu_T^2 +\frac{1}{4}\Big(\frac{\lambda_{hs}}{6}+\frac{y_t^2}{2}\Big)>0. 
\eea

After EW symmetry breaking $\phi=0$ is the maxima and we can find an epoch in between, where, a particular temperature $T_2$ is defined by demanding $V^{\prime \prime}(0)=0$. This will give a constraint as follows;
\bea \label{eqT2}
4A^2+9C^2K^2=0.
\eea
The $\phi=0$ is still the minimum above this temperature i.e. $T>T_2$, but there exist another maximum and minima at $\phi_{-}(T)$ and $\phi_{+}(T)$, respectively \cite{quiros}. This can be calculated by putting $V^\prime (\phi)=0$ and demanding that $\phi\neq 0$, which leads to,
\bea
\label{eq:2.24}
\phi_{\pm}(T)=\frac{1}{32 B^2}\Big(9C^2-16AB \pm |C|\sqrt{9C^2+32(2B^2K^2-AB)}\Big).
\eea
These two extrema can merge resulting $\phi_{-}(T)=\phi_{+}(T)$ at a particular temperature $T_1$ which is higher than $T_2$ but lower than the symmetric temperature ($T $). The $\phi_{-}(T)=\phi_{+}(T)$ condition from \autoref{eq:2.24} implies
\bea\label{eqT1s}
9C^2+32(2B^2K^2-AB)=0.
\eea

 Using the set of equations from \autoref{eq:2.19}-\autoref{eqT1s}, $T_1$ and $T_2$ are determined as

\bea
T_1^2  &=&  \frac{2\lambda_{T_1}(\lambda_{hs} \mu^2_{T_1}+ 2 \lambda_{T_1} m^2_S)}{{\lambda_{hs}\big((\frac{\lambda_{hs}}{6}+\frac{y^2_t}{2})\lambda_{T_1}-\frac{\lambda_{hs}^3}{64 \pi^2}-\frac{2 \lambda^2_{T1}}{3\lambda_{hs}}(\lambda_{hs}+2\lambda_s)\big)}},\\
T^2_2  &=&  \frac{1}{2\alpha}(\Lambda^2(T2)+\sqrt{\Lambda^4(T2)-16\alpha\mu^4_{T2}}),
\eea
where, 
\bea
\begin{split}
	&\alpha =  \Big(\frac{\lambda_{hs}}{6}+\frac{y^2_t}{2}\Big)^2-\frac{1}{24\pi^2}\lambda^2_{hs}\Big(\lambda_{hs}+2\lambda_s\Big),\\
	&\Lambda^2(T)  =  \frac{1}{4\pi^2}\lambda^2_{hs}m^2_S+4 \Big(\frac{\lambda_{hs}}{6}+\frac{y^2_t}{2}\Big)\mu^2_T .
\end{split}
\eea
\section{Calculation of finite temperature potential for  Inert Triplet scenario } \label{ITM}
We extend the SM with a Y=0 (hypercharge=0) real $SU(2)$ triplet which is odd under the $Z_2$ symmetry. The SM Higgs doublet $H$  as given below, transforms under $Z_2$ as $H \rightarrow H$, where as  the triplet $T$ goes to $-T$. The triplet has one complex charged component $T^\pm$ and one neutral component $T_0$ as shown below.  Being $Z_2$, the neutral component of the triplet $T_0$  becomes the dark matter candidate. The detailed tree-level mass spectrum and zero temperature vacuum stability analysis are given in  \cite{Jangid:2020qgo}. The corresponding scalar potential is given by
\be
  V=-\mu^2  H^\dagger H+m_T^2  Tr(T^\dagger T)+\lambda_1|H^\dagger H|^2+\lambda_t(Tr|T^\dagger T|)^2+\lambda_{ht}H^\dagger H Tr(T^\dagger T), \label{Eq:2.4}
\ee

\begin{center}
	$	H
	= \left(\begin{array}{c}
	G^+   \\
	\frac{1}{\sqrt{2}}(\phi+h)+i G^0  \end{array}\right) $, \qquad \qquad
	$T =\frac{1}{2} \left(
	\begin{array}{cc}
	T_0 & \sqrt{2} T^+ \\
	\sqrt{2} T^- & -T_0 \\
	\end{array}
	\right),$
\end{center}
where neutral component of SM Higgs doublet $H$, given by $\phi$, acts as the background field. However, the field dependent masses, which contribute to the effective potential in the SM includes Higgs field $h$, Goldstone bosons $G^0$, the gauge bosons ($W^{\pm}$ and Z boson) and the dominant top quark. The field dependent mass expressions contributing to effective potential from SM are calculated as follows;
\bea
\label{eq:2.2}
\begin{split}
&M_h^2(\phi)  =  3\lambda_1 \phi^2 - \mu^2,   \quad M_{G^0}^2  =   \lambda_1 \phi^2 -\mu^2  \\
& M_W^2(\phi) =  {g^2\over 4} \phi^2,  \quad 
M_Z^2(\phi)  =  {(g^2+g'^2)\over 4}\phi^2  \\
& M_{t}^2(\phi)  =  {y_t^2 \over 2}\phi^2,
\end{split}
\eea
where $M_t$ is the top-quark mass. As the triplet does not get vacuum expectation value (vev), the field dependent masses for triplet will be in terms of SM background field $\phi$ only. The neutral component $T_0$ and charged component $T^{\pm}$ both will contribute to the effective potential as we present their  field dependent masses:
\bea
\label{eq:2.3}
\begin{split}
& M_{T_0}^2 (\phi)  =  m_T^2 + \frac{\lambda_{ht}}{2}\phi^2,  \\
& M_{T^{\pm}}^2 (\phi)  =  m_T^2 + \frac{\lambda_{ht}}{2}\phi^2.
\end{split}
\eea

In this scenario, the one-loop contributions from bosons are given as;
\bea\label{eqv}
\begin{split}
&\Delta V_B  =  \sum_{i=h, G, W_L, Z_L, \gamma_L, W_T, Z_T, \gamma_T, T} n_i \Delta V_i,
\end{split}
\eea
where $G \in \{G^0, G^{\pm}\}$, $T \in \{T_0, T^{\pm}$\} and $W_L, Z_L, \gamma_L, W_T, Z_T, \gamma_T$ are defined as the longitudinal and transverse components for gauge bosons $W^{\pm}, Z$ and photon $\gamma$ and $\Delta V_i$ is given below
\bea
\label{eq:2.6}
\Delta V_i & = & {\frac{m_i^2(\phi)T^2}{24}-\frac{\mathcal{M}_i^3(\phi)T}{12\pi}-\frac{m_i^4(\phi)}{64\pi^2}\Big[log\frac{m_i^2(v)}{c_B T^2}-2\frac{m_i^2(v)}{m_i^2(\phi)}+\delta_{i G}log\frac{m_h^2(v)}{m_i^2(v)}\Big]}.
\eea 

In \autoref{eqv} the number of degrees of freedom for SM fields and triplet bosons are given as;
\bea
\label{eq:2.9}
n_h & = & 1, n_G=3, n_T=3, n_{t}=12  \nn \\
n_{W_L} & = & n_{Z_L}=n_{\gamma_L}=1, n_{W_T}=n_{Z_T}=n_{\gamma_T}=2,
\eea 
The Debye masses used in Eq: \eqref{eq:2.6} for inert Triplet scenario,  $\mathcal{M}_i^2(\phi)$ for $i=h, G, T, W_L, W_T, Z_T, \gamma_T$ are as follows;
\bea
\mathcal{M}_i^2 & = & m_i^2 (\phi) + \Pi_i(\phi, T),
\eea
where the field-dependent masses, $m_i^2 (\phi)$ and the self-energy contributions, $\Pi_i(\phi,T)$ are given by;
\bea
\begin{split}
& \Pi_h(\phi,T)  =  \Big(\frac{3g_2^2+g_1^{2}}{16}+\frac{\lambda_1}{2}+\frac{y_t^2}{4}+\frac{\lambda_{ht}}{12}\Big)T^2, \\
& \Pi_{G}(\phi,T)  =  \Big(\frac{3g_2^2+g_1^{2}}{16}+\frac{\lambda_1}{2}+\frac{y_t^2}{4}+\frac{\lambda_{ht}}{12}\Big)T^2, \\
 & \Pi_T(\phi,T) =  \frac{2\lambda_t+\lambda_{ht}}{6}T^2, \\
& \Pi_{W_L}(\phi,T)  =  \frac{11}{6}g_2^2T^2, \\
& \Pi_{W_T}(\phi,T)  =  \Pi_{Z_T}(\Phi,T)=\Pi_{\gamma_T}=0.
\end{split}
\eea
Similar to the previous scenario, only the longitudinal components get the self energy contribution while the self energy contribution to the transverse component of gauge bosons $W_T, Z_T$ and $\gamma_T$ is zero. The Debye mass expressions for $Z_L$ and $\gamma_L$ are same as the earlier and are written as follows;
\bea
\begin{split}
\mathcal M_{Z_L}^2 = \frac{1}{2}[m_Z^2(\phi)+ \frac{11}{6}\frac{g^2}{\cos^2\theta_W}T^2 + \Delta(\phi,T)], \\
\mathcal M_{\gamma_L}^2 = \frac{1}{2}[m_Z^2(\phi)+ \frac{11}{6}\frac{g^2}{\cos^2\theta_W}T^2 - \Delta(\phi,T)],
\end{split}
\eea
where $\Delta$ is given as
\bea
\Delta^2(\phi,T)= m_Z^4(\phi) + \frac{11}{3} \frac{g^2 \cos^2 2\theta_W}{\cos^2\theta_W}\Big[M_Z^2(\phi)+\frac{11}{12}\frac{g^2}{\cos^2\theta_W}T^2\Big]T^2.
\eea

Here the temperature dependent coefficients in are now given as;
\bea
\begin{split}
\label{eq:3.11}
& A(T) =  -\frac{1}{2}\mu_T^2 +\frac{1}{4}\Big(\frac{\lambda_{ht}}{4}+\frac{y_t^2}{2}\Big)T^2,\\
& B(T)  =  \frac{1}{4}\lambda_T ,\\
& C(T)  =  -\Big({\frac{\lambda_{ht}}{2}}\Big)^\frac{3}{2}\frac{T}{4\pi},\\
& K^2(T)  =  \frac{(2\lambda_{ht}+4\lambda_t)T^2+ 6 m_T^2}{3 \lambda_{ht}},
\end{split}
\eea
where,
\bea
\mu_T^2 & = & \mu^2-\frac{3\lambda_{ht}}{32\pi^2}\Big\{\sum_{i = T_0,T^{\pm}}m_{i}^2(v)+m_{T}^2 \sum_{i = T_0,T^{\pm}} \log\frac{c_BT^2}{m_i^2(v)}\Big\} + \frac{3}{8 \pi^2}y_t^2 m_{top}^2(v)\log\frac{m_{top}^2(v)}{c_FT^2},\\
\lambda_T & = & \lambda_1+\frac{3\lambda_{ht}^2}{64 \pi^2}\log\frac{c_BT^2}{m_T^2(v)}+\frac{3}{16\pi^2}y_t^4\log\frac{m_{top}^2(v)}{c_FT^2}.
\eea
It is clear from \autoref{eq:3.11} that at very earlier epoch, $\phi=0$ is the local minima if $A(T)>0$, which leads to following condition;
\bea
-\frac{1}{2}\mu_T^2 +\frac{1}{4}\Big(\frac{\lambda_{ht}}{4}+\frac{y_t^2}{2}\Big)>0. 
\eea
\begin{figure}[hbt]
	\begin{center}
		\mbox{\subfigure[]{\includegraphics[width=0.5\linewidth,angle=-0]{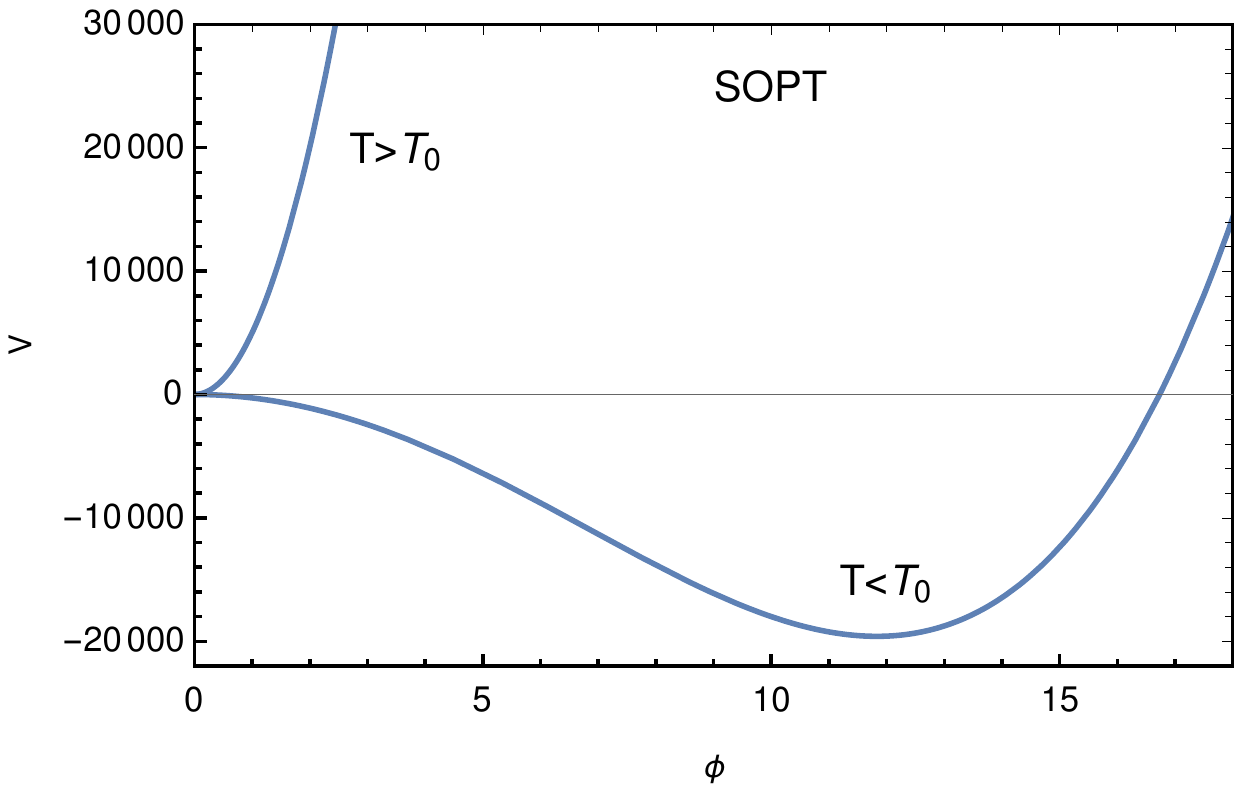}\label{f1a}}
/			\subfigure[]{\includegraphics[width=0.5\linewidth,angle=-0]{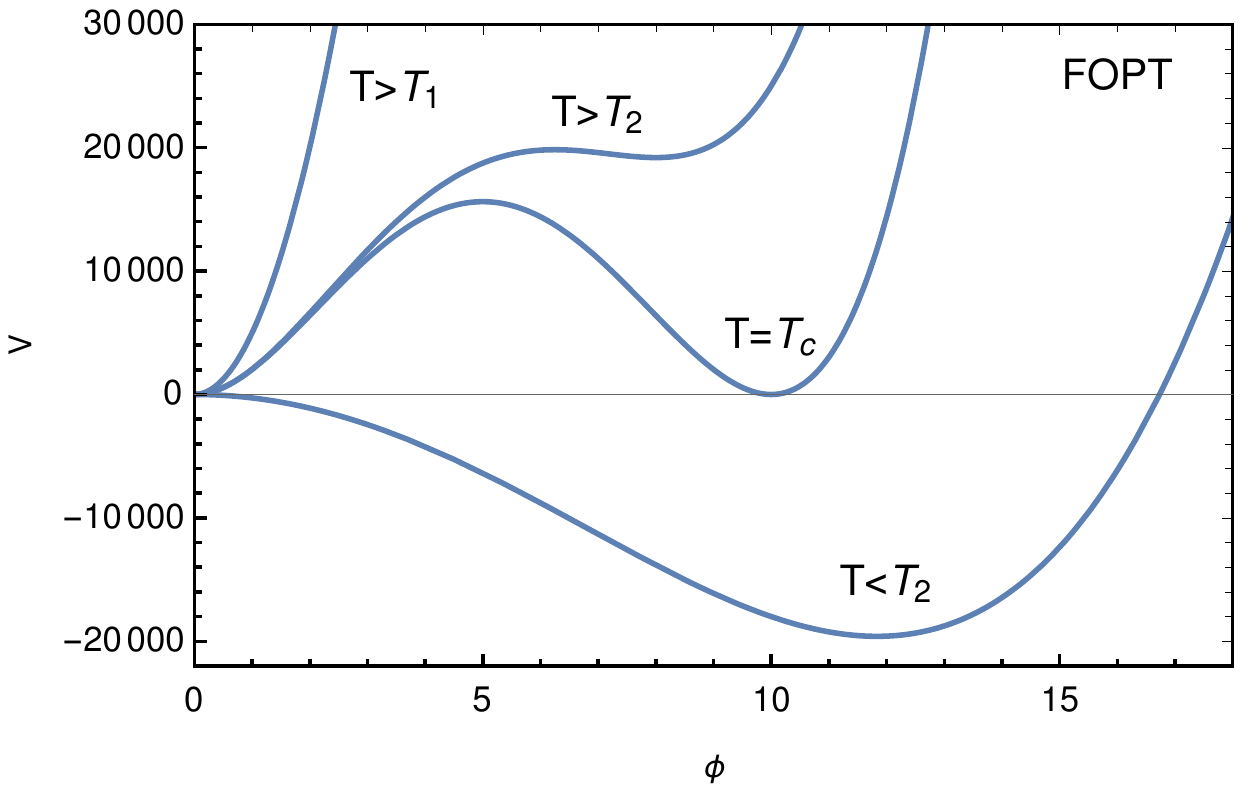}\label{f1b}}}
	\caption{(a) Describes second order phase transition and (b) shows the steps of first order phase transition.}\label{fig:PTs}
\end{center}
\end{figure}
After symmetry breaking $\phi=0$ will be maxima and in between we can find a epoch, where, 
we can define a particular temperature $T_2$ by demanding $V^{\prime \prime}(0)=0$. This will give a condition as follows;
\bea \label{eqT2}
4A^2+9C^2K^2=0.
\eea
If we go above this temperature i.e. $T>T_2$, then $\phi=0$ is still the minimum but there exist other maximum  at $\phi_{-}(T)$ and minima at $\phi_{+}(T)$, respectively \cite{quiros}. This can be achieved by putting $V^\prime (\phi)=0$ and demanding $\phi\neq 0$, that give to,
\bea
\label{eq:2.21}
\phi_{\pm}(T)=\frac{1}{32 B^2}\Big(9C^2-16AB \pm |C|\sqrt{9C^2+32(2B^2K^2-AB)}\Big).
\eea
At temperature higher than $T_2$  but lower than the symmetric temperature ($T $)  these two extrema can merge  resulting  $\phi_{-}(T)=\phi_{+}(T)$, which is defined as $T_1$. The condition for $\phi_{-}(T)=\phi_{+}(T)$ from  \autoref{eq:2.21} implies
\bea\label{eqT1}
9C^2+32(2B^2K^2-AB)=0.
\eea
Just to remind ourselves, that the temperatures higher than $T_1$, i.e. $T>T_1$ which designates the symmetric phase, has just one minimum, i.e. $\phi=0$. \autoref{fig:PTs}(b) shows the shapes of the  potential at  different thermal epoch. We shall see that these transitions can lead to the first order phase transitions as compared to the smooth second order phase transition as shown in \autoref{fig:PTs}(a).  

For $T<T_2$, $\phi=0$ is the maximum and there exits minimum at $\phi \neq 0$ which evolves towards the zero temperature minimum.  Using the set of equations from  \autoref{eq:3.11}-\autoref{eqT1} we determine  $T_1$ and $T_2$

\bea
T_1^2  &=&  \frac{6144\pi^2\lambda_{T_1}(\lambda_{ht} \mu^2_{T_1}+ 2 \lambda_{T_1} m^2_T)}{{\lambda_{ht}\big(3072\pi^2(\frac{\lambda_{ht}}{4}+\frac{y^2_t}{2})\lambda_{T_1}-27\lambda^3_{ht}-\frac{2048\pi^2\lambda^2_{T1}}{\lambda_{ht}}(2\lambda_{ht}+4\lambda_t)\big)}},\\
T^2_2  &=&  \frac{1}{\alpha}(\Lambda^2(T2)+\sqrt{\Lambda^4(T2)-65536\alpha\mu^4_{T2}}),
\eea
where, 
\bea
\begin{split}
&\alpha =  \Big(\frac{\lambda_{ht}}{4}+\frac{y^2_t}{2}\Big)-\frac{3}{128\pi^2}\lambda^2_{ht}\Big(2\lambda_{ht}+4\lambda_t\Big),\\
&\Lambda^2(T)  =  9\lambda^2_{ht}m^2_T+256 \Big(\frac{\lambda_{ht}}{4}+\frac{y^2_t}{2}\Big)\mu^2_T .
\end{split}
\eea
From temperature $T_1$ and $T_2$, we can get the idea about the nature of the phase transition.  The condition, when $T_1=T_2$ for particular value of parameters $(\lambda_{ht},\lambda_t,m_t)$, the nature of phase transition becomes second-order to first-order. The first-order phase transition happens via bubble nucleation, when the bubbles of broken phase starts nucleating in the sea of symmetric phase.  This process requires 
$T_1>T_2$, when at lower temperature $T_2$, $\phi=0$ is the maximum and there exists an deeper $\phi \neq 0$ minimum. While for $T_1 < T_2$, at lower temperature $T_1$ there is no second minima deeper than $\phi=0$ and this gives second-order phase transition. We considered only the direct one-step transitions from EW symmetric and broken minima. There is also two-step phase transition possible in a $Z_2$ symmetric scenario, where the electroweak phase transition proceeds by the spontaneous breaking of the $Z_2$ symmetry. \footnote{The spontaneous breakdown of $Z_2$ symmetry gives rise to the domain wall problem and $Z_2$ breaking transition is expected to be of second order, but not possible to  verify within perturbative effective theory \cite{Niemi:2021qvp}.} In the following subsection we investigate such effect of Higgs quartic coupling and bare masses of the extra scalars in determining the order of phase transition. 
\subsection{Effect of  scalar quartic couplings in phase transition   } 
\begin{figure}[hbt]
	\begin{center}
	\mbox{\subfigure[Singlet ]{\includegraphics[width=0.5\linewidth,angle=-0]{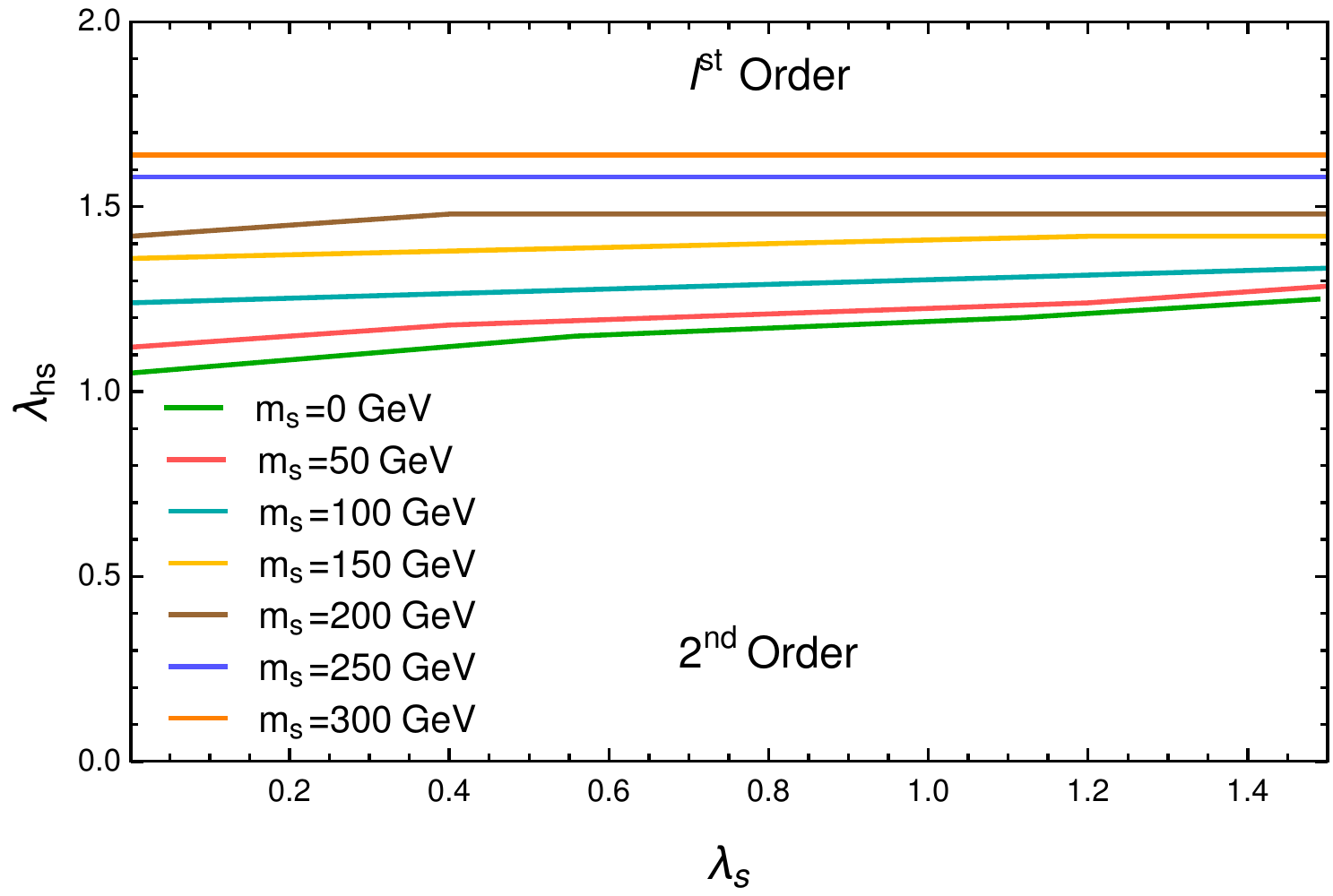}\label{f3a}}
	\subfigure[Triplet ]{\includegraphics[width=0.5\linewidth,angle=-0]{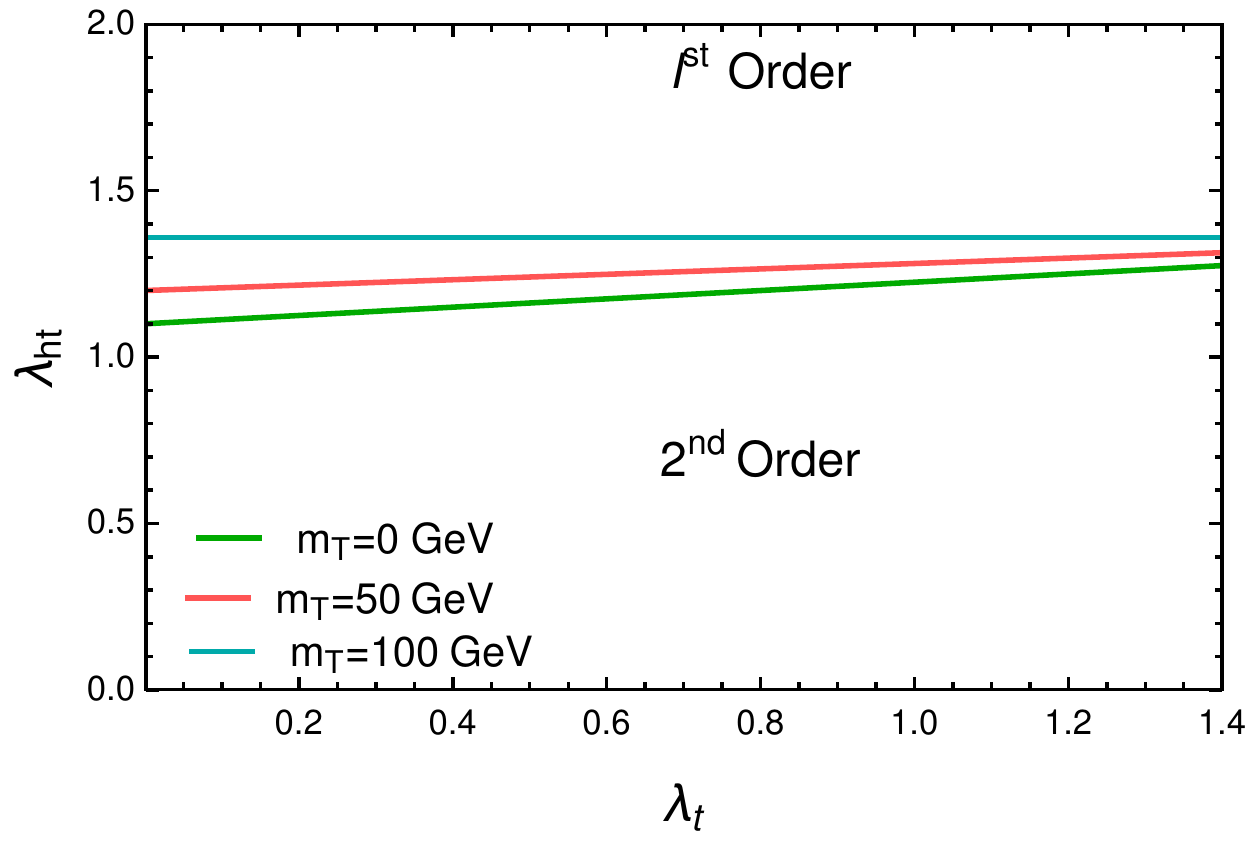}\label{f4a}}}
		\caption{Plot for the condition $T_1=T_2$ by varying parameters $\lambda_s/\lambda_t$ vs $\lambda_{hs}/\lambda_{ht}$ for singlet and triplet, respectively. The mass parameter $m_S/m_T$ is varied from 0-300 GeV and 0-100 GeV with a gap of 50 GeV for singlet and triplet, respectively. We considered the current experimental values $m_h=125.5$ GeV, $m_t=173.2$ GeV. The region above condition $T_1=T_2$ corresponds to first-order and below region for second-order phase transition. The lower green curve is for $m_S/m_T$ = 0 GeV and the upper one is for $m_S/m_T$ = 300/ 100 GeV with a gap of 50 GeV in case of singlet and triplet respectively. }\label{fig:t1=t2}
	\end{center}
\end{figure}

Here we explore the dependency of the scalar quartic couplings by presenting $T_1=T_2$ lines in $\lambda_t -\lambda_{ht}$ plane to segregate regions of first and second order phase transition. For a comparison with the complex singlet  we consider the potential of a complex singlet ($S$) extended SM as given in \cite{quiros}, where the Higgs-singlet quartic coupling $\lambda_{hs}$,  $\lambda_s$ is the self quartic coupling for the singlet and $m_S=M$ is the bare mass term for the singlet. The nature of phase transition is discussed in \autoref{fig:t1=t2} by varying the parameters $\lambda_s/\lambda_t$ vs  $\lambda_{hs}/\lambda_{ht}$ for singlet and triplet, respectively.  The coloured lines correspond to the condition $T_1=T_2$ for different values of mass parameter, which defines the cross-over from first-order to second-order phase transition.  The region above the $T_1=T_2$ condition is first-order and below one is second-order.
For this analysis we considered  the current experimental values $m_h=125.5$ GeV, $m_t=173.2$ GeV respectively \cite{10.1093/ptep/ptaa104}. The mass parameters $m_S/m_T$ are varied from 0-300 GeV and 0-100 GeV with a gap of 50 GeV for singlet and triplet, respectively. The lower lines denote $m_S/m_T$ = 0 GeV and the uppermost lines  correspond to  300 GeV and 100 GeV for singlet and triplet case.
It is evident from both \autoref{fig:t1=t2}(a) \& (b)  that as we enhance the value of the mass parameter $m_S/m_T$, the required Higgs quartic couplings  $\lambda_{hs}/\lambda_{ht}$  for $T_1=T_2$ are also enhanced, i.e.  the first order phase transition now needs higher quartic couplings. The effect of self quartic coupling is very minimal and reduces further as we increase the bare mass parameter.


In \autoref{cricT} we analyse both singlet and triplet scenarios considering all the bosonic degrees of freedom, coupling constants  within the perturbativity at  two-loop and calculating the exact critical temperature $T_c$ . 
\section{Critical Temperature and Electroweak Baryogenesis}\label{cricT}
In this section, we focus  on  electroweak baryogenesis and critical temperature during electroweak phase transition caused by  strongly first-order phase transition and the out of equilibrium condition. Inside the bubble walls a net baryon number is generated due to the first order phase transition as well as the suppressed sphaleron transition. Such  B-violating  interactions inside the bubble walls also achieve  the out of equilibrium which helps in baryogenesis. The required criteria for the strongly first-order phase transition can be defined as follows \cite{cohen,Rubakov:1996vz};
\bea
\frac{\phi_{+}(T_c)}{T_C} \geq 1,
\eea
 where $T_C$ is defined as the critical temperature and $\frac{\phi_{+}(T_c)}{T_C}$ is the parameter which defines the strength of phase transition. At critical temperature, different two minima of the potential are degenerate i.e., the same depth and such condition defines the critical temperature as;
\bea
V(0;T_C)=V(\Phi_{+}(T_C);T_C),
\eea
where $V(0;T_C)$ is the potential at $\phi=0$ minima and $V(\Phi_{+}(T_C);T_C)$ is the second minima at $\phi_{+}$.
\begin{figure}[hbt]
	\begin{center}
		\mbox{\subfigure[Singlet]{\includegraphics[width=0.5\linewidth,angle=-0]{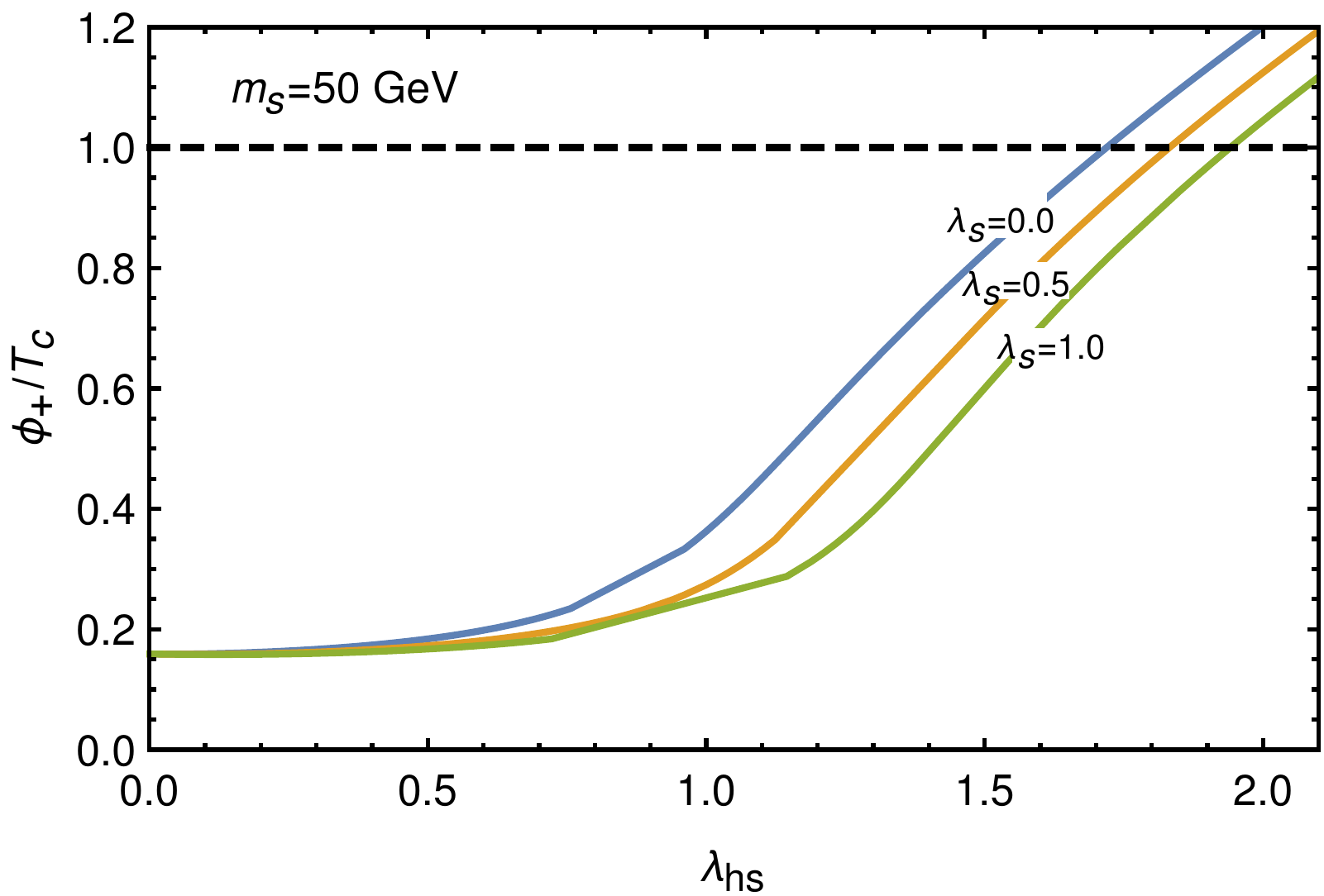}\label{f7a}}
			\subfigure[Triplet]{\includegraphics[width=0.5\linewidth,angle=-0]{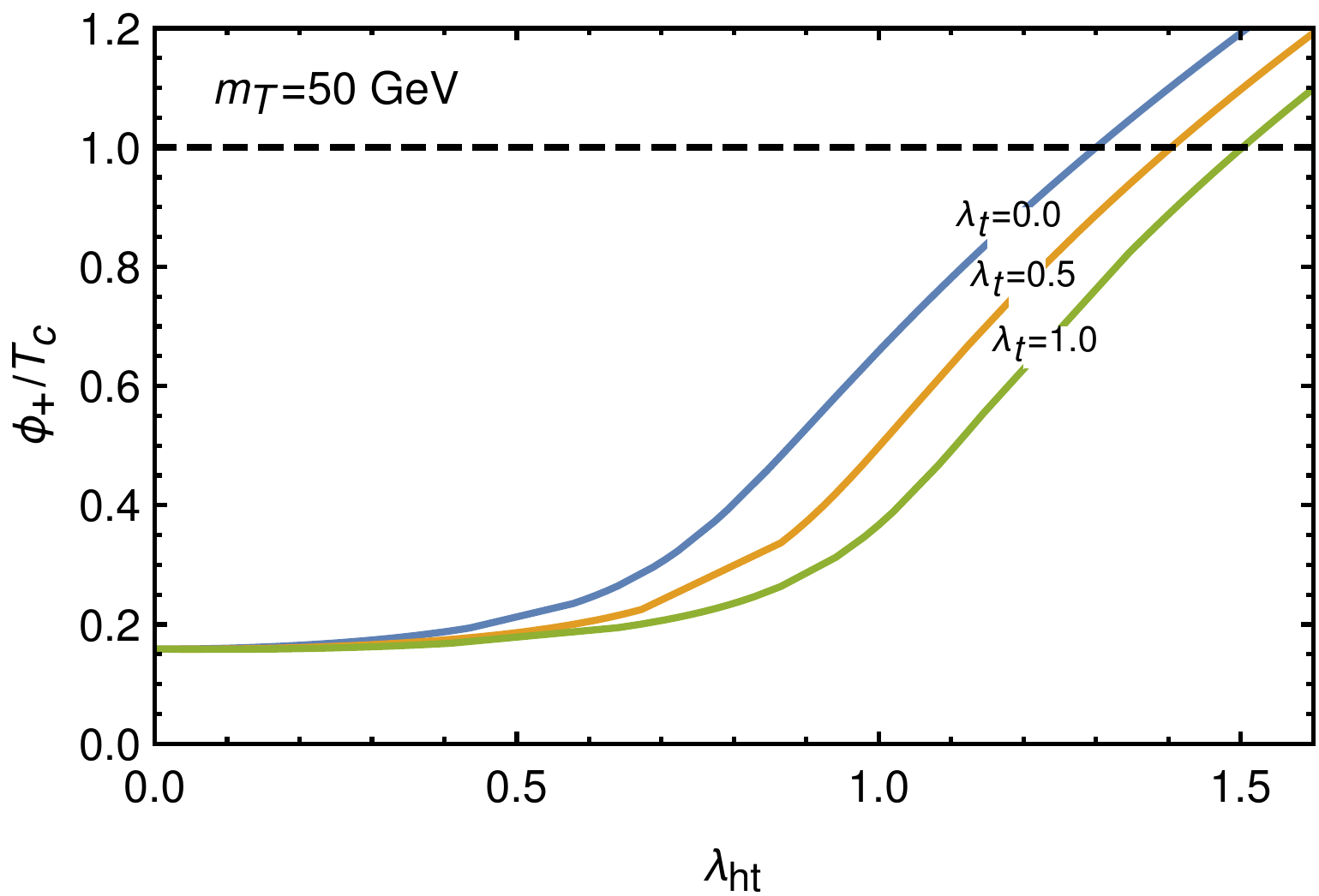}\label{f8a}}}
		\caption{Variation of $\frac{\phi_{+}(T_C)}{T_C}$  with respect to the quartic couplings $\lambda_{hs}/\lambda_{ht}$ are shown for the singlet and the triplet, respectively. The self quartic couplings of singlet and triplet i.e. $\lambda_s/\lambda_t$ are assigned three different values 0, 0.5 and 1.0, which are depicted by blue, orange and green curves, respectively for fixed mass parameter $m_S/m_T=50$ GeV with the current experimental values i.e $m_h=125.5$ GeV, $m_t=173.2$ GeV.}\label{fig:ltvar}
	\end{center}
\end{figure}
In order to calculate $\frac{\phi_{+}(T_c)}{T_C}$ we take the contributions from all the bosons i.e., SM plus the triplet Higgs boson. The variation of  $\frac{\phi_{+}(T_C)}{T_C}$  with respect to the quartic coupling $\lambda_{hs}/\lambda_{ht}$ are considered for $m_S/m_T=50$ GeV in \autoref{fig:ltvar} for the singlet and the triplet scenarios, respectively. Here self quartic coupling $\lambda_s/\lambda_t$ are set to 0, 0.5 and 1.0,  which are delineated by blue, orange and green
curves and with the  current experimental values of $m_h=125.5$ GeV, $m_t=173.2$ GeV. 

For lower values of $\lambda_{hs}/\lambda_{ht}$, the dominant contributions are mainly  from the SM fields. For the singlet case \autoref{fig:ltvar}(a) as we cross $\lambda_{hs} \gsim 1$ the effect of the singlet filed starts showing up and for $\lambda_{hs} \gsim 1.65$, we attain the regions with  
$\frac{\phi_{+}(T_c)}{T_C}>1$ . On the contrary, due to more degrees of freedom in the case of the triplet, we see such transitions much earlier i.e., $\lambda_{ht} \gsim 1.3$. One interesting point to note that with the increase of the self couplings i.e. $\lambda_s/\lambda_t$, the $\frac{\phi_{+}(T_c)}{T_C}>1$ requires higher values of the interactive Higgs couplings  i.e. $\lambda_{hs}/\lambda_{ht}$.

\begin{figure}[h!]
	\begin{center}
		\mbox{\subfigure[Singlet]{\includegraphics[width=0.5\linewidth,angle=-0]{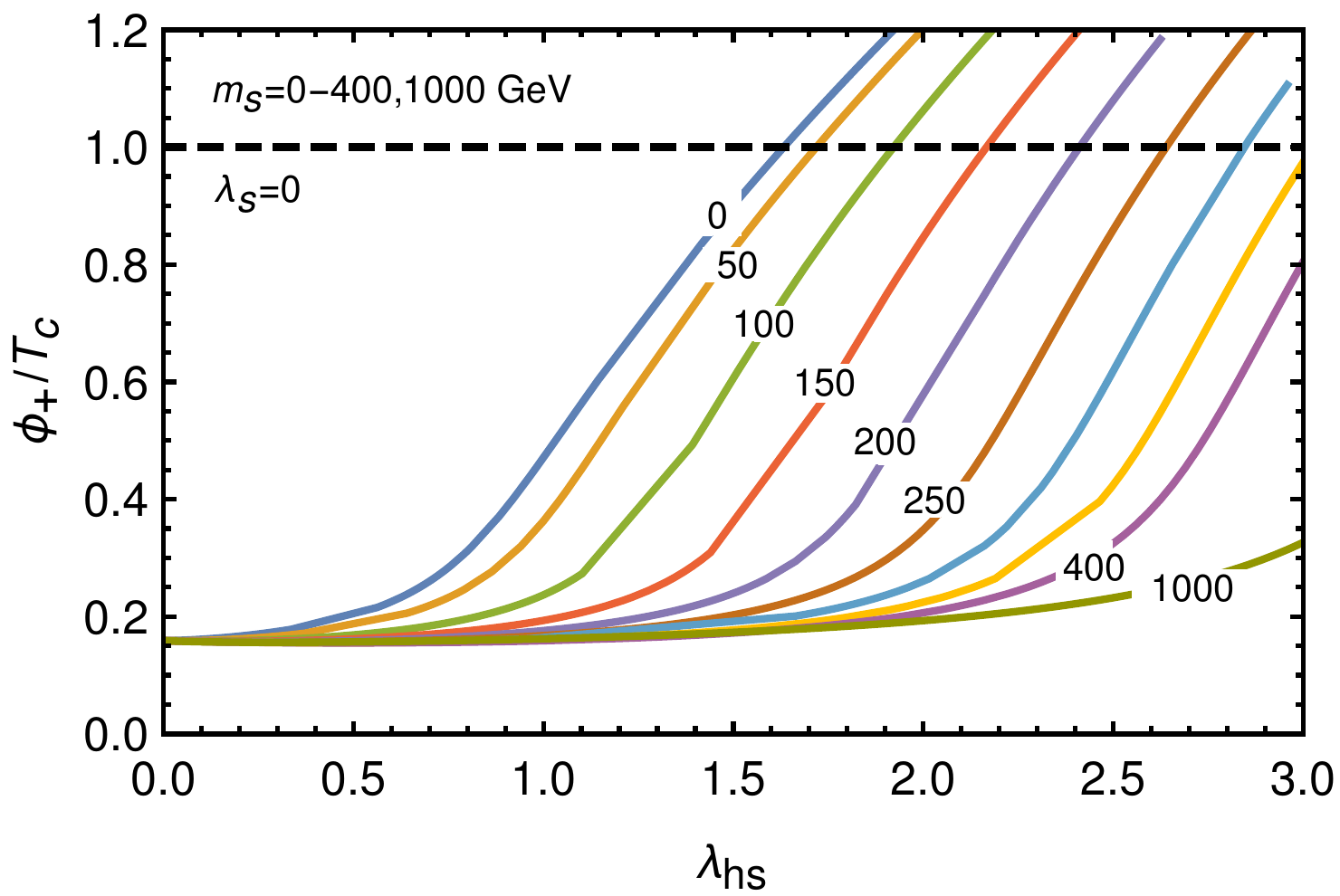}\label{f11a}}
			\subfigure[Triplet]{\includegraphics[width=0.5\linewidth,angle=-0]{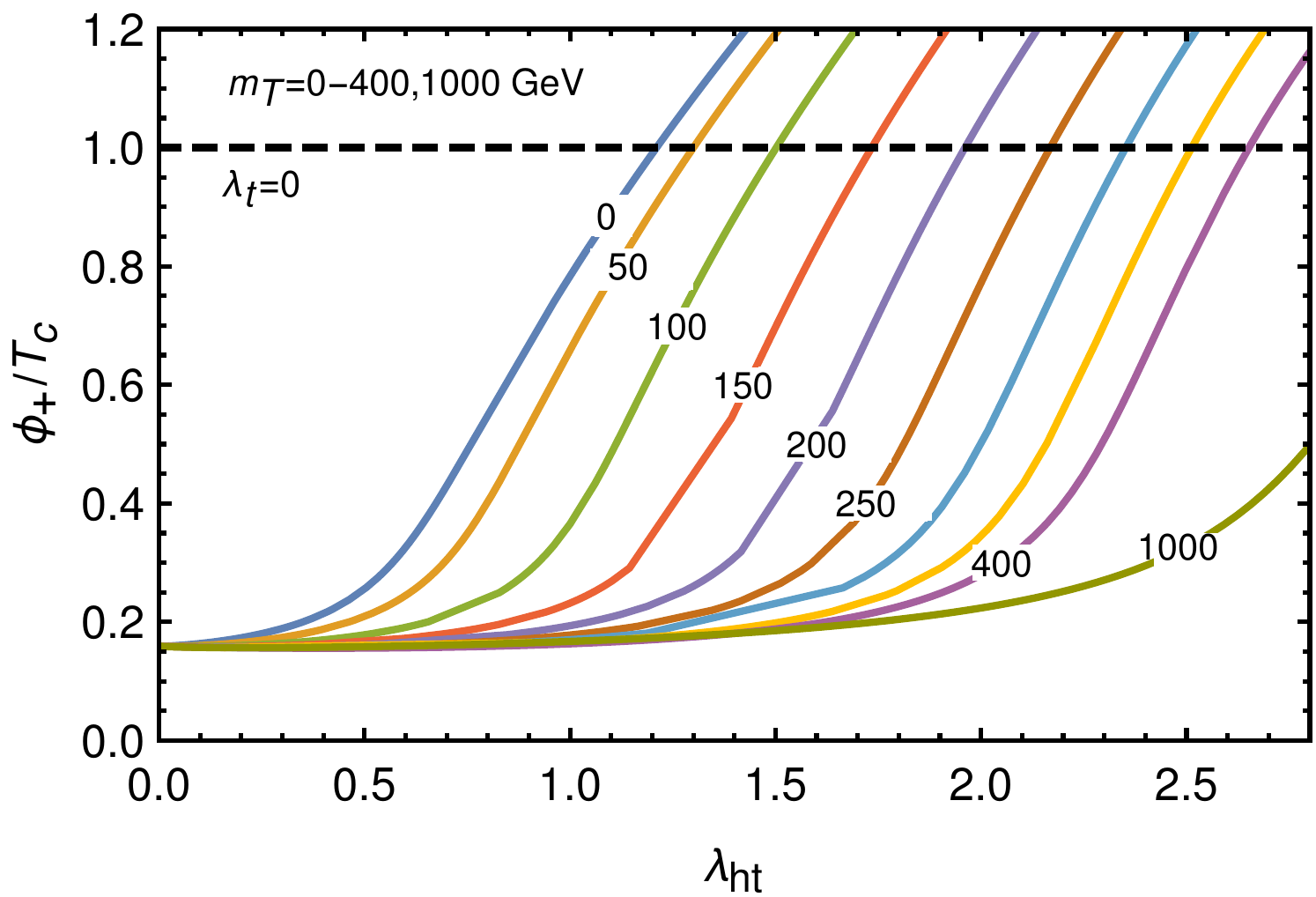}\label{f12a}}}
		\caption{Variation of  $\frac{\phi_{+}(T_C)}{T_C}$ with quartic coupling $\lambda_{hs}/\lambda_{ht}$ for the fixed values of self quartic couplings $\lambda_s/\lambda_t$=0 for the singlet and the triplet, respectively. The mass parameter $m_S/m_T$ are varied for $0-400$ GeV with a gap of 50 GeV and 1000 GeV,  respectively.}\label{fig:Mvar}
	\end{center}
\end{figure}

In  \autoref{fig:Mvar} we describe the similar variations with respect to $\lambda_{hs/ht} $for the fixed values of self quartic couplings i.e., $\lambda_s/\lambda_t$=0  to maximize $\frac{\phi_{+}(T_c)}{T_C}$ for the singlet and the triplet, respectively. We also check the dependency over the soft mass parameter $m_S/m_T$  by varying them for $0-400$ GeV with a gap of 50 GeV and 1000 GeV, respectively and are denoted by blue, orange, green, red curves and so on. It can be seen that as we increase the soft mass $m_S/m_T$ the value of  $\frac{\phi_{+}(T_c)}{T_C}$  decreases for a fixed value of  $\lambda_{hs}/\lambda_{ht}$. From  \autoref{fig:Mvar} (a) we see that after $m_S\geq 350$ GeV  getting $\frac{\phi_{+}(T_c)}{T_C}>1$ 
will require $\lambda_{hs} > 3.0$.  However, for the triplet scenario in \autoref{fig:Mvar} (b) $m_T=400$ GeV  can still give rise to $\frac{\phi_{+}(T_c)}{T_C}>1$ with $\lambda_{ht} \geq 2.6$. The couplings $\lambda_{hs}/\lambda_{ht}$  are restricted differently for the singlet and the triplet case from the perturbative unitarity as we will see in \autoref{RGb}. It is clear from \autoref{fig:ltvar}  and \autoref{fig:Mvar} that $\frac{\phi_{+}(T_C)}{T_C}$ parameter is maximum for mass parameter $m_S/m_T$=0 for fixed value of self quartic coupling of singlet/triplet and is also maximum for $\lambda_s/\lambda_t$=0 for fixed value of mass parameter.

\section{RG evolution of Scalar Quartic Couplings}\label{RGb}
The RG evaluation of the scalar quartic couplings can give sufficient constraints to the regions responsible for the first order phase transition from their perturbative unitarity. We explore such possibility via considering both one- and two-loop beta functions as explained in the following subsections.

\subsection{Constraints from one-loop perturbativity}
In this section, we study the RG evolution of  the scalar quartic couplings $\lambda_1, \lambda_t$ and $\lambda_{ht}$ with their  one-loop $\beta-$functions generated by SARAH \cite{Staub:2013tta} as given below;

\bea
\beta_{\lambda_1} &   =  &  \beta_{\lambda_1}^{\rm SM}+	\Delta \beta_{\lambda_1}^{\rm ITM}, 
\eea
\bea
\beta_{\lambda_1}^{\rm SM}  &  =   & \frac{1}{16\pi^2}\Bigg[
\frac{27}{200} g_{1}^{4} +\frac{9}{20} g_{1}^{2} g_{2}^{2} +\frac{9}{8} g_{2}^{4} -\frac{9}{5} g_{1}^{2} \lambda_1 -9 g_{2}^{2} \lambda_1 +24 \lambda_1^{2}\nonumber \\
&& +12 \lambda_1 \mbox{Tr}\Big({Y_u  Y_{u}^{\dagger}}\Big) +12 \lambda_1 {\rm Tr}\Big({Y_d  Y_{d}^{\dagger}}\Big) +4 \lambda_1 \mbox{Tr}\Big({Y_e  Y_{e}^{\dagger}}\Big)  \nonumber\\
&& -6 \mbox{Tr}\Big({Y_u  Y_{u}^{\dagger}  Y_u  Y_{u}^{\dagger}}\Big)-6 \mbox{Tr}\Big({Y_d  Y_{d}^{\dagger}  Y_d  Y_{d}^{\dagger}}\Big) -2 \mbox{Tr}\Big({Y_\ell  Y_{\ell}^{\dagger}  Y_\ell  Y_{\ell}^{\dagger}}\Big) \Bigg].\\
\rm \Delta \beta_{\lambda_1}^{ITM} & = &  8 \lambda_{ht}^2 \\
\beta_{\lambda_t} &  = &  \frac{1}{16\pi^2}\big[-24 g_2^2 \lambda_t +88 \lambda_t^2+8 \lambda_{ht}^2+\frac{3}{2}g_2^4\big],\\
\beta_{\lambda_{ht}} & = & \frac{1}{16\pi^2} \big[\frac{3}{4}g_2^4-\frac{9}{10}g_1^2\lambda_{ht}-\frac{33}{2}g_2^2\lambda_{ht}+12\lambda \lambda_{ht}+16\lambda_{ht}^2+24 \lambda_{ht}\lambda_{t}+6 y_t^2\lambda_{ht}\big],
\eea
where $\Delta \beta^{\rm ITM}_{\lambda}$ is the additional contribution to SM $\beta_{\lambda}$ from inert triplet. Since  $\frac{\phi_{+}(T_C)}{T_C}$ is maximum, where both mass parameter $m_S/m_T$ and the self quartic coupling $\lambda_s/\lambda_t$ are zero. We have chosen $\lambda_s/\lambda_t=0$ at the EW scale for our analysis and RG evolutions at one- and two-loops govern the couplings at any other scales. Hence, to maximize $\frac{\phi_{+}(T_C)}{T_C}$, we choose $\lambda_s/\lambda_t=0$ at the EW scale for further analysis. One point to note here is that the mass parameter does not enter in the running of quartic couplings thus the choice of  $\lambda_s/\lambda_t=0$ is sufficient for the perturbative unitarity. To keep the SM Higgs mass around $125.5$ GeV, we keep the SM quartic coupling  $\lambda_1=0.13$ at the EW scale. In  \autoref{tab:table1}, $\Lambda$ designates  the perturbative scale where any of the coupling crosses the perturbativity ($4\pi$). We fix quartic coupling $\lambda_{hs}/\lambda_{ht}$ at the EW scale and check the  perturbative unitarity till a particular scale $\Lambda$.  To show the effect of the top quark mass, we present the maximum values of the quartic couplings at the EW scale allowed for two different top quark masses i.e. $120.0, 173.2$ GeV, respectively for the singlet and the triplet scenarios. We see that due to larger scalar degrees of freedom triplet scenario gets more restriction than the singlet one. For example considering Planck scale perturbativity  the singlet can have  a $\lambda^{\rm max}_{hs}=0.237(0.248)$, whereas the triplet scenario gets   $\lambda^{\rm max}_{ht}=0.2180(0.2202)$ for $m_t=173.2(120.0)$ GeV.
For lower top mass the large negative contribution from the top quark slows  down the running of scalar quartic coupling  towards perturbative limit.  It can also be observed that as we demand lower scale for the perturbativity, higher values of $\lambda^{\rm max}_{hs}/\lambda^{\rm max}_{ht}$ at the EW scale can be attained. In the next subsection we would discuss such effects at the two-loop level.

\begin{figure}[t]
	\begin{center}
		\mbox{\subfigure[Singlet]{\includegraphics[width=0.5\linewidth,angle=-0]{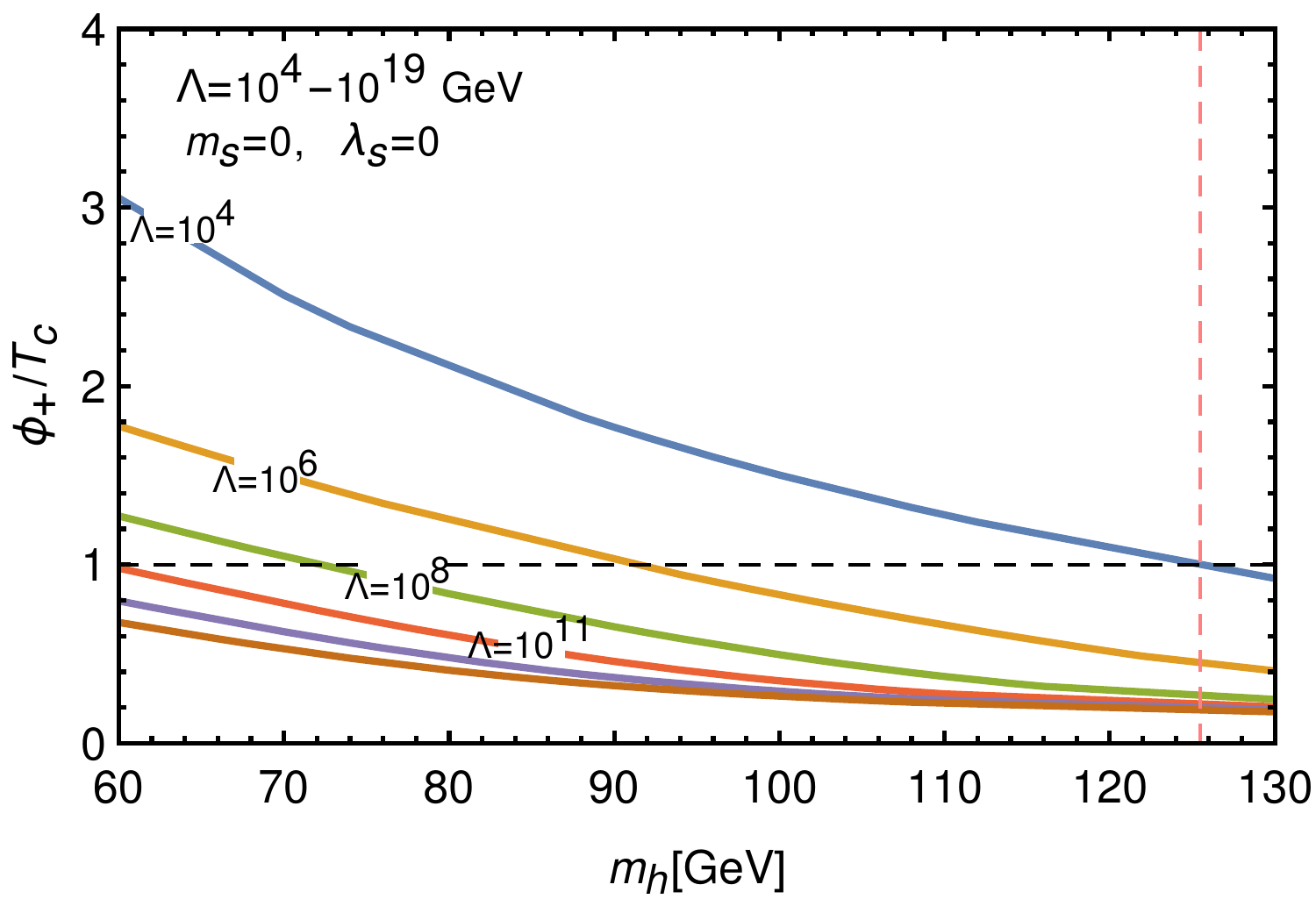}\label{f15a}}
			\subfigure[Triplet]{\includegraphics[width=0.5\linewidth,angle=-0]{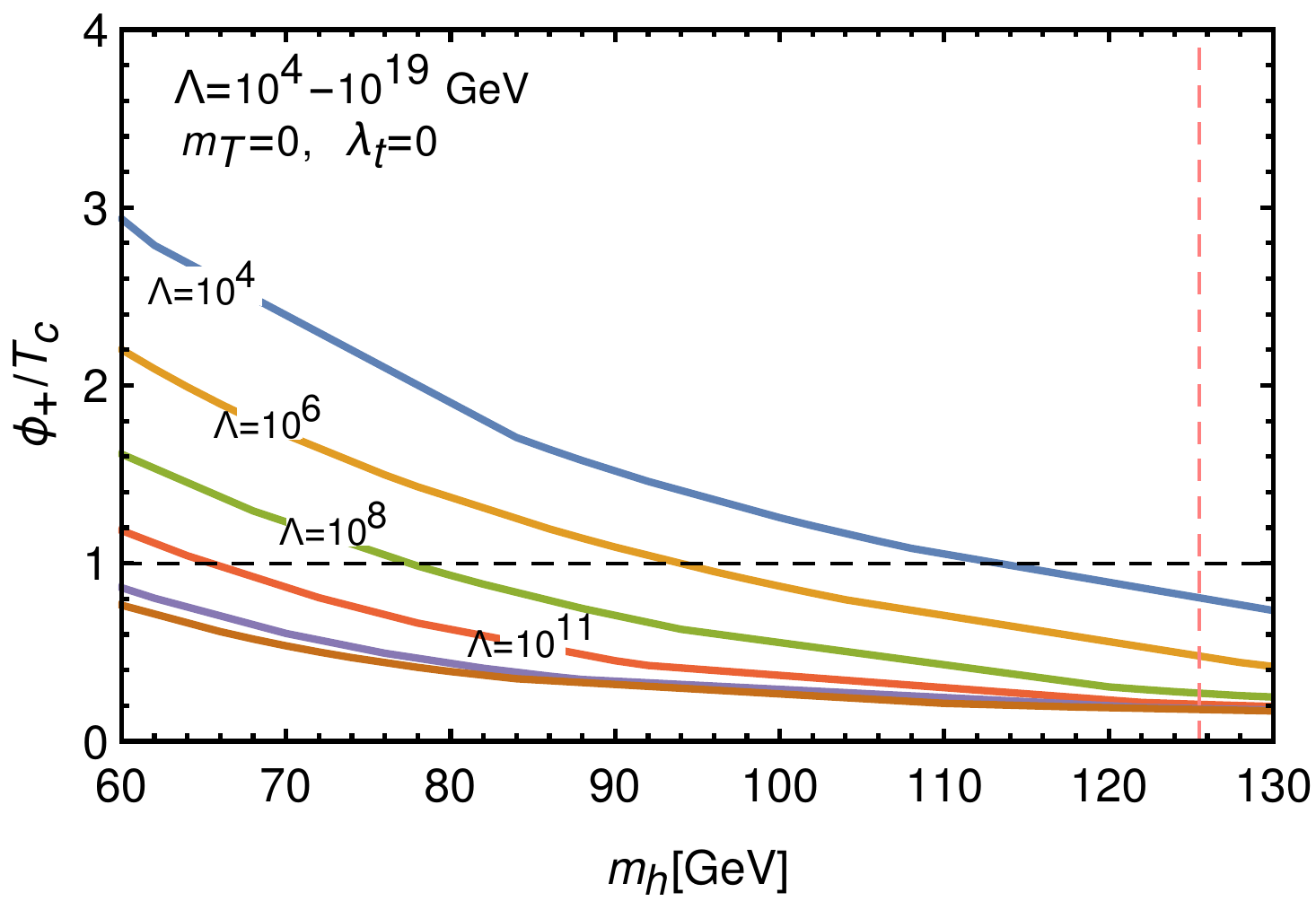}\label{f16a}}}
		\caption{Variation of $\frac{\phi_{+}(T_C)}{T_C}$ with respect to the Higgs boson mass $m_h$ in GeV for fixed initial values of $\lambda_{hs}^{max}/\lambda_{ht}^{max}$ at different perturbative scales as shown in \autoref{tab:table1}. The mass parameter $m_S/m_T$ and self quartic coupling $\lambda_s/\lambda_t$ is chosen to be zero to maximize the strength of phase transition with $m_t=173.2$ GeV.}\label{fig:mhupp}
	\end{center}
\end{figure}

In \autoref{fig:mhupp} we present the variation of $\frac{\phi_{+}(T_C)}{T_C}$ i.e. the strength of phase transition with SM Higgs boson mass for the singlet and the triplet scenario, where we consider $\lambda_{hs}^{max}/\lambda_{ht}^{max}$ as given in \autoref{tab:table1} for a given scale $\Lambda$. \autoref{fig:mhupp}(a) depicts the situation for the complex singlet extension, where  it is  evident that higher values of   $\frac{\phi_{+}(T_C)}{T_C}$ are possible with lower perturbativity scale and lower  SM Higgs boson mass. It is interesting to note that $m_h =125.5$ GeV and  $\frac{\phi_{+}(T_C)}{T_C}> 1$ is not possible even for the perturbative scale $\Lambda=10^6$ GeV and only $\Lambda=10^4$ GeV can barely satisfy the condition of the first order phase transition.  The values of $\lambda_{hs}^{max}/\lambda_{ht}^{max}$ are similar for 
$\Lambda=10^6$ GeV, however due to more degrees of freedom the triplet scenario guarantees larger  $\frac{\phi_{+}(T_C)}{T_C}$ for a given $m_h$. The perturbative scale $\Lambda=10^4$ GeV allows larger $\lambda_{hs}^{max}$ compared to $\lambda_{ht}^{max}$ resulting an enhancement of $\frac{\phi_{+}(T_C)}{T_C}$  in favour of the singlet and it barely makes it  for  $\frac{\phi_{+}(T_C)}{T_C}\simeq 1$ at $m_h=125.5$ GeV,  however the triplet case fails to achieve that at one-loop level.
\begin{figure}[h]
	\begin{center}
		\mbox{\subfigure[Singlet]{\includegraphics[width=0.5\linewidth,angle=-0]{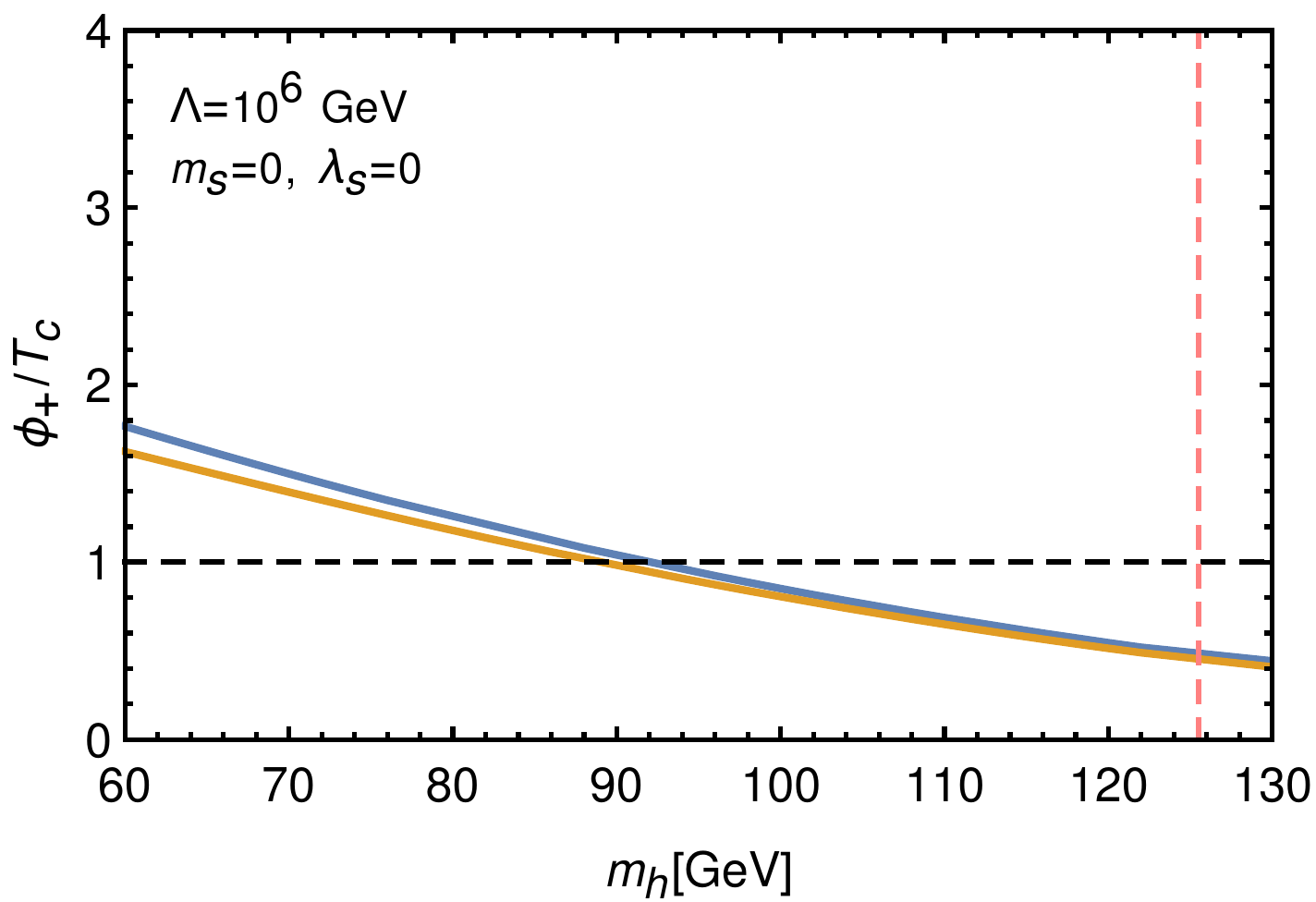}\label{f17a}}
			\subfigure[Triplet]{\includegraphics[width=0.5\linewidth,angle=-0]{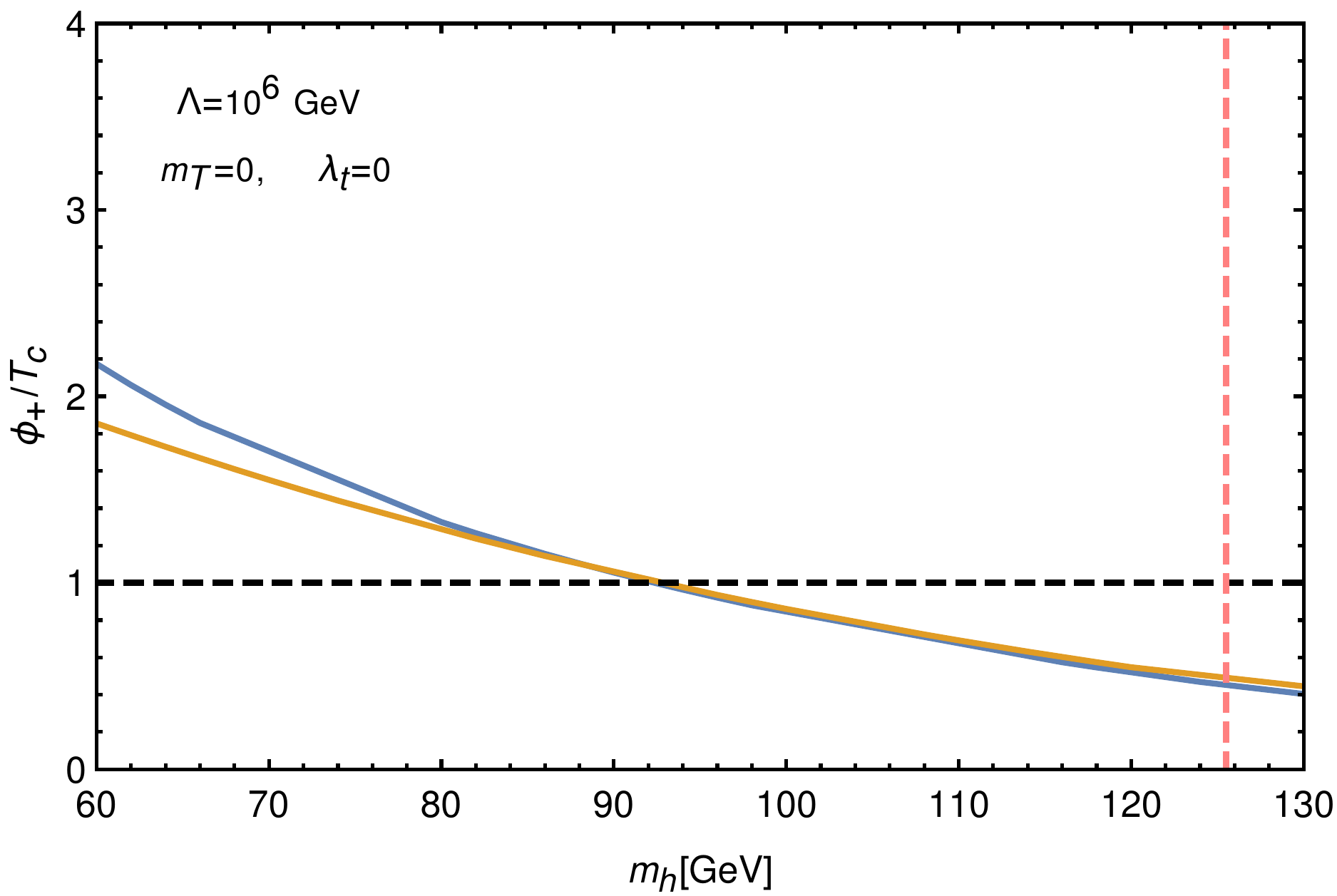}\label{f18a}}}
		\caption{Variation of $\phi_+(T_c)/T_c$ with Higgs mass $m_h$ for two different values of top mass $m_t=120.0,\, 173.2$ GeV  designated by blue and orange curves for the perturbative scale of $10^6$ GeV.}\label{fig:mh106}
	\end{center}
\end{figure}

The dependence of the top quark mass is explored in \autoref{fig:mh106} for the variation of $\frac{\phi_{+}(T_C)}{T_C}$ with the Higgs boson mass for the choices of the mass parameters $m_S/m_T$ and self quartic coupling $\lambda_s/\lambda_t$ equal to zero for the perturbative scale $\Lambda={10^6}$ GeV. The maximum allowed quartic couplings, $\lambda_{hs}^{max}/\lambda_{ht}^{max}$ are estimated using $m_t=120.0, \, 173.2$ GeV and $m_h=125.5$ GeV at the electroweak scale for the perturbative scale of $\Lambda=10^6$ GeV and $m_t=173.2$ values are described in \autoref{tab:table1}. The  blue and orange curves present  $m_t=120.0, \, 173.2$ GeV  cases, respectively for the singlet (\autoref{fig:mh106}(a)) and the triplet scenario (\autoref{fig:mh106}(b)). The maximum allowed quartic coupling $\lambda_{ht}^{max}$ is lower for the triplet due to more degrees of freedom which catalyses an early perturbative restriction. Nevertheless, the slight decrement of $\lambda_{ht}^{max}$ compared to $\lambda_{hs}^{max}$ is over powered by more degrees of freedom giving little higher values of  the $\frac{\phi_{+}(T_C)}{T_C}$ for a given $m_h$.  The upper bound on Higgs mass to avoid Baryon asymmetry washout i.e. $\frac{\phi_{+}(T_C)}{T_C}>1$ is $91.0$ GeV and $93.0$ GeV for the singlet and the triplet, respectively.  Thus, we can conclude that these upper bounds on Higgs mass from Baryon asymmetry for both cases, considering one-loop perturbativity of the quartic couplings, are not consistent with the current observed experimental Higgs mass of $125.5$ GeV.

\begin{figure}[h]
	\begin{center}
		\mbox{\subfigure[Singlet]{\includegraphics[width=0.4\linewidth,height=0.3\textheight,angle=-0]{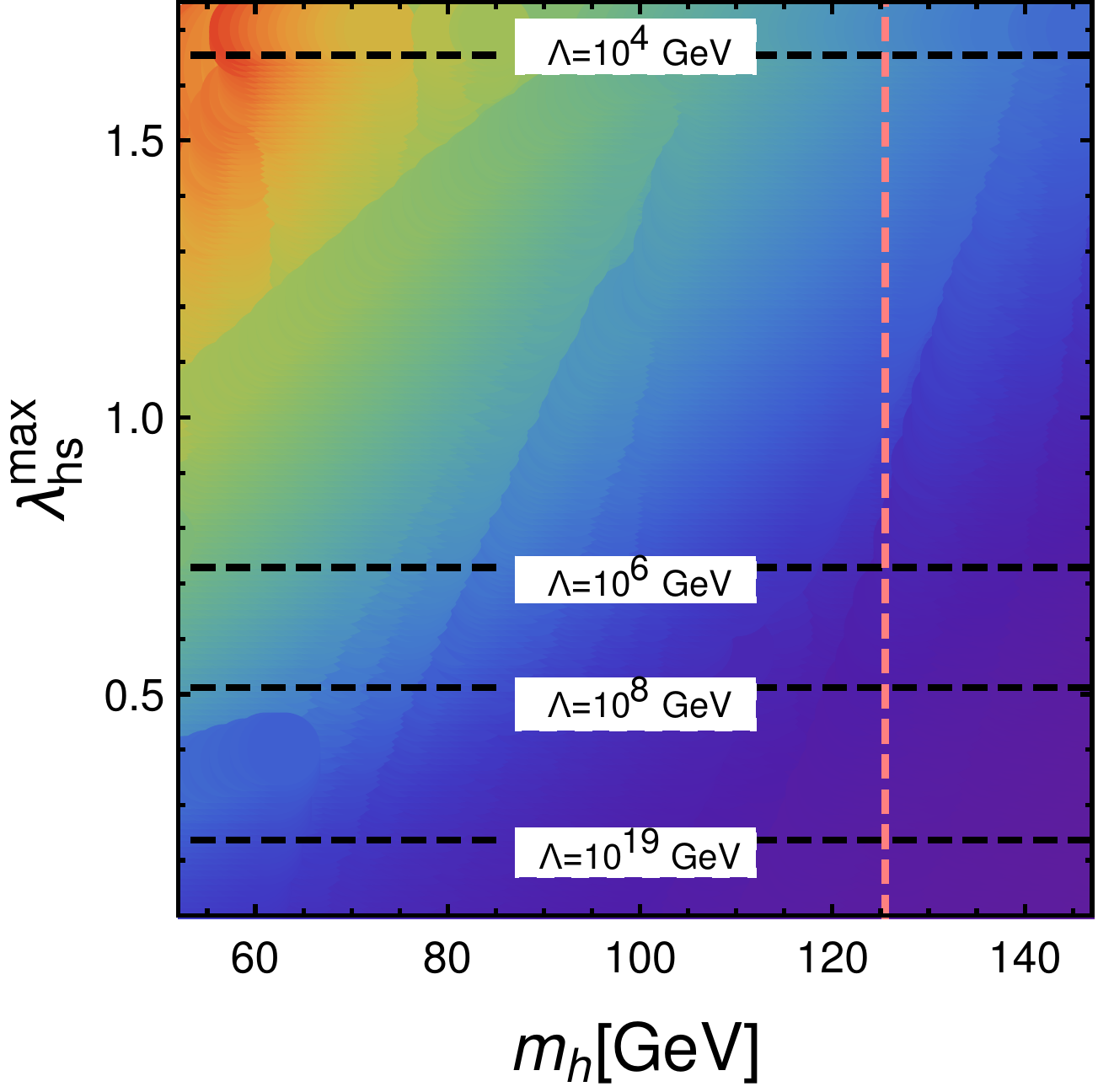}\label{f23a}}
			\subfigure[Triplet]{\includegraphics[width=0.52\linewidth,height=0.3\textheight,angle=-0]{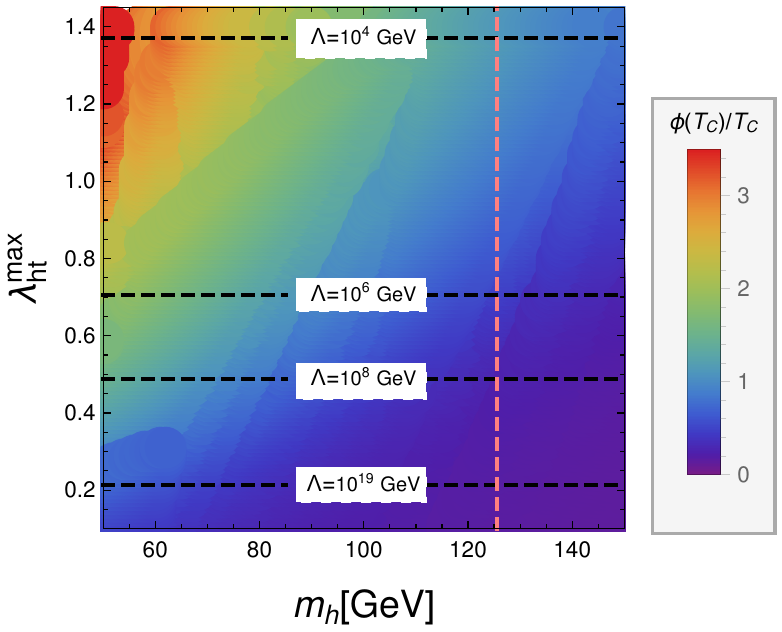}\label{f24a}}}
		\caption{Variation of $\phi_+(T_c)/T_c$ in $\lambda_{hs}^{max}/\lambda_{ht}^{max}-m_h$ plane, where $\lambda_{hs}^{max}/\lambda_{ht}^{max}$ are the maximum allowed values of quartic coupling  at different perturbative scale in GeV for the singlet and the triplet scenarios, respectively. The colour band from deep blue to red regions signify $\phi_+(T_c)/T_c$ in $0-3.5$ for both scenarios.}\label{fig:3d}
	\end{center}
\end{figure}

\begin{table}[h]
	\begin{center}
		\renewcommand{\arraystretch}{1.2}
		\begin{tabular}{|c|c|c|}
			\hline
			\multirow{3}{*}{$\Lambda$ (GeV)}&
			\multicolumn{1}{|c|}{$\lambda_{hs}=\lambda_{hs}^{max}$  } & \multicolumn{1}{|c|}{$\lambda_{ht}=\lambda_{ht}^{max}$}\\
			\cline{2-3}
			\multirow{1}{*}{}&	\multicolumn{1}{|c|}{$m_t$ (GeV) } & \multicolumn{1}{|c|}{$m_t$ (GeV)}\\
			\cline{2-3}
			&
			173.2&
			173.2 \\
			\hline \hline
			$10^4$ &  1.6545 &1.3710\\
			$10^6$ & 0.7290&0.7067\\
			$10^8$ &  0.5120 &0.4873\\
			$10^{11}$ &  0.4780&0.3477\\
			$10^{16}$ &   0.3090&0.2490\\
			$10^{19}$ &  0.2370&0.2180\\
			\hline
		\end{tabular}
		\caption{Maximum allowed values of $\lambda_{hs}, \lambda_{ht}$ from the perturbativity at one-loop for top mass i.e., $m_t=173.2$ GeV and  for the Higgs boson mass $m_h= 125.5$ GeV.}\label{tab:table1}
	\end{center}
\end{table}

Before ending the discussion of one-loop perturbativity and move on to two-loop results, we present the results in a 3-dimensional graph in \autoref{fig:3d}, where we study the variation of  $\phi_+(T_c)/T_c$ in  $\lambda_{hs}^{max}/\lambda_{ht}^{max}- m_h$ plane. The colour band of  $\phi_+(T_c)/T_c$ from deep blue to red regions signify $\phi_+(T_c)/T_c$ in $0-3.5$ for both scenarios. The self couplings  for the singlet and the triplet, and their corresponding soft masses are chosen to zero in order to enhance $\phi_+(T_c)/T_c$ . It is very apparent from \autoref{fig:3d}  that a much lower perturbative scale $\Lambda$ and lighter SM Higgs boson are preferred in order to achieve first order phase transition i.e. $\phi_+(T_c)/T_c > 1$.  Only for singlet case, $\Lambda=10^4$ GeV scale can have a first order phase transition with SM Higgs boson mass around $125.5$ GeV. The choice of zero soft masses in order to have first order phase transition for both scenarios may restrict the physical singlet and triplet scalars. However, as we explore in the following  subsection the two-loop perturbativity  gives little breather and such upper limits on the physical singlet and triplet masses are enhanced.

\subsection{Constraints from two-loop perturbativity}
For the given values of quatric couplings at the electroweak scale i.e. $\lambda_{1, },\,\lambda_{hs/ht}, \,\lambda_{s/t}$, they hit the Landau pole at the same scale considering one-loop RG-evolution. Depending on the validity scale of perturbativity certain constraints come for the maximum electroweak values of the couplings, as we have seen in \autoref{tab:table1}. For example if we choose the perturbativity scale as Planck scale i.e. $\Lambda=10^{19}$ GeV,\, $\lambda_{hs}^{max}$ and $\lambda_{ht}^{max}$ are restricted to 0.23 and 0.21 at one-loop level. The slight difference comes due to the variation of $\lambda_{t}$ affecting $\lambda_{1}$.
	
The situation changes a lot as we move to two-loop RG-evolution \autoref{betaf1}. Contrary to one-loop case, here $\lambda_1$ hits the Landau pole before $\lambda_{hs/ht},\, \lambda_{s/t}$. However, the growth of $\lambda_1$ coupling slows down at two-loop as compared to one-loop, due to the negative contributions i.e. $-312 \lambda_1^3,\, -80 \lambda_1 \lambda_{ht}^2,\, -128 \lambda_{ht}^3$ as shown in \autoref{A1}. Similarly, other quatric couplings i.e. $\lambda_{ht}, \,\lambda_{t}$ also slow down due to some extra negative contributions appearing at two-loop (see \autoref{A1}). In comparison, the singlet also suffers from the negative contributions of $-312 \lambda_1^3,\, -40 \lambda_1 \lambda_{hs}^2,\, -32 \lambda_{hs}^3$ as can be read from \autoref{B1}. However, if we look at the maximum allowed value ($\lambda_{hs}^{max}/\lambda_{ht}^{max}$) at the electroweak scale at two-loop level in \autoref{tab:table2} for the Planck scale perturbativity, the singlet can access double the value that of the triplet one. The can be understood as the triplet has more positive contributions in terms of $10g_2^4 \lambda_{ht}, \,32 g_2^2 \lambda_{ht}^2$ in $\lambda_1$, which are absent in the singlet one. Thus the growth of $\lambda_1$ in the triplet case is faster hitting the Landau pole much earlier, as compared to the singlet one. Presence of such extra postive contributions at the two-loop level for the triplet, explains the larger difference in $\lambda_{ts}^{max}$ and $\lambda_{hs}^{max}$ (See \autoref{tab:table2}) as compared to one-loop (see \autoref{tab:table1}).

In order to examine the situation  of  the possibility of the first order phase transition with the perturbativity at the two-loop level, we calculate the maximum allowed values of the quartic couplings i.e. $\lambda_{hs}^{max}/\lambda_{ht}^{max}$ with the Planck scale perturbativity ($\Lambda=10^{19}$ GeV)  as given in \autoref{tab:table2}. The slow-growing quartic coupling at the two-loop compared to one-loop enhanced the allowed couplings for $\lambda_{hs}^{max}=4.00$ and $\lambda_{ht}^{max}=1.95$ at the electroweak scale. These are now enormously amplified compared to the corresponding one-loop values $\lambda_{hs}^{max}=0.25670$ and $\lambda_{ht}^{max}=0.2180$, which result in higher values of  $\phi_+(T_c)/T_c$ strengthening the possibility of first order phase transition for both scenarios.

\begin{table}[h]
	\begin{center}
		\renewcommand{\arraystretch}{1.4}
		\begin{tabular}{|c|c|c|c|c|c|c|}
			\hline
			\multirow{1}{*}{$\Lambda$ (GeV)}&
			\multicolumn{1}{|c|}{$\lambda_{hs}^{max}$}&	\multicolumn{1}{|c|}{$\lambda^{max}_{ht}$}\\
			\hline 
			$10^{19}$ &4.00&1.95\\
			\hline
		\end{tabular}
		\caption{Maximum allowed value of quartic couplings i.e.  $\lambda_{hs}^{max}/\lambda_{ht}^{max}$ allowed at the electroweak scale for the perturbativity till  Planck scale at two-loop, for the singlet and the triplet scenarios. }\label{tab:table2}
	\end{center}
\end{table}

\begin{figure}[h]
	\begin{center}
		\mbox{\subfigure[Singlet]{\includegraphics[width=0.5\linewidth,angle=-0]{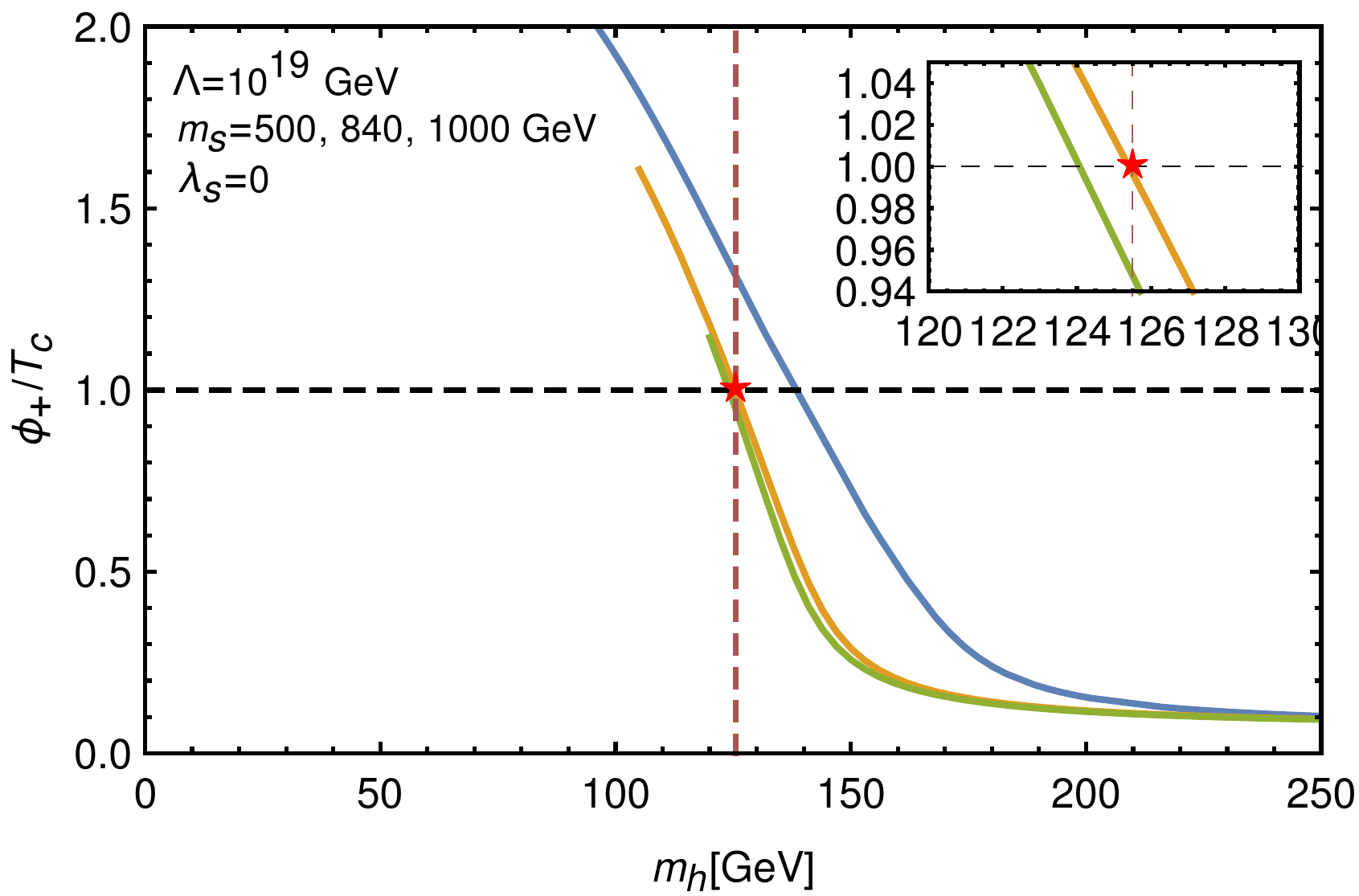}\label{f21a}}
			\subfigure[Triplet]{\includegraphics[width=0.5\linewidth,angle=-0]{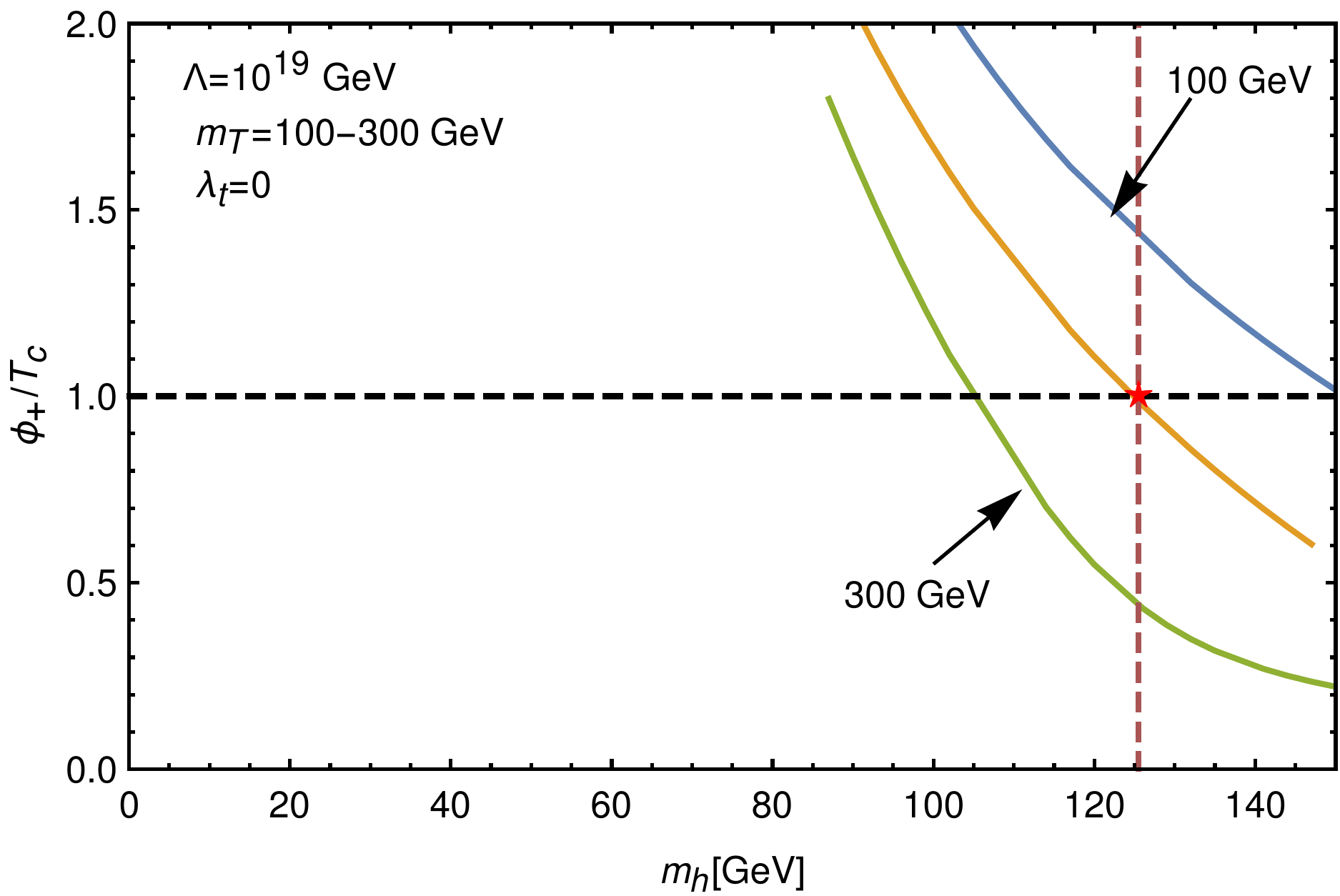}\label{f22a}}}
		\caption{Variation of $\phi_+(T_c)/T_c$ with respect to the SM Higgs boson mass $m_h$ (in GeV) for the singlet and the triplet scenarios. Three different values of mass parameters  for $m_S(m_T)=500(100), 840(200), 1000(300)$ are denoted by blue, orange and green curves, respectively.  The quartic coupling $\lambda_{hs}/\lambda_{ht}$ are fixed to their respective $\lambda_{hs}^{max}/\lambda_{ht}^{max}$ with the perturbativity at the Planck scale ( $10^{19}$ GeV) and the self coupling are chosen to be zero. }\label{fig:mtriplet}
	\end{center}
\end{figure}

Equipped with relatively larger $\lambda_{hs}^{max}/\lambda_{ht}^{max}$ for $\Lambda=10^{19}$ GeV, we now perform the variation  of   $\frac{\phi_{+}(T_C)}{T_C}$  with respect to $m_h$ in \autoref{fig:mtriplet}, where the scalar self coupling are chosen to be zero.  The mass parameters  varied for $m_S(m_T)=500(100), 840(200), 1000(300)$  are denoted by blue, orange and green curves, respectively. The red star in both the cases denotes $\frac{\phi_{+}(T_C)}{T_C}=1$ and $m_h=125.5$ GeV point.  Higher mass values diminish the 
$\frac{\phi_{+}(T_C)}{T_C}$ and push for second order phase transition for both scenarios. However, for the singlet scenario we see a maximum of $m_S=840$ GeV can still be consistent with SM Higgs boson mass as well as first order phase transition, whereas, for the triplet scenario such bounds comes for rather low mass i.e. $m_T\simeq 193$ GeV. We see an order of magnitude difference in the upper bound on the soft mass parameter in the singlet and the triplet scenarios.

\begin{figure}
	\begin{center}
		\mbox{\subfigure[Singlet]{\includegraphics[width=0.38\linewidth,angle=-0]{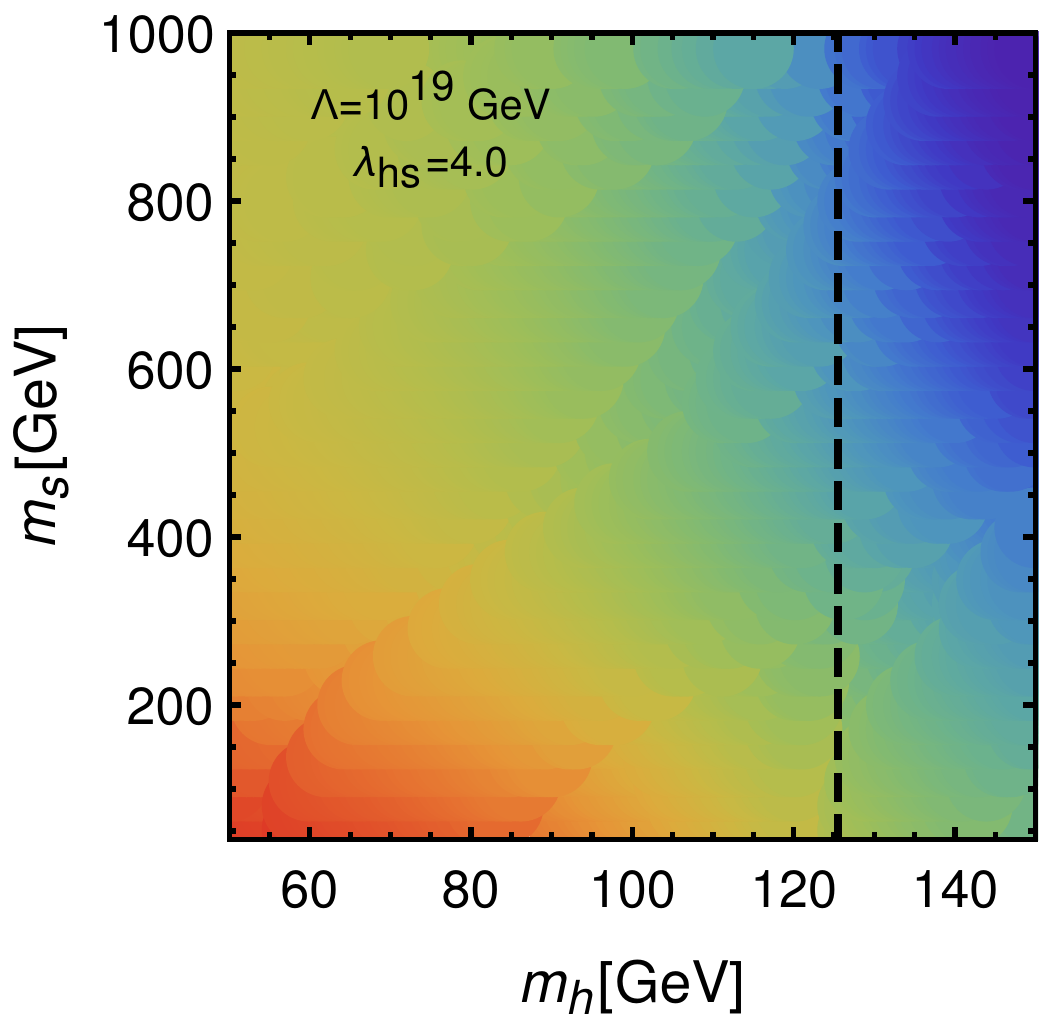}\label{f23a}}
			\subfigure[Triplet]{\includegraphics[width=0.5\linewidth,angle=-0]{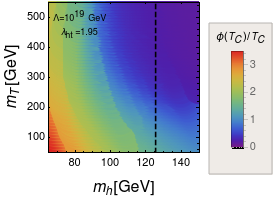}\label{f24a}}}
		\caption{Variation of $\phi_+(T_c)/T_c$ in $m_S/m_T - m_h$ for the maximum allowed values of quartic coupling $\lambda_{hs}^{max}/\lambda_{ht}^{max}$ at the electroweak scale for the perturbativity till  Planck scale ($10^{19}$) GeV for the singlet and the triplet scenarios, respectively.  The colour band from deep blue to red regions signify $\phi_+(T_c)/T_c$ in $0-3.5$ for both scenarios.}\label{fig:3d2l}
	\end{center}
\end{figure}

In \autoref{fig:3d2l} we present $\phi_+(T_c)/T_c$ in $m_S/m_T - m_h$  plane for the maximum allowed values of quartic coupling $\lambda_{hs}^{max}/\lambda_{ht}^{max}$ at the electroweak scale for the perturbativity till  Planck scale ($10^{19}$) GeV for the singlet and the triplet scenarios, respectively.  The colour band from deep blue to red regions signify $\phi_+(T_c)/T_c$ in $0-3.5$ for both scenarios. It is evident that for higher mass values $m_T \geq 193$ GeV in the triplet case stay in the deep blue region for $m_h=125.5$ GeV conferring a second order phase transition.  On the  contrary, the  singlet case one can obtain regions up to  $m_S\simeq 840$ GeV satisfying first order phase transition at $m_h=125.5$ GeV.

\subsection{Two-loop resummed potential}
The field-dependent terms in the effective potential from one-loop daisy resummation is $\mathcal{O}(g^3)$ but achieving accuracy of $\mathcal{O}(g^4)$ requires two-loop corrections when two-loop $\beta$-functions are analysed. The most efficient two-loop contributions are of the form $\phi^2 \rm log(\phi)$, which are induced by the Standard Model weak gauge boson loops\cite{Quiros:1999jp}. The diagrams contributing to the two-loop potential for the minimal standard model are given in \cite{Arnold:1992rz,Carena:2008vj,Espinosa:1996qw}. In case of inert singlet, there is no additional diagram, which contributes to the two-loop potential. Therefore, the two-loop correction for inert singlet comes only from the Standard Model and is given as follows 
	\bea
	V_2 \simeq {\log \frac{T}{\phi}}\frac{\phi^2 T^2}{32 \pi^2 }\Big[\frac{51}{16}g_2^4\Big].\label{eq:5.6}
	\eea

Similarly, in case of inert triplet, there are diagrams which give additional contributions to the two-loop potential along with the SM part. The two-loop resumed potential for the inert triplet scenario is given as
 \bea
 V_2 \simeq {\log \frac{T}{\phi}}\frac{\phi^2 T^2}{32 \pi^2  }\Big[\frac{51}{16}g_2^4 + \frac{3}{16}g_2^4\Big], \label{eq:5.7}
 \eea
 where the first term comes from SM and the second term comes from the inert triplet.

After adding these two-loop contributions to the full one-loop effective potential, the strength of phase transition enhances\cite{Carena:1997ki,Laine:2017hdk} in both cases, which actually changes the mass bounds. However, for  the singlet one-loop maximum mass required for first order phase  transition  $\simeq 909$ GeV, which already decouples and does not alter the phase transition.  With inclusion of two-loop correction, this bound is still consistent with Planck scale perturbativity and satisfies the Higgs boson mass bound within 1$\sigma$ uncertainity because of increase in the strength of phase transition. Thus, the singlet mass bound still remains same at the two-loop resumed potential. On the contrary, the effect is visible  in the case of inert triplet, owning to lower mass bound of $\simeq 310$ GeV at one-loop potential.  Inclusion of the two-loop resumed potential  inflate the mass bound slightly to $\simeq 320$ GeV, satisfying the Planck scale perturbativity and the current experimental Higgs boson mass bound.

\section{Dimensional reduction}
The effective potential at finite temperature has residual scale dependence at $\mathcal{O}(g^4)$. The cancellation of this scale dependence at $\mathcal{O}(g^4)$ requires the inclusion of two-loop thermal masses to bare masses for Higgs, singlet and the triplet i.e. $\mu, m_S$ and $m_T$, respectively. The most common way is to utilise high-temperature dimensional reduction to a three-dimensional effective field theory (3d EFT) in order to derive the full $\mathcal{O}(g^4)$ thermal effective potential. The dimensional reduction technique is an systematic approach required at high temperature to the resummations done order-by-order in power of couplings \cite{Gould:2021oba,Farakos:1994kx}. The $\mathcal{O}(g^4)$ result for the $Z_2$-symmetric real scalar theory with the two-loop results has been derived long ago. The one-loop potential to this order reads as:
\bea
V^{g^4}_{thermal}(v) & = & \frac{1}{(4\pi)^2}\Big[\frac{1}{12}g_2^4 T^2 v^2 \Big(\frac{1}{2}\log \Big(\frac{M^2(v)+ \Pi_T}{T^2}\Big)-\frac{3}{8}L_b(\Lambda)-c + \frac{1}{4}\Big)\Big] \nn \\
&& + \frac{1}{(4\pi)^2}\Big(-\frac{1}{4}g_2^2\mu^2v^2L_b(\Lambda)-\frac{1}{16}g_2^4v^4L_b(\Lambda)\Big) + \rm const,
\eea
where, we have introduced the notation using following Refs.,
\bea
c & = & - \log\Big(\frac{3 e^{\gamma/2}A^6}{4\pi}\Big) = -0.348723...,\\
L_b(\Lambda) & = & 2 \log \Big(\frac{e^{\gamma}\Lambda}{4 \pi T}\Big).
\eea
where, A is the Glaisher-Kinkelin constant and $\gamma$ is the Euler-Macheroni constant. The full two-loop effective potential expression is as follows:
\bea
\label{eq:6.4}
V_{thermal} & = & T V^{3d}_{eff} = T \Big[\frac{1}{2}\mu_3^2v_3^2 + \frac{1}{4!}g_{2,3}^2 v_3^4 - \frac{1}{3(4\pi)}(M_3^2)^{3/2} \nn \\
&& + \frac{1}{(4 \pi)^2} \Big(\frac{1}{8}g_{2,3}^2 M_3^2 - \frac{1}{24}g_{2,3}^4 v_3^2 \Big[1+2 \ln \Big(\frac{\mu_3}{3 M_3}\Big)\Big]\Big)\Big],
\eea
where, the one-to-one correspondence between Higgs quartic coupling $\lambda_3$ and $g_{2,3}$ is $\lambda_3=\frac{1}{6}g_{2,3}^2$ and the expressions for two-loop thermal masses are as follows:
\bea
M_3^2 = \mu_3^2  + \frac{1}{2} g_{2,3}^2 v_3^2.
\eea
The 3d effective parameters to the same $\mathcal{O}(g^4)$ order are given as:
\bea
g_{2,3}^2  & = &  T \Big(g_2^2(\Lambda)- \frac{3}{2 (4 \pi)^2}g_2^4 L_b(\Lambda)\Big),\\
\mu_3^2 &  = &  \mu^2(\Lambda) + \frac{1}{24} g_2^2(\Lambda) T^2 
 -\frac{1}{(4\pi)^2} \Big(\frac{1}{2}g_2^2 \mu^2 L_b(\Lambda) + \frac{1}{16} g_2^4 T^2 L_b(\mu) + \frac{1}{6} g_{2,3}^4\Big[c+ \ln\Big(\frac{3T}{\Lambda_{3d}}\Big)\Big]\Big),\\
            v_3  & = &  \frac{v}{\sqrt{T}}.
\eea
In the next section, we present the similar expressions for dimensionally reduced 3d theory for the SM extended with a inert singlet and an inert triplet.
\subsection{Singlet extension}
The scalar potential given in \autoref{Eq:2.1} for the inert singlet scenario in the dimensionally reduced 3D effective theories (DR3EFTs) is given as:
\be
 V=-\mu_3^2  H^\dagger H+m_{S,3}^2  S^*S+\lambda_{1,3}|H^\dagger H|^2+\lambda_{s,3} |S^*S|^2+\lambda_{hs,3}(H^\dagger H) (S^*S). \label{Eq:6.9}
\ee
Now, the thermal contributions to the self-energy in \autoref{eq:2.15} has the schematic form $\Pi_T \sim g^2 T^2$. If we consider the running of $\Pi_T v^2$ term, the effect is of $\mathcal{O}(g^4 T^2 v^2)$ which is not cancelled by any resummed one-loop or any other one-loop contribution to the effective potential.
This cancellation of renormalization scale dependence is only cancelled by the inclusion of explicit logarithms of the renormalization scale which only appear at two-loop level. 
At high temperature, the order of the running of tree-level parameters is same as the running of one-loop thermal mass, and this order is also similar to explicit logarithms of the renormalization scale appearing at two-loop order. All these three terms are of order $\mathcal{O}(g^4T^2)$.
The third and the fourth term in the full two-loop potential at $\mathcal{O}(g^4)$ for 3d theory will include contributions from the bosons and the fermions similar to \autoref{eqvs}.
Hence, the matching relations for the quartic couplings and the bare masses (which are the tree-level parameters) are computed as follows \cite{Schicho:2021gca,Niemi:2021qvp}:
\bea
\lambda_{1,3} & = & T\Big[\lambda_1 (\Lambda) + \frac{1}{(4\pi)^2}\Big(\frac{2-3L_b}{16}(3g_2^4+2g_2^2g_1^2+g_1^4)+ N_c L_f(y_t^4-2\lambda_1y_t^2)+L_b\Big(\frac{3}{2}(3g_2^2+g_1^2)\lambda_1-12\lambda_1^2-\frac{1}{4}\lambda_{hs}\Big)\Big)\Big] \nn ,\\
\lambda_{s,3} & = & T\Big[\lambda_s(\Lambda)-\frac{1}{(4\pi)^2}L_b(\lambda_{hs}^2+9\lambda_s^2)\Big], \nn\\
\lambda_{hs,3} & = & T\Big[\lambda_{hs}(\Lambda)+ \frac{\lambda_{hs}}{(4\pi)^2}\Big(L_b\Big(\frac{3}{4}(3g_2^2+g_1^2)-6\lambda_1-2\lambda_{hs}-3\lambda_s\Big)-N_cL_f y_t^2)\Big],
\eea
where 
\bea
L_b & = & \ln\Big(\frac{\Lambda^2}{T^2}\Big)-2[\ln(4\pi)-\gamma],\\
L_f & = & L_b + 4 \ln2.
\eea
Here, $L_b$ and $L_f$ are logarithms that arise frequently from one-loop bosonic and fermionic sum integrals with $\Lambda$ is the $\overline{MS}$ scale and $\gamma$ is the Euler-Mascheroni constant. The expressions for the two-loop mass parameters are computed as follows:
\bea
\mu_3^2 & = & (\mu_3^2)_{\rm SM} + \frac{T^2}{24}\lambda_{hs}(\Lambda)-\frac{L_b}{(4\pi)^2}\Big(\frac{1}{2}\lambda_{hs}\mu_S^2(\Lambda)\Big)+\frac{1}{(4\pi)^2}\Big(\frac{3}{4}(3g_2^2+g_1^2)L_b-N_c y_t^2L_f\Big)\Big(\frac{T^2}{24}\lambda_{hs}\Big)-\frac{T^2}{(4\pi)^2}L_b\lambda_{hs}\Big(\frac{1}{4}\lambda_1 + \frac{5}{24}\lambda_{hs} \nn \\
&& +\frac{1}{8}\lambda_s\Big)-\frac{1}{(4\pi)^2}\frac{1}{2}\lambda_{hs,3}^2\Big(c+ \ln\Big(\frac{3T}{\Lambda_{3d}}\Big)\Big),
\eea
where
\bea
(\mu_3^2)_{\rm SM} & = & - \mu^2(\Lambda) + \frac{T^2}{12}\Big(\frac{3}{4}(3g_2^2(\Lambda)+g_1^2(\Lambda))+N_cy_t^2(\Lambda)+6\lambda_1(\Lambda)\Big)+ \frac{\mu^2(\Lambda)}{(4\pi)^2}\Big(\Big(\frac{3}{4}(3g_2^2+g_1^2)-6\lambda_1\Big)L_b-N_cy_t^2L_f\Big) \nn \\
&& + \frac{T^2}{(4\pi)^2}\Big[\frac{167}{96}g_2^4 + \frac{1}{288}g_1^4 -\frac{3}{16}g_2^2g_1^2 + \frac{(1+3L_b)}{4}\lambda_1(3g_2^2+g_1^2)+L_b\Big(\frac{17}{16}g_2^4 - \frac{5}{48}g_1^4-\frac{3}{16}g_2^2g_1^2-6\lambda_1^2\Big) \nn \\
&& +\frac{1}{T^2}\Big(c+ \ln \Big(\frac{3T}{\Lambda_{3d}}\Big)\Big)\Big(\frac{39}{16}g_{2,3}^4+ 12 g_{2,3}^2h_3 - 6 h_3^2 + 9 g_{2,3}^2\lambda_{1,3}-12\lambda_{1,3}^2-\frac{5}{16}g_{1,3}^4 -\frac{9}{8}g_{2,3}^2g_{1,3}^2-2h_3'^2-3h_3''^2 \nn \\
&& +3g_{1,3}^2\lambda_{1,3}\Big) -\frac{1}{96}\Big(9L_b-3L_f-2\Big)\Big((N_c+1)g_2^4 + \frac{1}{6}Y_{2f}g_1^4\Big)n_f + \frac{N_c}{32}\Big(7L_b-L_f-2\Big)g_2^2y_t^2 \nn \\
&& -\frac{N_c}{4}(3L_b+L_f)\lambda_1y_t^2 + \frac{N_c}{96}\Big(\Big(9(L_b-L_f)+4\Big)Y_{\phi}^2-2\Big(L_b-4L_f+3\Big)(Y_q^2+Y_u^2)\Big)g_1^2 y_t^2 \nn \\
&& -\frac{N_cC_F}{6}\Big(L_b-4L_f+3\Big)g_s^2y_t^2+\frac{N_c}{24}\Big(3L_b-2(N_c-3)L_f\Big)y_t^4\Big],
\eea
with $C_F=\frac{N_c^2-1}{2N_c}=\frac{4}{3}$ and $c \sim -0.348723.$ is the fundamental quadratic Casimir of $SU(3)$. And the two-loop mass paramter for singlet is given as:
\bea
m_{S,3}^2 & = & m_S^2(\Lambda)+ T^2 \Big(\frac{1}{6}\lambda_{hs}(\Lambda)+\frac{1}{4}\lambda_s(\Lambda)\Big) -\frac{L_b}{(4\pi)^2}\Big(2\lambda_{hs}\mu^2+3 \lambda_s m_S^2\Big)+ \frac{1}{(4\pi)^2}\Big((3g_{2,3}^2+g_{1,3}^2)\lambda_{hs,3}-2\lambda_{hs,3}^2-6\lambda_{s,3}^2\Big)\nn\\
&&\Big(c+\ln\Big(\frac{3T}{\Lambda_{3d}}\Big)\Big)+\frac{T^2}{(4\pi)^2}\Big[\frac{(2+3L_b)}{24}(3g_2^2+g_1^2)\lambda_{hs}-L_b\Big(\Big(\lambda_1 + \frac{7}{12}\lambda_{hs}+\frac{1}{2}\lambda_s\Big)\lambda_{hs}+\frac{9}{4}\lambda_s^2\Big)-\frac{N_c}{12}(3L_b-L_f)y_t^2\lambda_{hs}\Big] \nn .
\eea
The other parameters which are used in the above expressions are computed as follows:
\bea
g_{2,3}^2 & = & g_2^2(\Lambda)T\Big[1+\frac{g_2^2}{(4\pi)^2}\Big(\frac{43}{6}L_b+\frac{2}{3}-\frac{(N_c+1)n_f}{3}L_f\Big)\Big], \nn \\
g_{1,3}^2 & = & g_1^2(\Lambda)T\Big[1-\frac{g_1^2}{(4\pi)^2}\frac{1}{6}\Big(L_bY_{\phi}^2+L_f Y_{2f}n_f\Big)\Big],\nn \\
h_3 & = & \frac{g_2^2(\Lambda)T}{4}\Big[1+\frac{1}{(4\pi)^2}\Big(\Big(\frac{43}{6}L_b + \frac{17}{2}-\frac{(N_c+1)n_f}{3}(L_f-1)\Big)g_2^2 +\frac{g_1^2}{2}-2N_cy_t^2+12\lambda_1\Big)\Big], \nn \\
h_3' & = & \frac{g_1^2(\Lambda)T}{4}\Big[1+\frac{1}{(4\pi)^2}\Big(\frac{3g_2^2}{2}-\frac{1}{6}\Big((L_b-1)Y_{\phi}^2+(L_f-1)Y_{2f}n_f\Big)g_1^2-2(Y_q^2+Y_u^2)N_cy_t^2+12\lambda_1\Big)\Big], \nn \\
h_3''^2 & = & \frac{g_2(\Lambda)g_1(\Lambda)T}{2}\Big[1+\frac{1}{(4\pi)^2}\Big(\Big(\frac{43}{12}L_b-1\Big)g_2^2 - \frac{Y_{\phi}^2}{3}\Big(\frac{1}{4}L_b-1\Big)g_1^2 + 4\lambda_1 +\frac{2}{3}N_cy_t^2-(L_f-1)\Big(\frac{N_c+1}{6}g_2^2 +\frac{Y_{2f}}{12}g_1^2\Big)n_f\Big)\Big].\nn \\
\eea
With the inclusion of two-loop corrections to the thermal masses specially to $m_S$ and the Higgs, the upper bound on the singlet mass coming from the first-order phase transition and the current Higgs mass bound remains the same. Since, the two-loop corrections are less for the chosen benchmark point from Planck scale perturbativity, the strength of the order of phase transition does not change significantly and the Higgs mass bound is now satisfied in the 1$\sigma$ limit with this slight change. 
\subsection{Triplet extension}
In the similar way, the scalar potential given in \autoref{Eq:2.4} for the inert triplet in the dimensionally reduced 3D effective theories (DR3EFTs) is given as:
\be
 V=-\mu_3^2  H^\dagger H+m_{T,3}^2  Tr(T^\dagger T)+\lambda_{1,3}|H^\dagger H|^2+\lambda_{t,3}(Tr|T^\dagger T|)^2+\lambda_{ht,3}H^\dagger H Tr(T^\dagger T), \label{Eq:6.10}
\ee
The matching relations for the corresponding quartic couplings are given as:
\bea
\lambda_{1,3} & = & T\Big[\lambda_1(\Lambda)+ \frac{1}{(4\pi)^2}\Big[\frac{1}{8}\Big(3g_2^4 +g_1^4 + 2 g_2^2 g_1^2 \Big)+3 L_f\Big(y_t^4 - 2 \lambda_1 y_t^2\Big)-L_b \Big(\frac{3}{16}\Big(3 g_2^4+g_1^4+2g_1^2 g_2^2\Big)\nn \\
&& -\frac{3}{2}\Big(3g_2^2 + g_1^2 - 8 \lambda_1\Big)\lambda_1 + \frac{3}{4}\lambda_{ht}^2\Big)\Big]\Big],\\
\lambda_{ht,3} & = & T \Big[\lambda_{ht}(\Lambda) + \frac{1}{(4\pi)^2}\Big[2g_2^4-3\lambda_{ht}y_t^2 L_f -L_b\Big(2 \lambda_{ht}^2 + 5 \lambda_{ht}\lambda_t + 3 g_2^4 + 6 \lambda_{ht}\lambda_1 -\frac{3}{4}\lambda_{ht}(g_1^2 + 11 g_2^2)\Big)\Big]\Big],\\
\lambda_{t,3} & = & T\Big[\lambda_t(\Lambda) + \frac{1}{(4\pi)^2} \Big[4 g_2^4 - L_b(\lambda_{ht}^2 + 11 \lambda_t^2 -12 g_2^2 \lambda_t + 6 g_2^4)\Big]\Big],
\eea
where 
\bea
L_b & = & \ln\Big(\frac{\Lambda^2}{T^2}\Big)-2[\ln(4\pi)-\gamma],\\
L_f & = & L_b + 4 \ln2.
\eea
 The matching relations for the corresponding bare mass parameters are given as \cite{Niemi:2018asa}:
 \bea
 \mu_3^2 & = & (\mu_3^2)_{\rm SM} + \frac{T^2}{8}\lambda_{ht}(\Lambda) + \frac{1}{16 \pi^2}\Big[+\frac{3}{2}\lambda_{ht} m_T^2L_b + T^2 \Big(\frac{5}{24}g_2^4 +\frac{1}{2}\lambda_{ht}g_2^2 - \frac{3}{8}\lambda_{ht}y_t^2 L_f + L_b \Big(-\frac{7}{16}g_2^4 - \frac{5}{8}\lambda_{ht}^2 \nn \\
 && - \frac{5}{8}\lambda_{ht}\lambda_t + \frac{33}{32}\lambda_{ht}g_2^2 + \frac{3}{32}\lambda_{ht}g_1^2 - \frac{3}{4}\lambda_{ht}\lambda_1\Big)+\Big(c+ \ln(\frac{3T}{\Lambda_{3d}})\Big)\Big(-\frac{3}{2}\lambda_{ht,3}^2+6\lambda_{ht,3}g_{2,3}^2- \frac{3}{4}g_{2,3}^4\Big)\Big)\Big], 
 \eea
 where 
 \bea
 (\mu_3^2)_{\rm SM} & = & -\mu^2 (\Lambda) + \frac{T^2}{16}\Big(3 g_2^2(\Lambda)+ g_1^2(\Lambda) + 4 y_t^2 (\Lambda) + 8 \lambda_1(\Lambda)\Big)+ \frac{1}{16\pi^2}\Big[-\mu^2 \Big((\frac{3}{4}(3g_2^2 + g_1^2)-6\lambda_1)L_b-3 y_t^2 L_f)\nn \\
 && +T^2 \Big(\frac{167}{96}g_2^4 + \frac{1}{288}g_1^4 - \frac{3}{16}g_2^2g_1^2+\frac{1}{4}\lambda_1(3g_2^2+g_1^2)+L_b\Big(\frac{17}{16}g_2^4 - \frac{5}{48}g_1^4 -\frac{3}{16}g_2^2g_1^2 + \frac{3}{4}\lambda_1(3g_2^2 + g_1^2)-6\lambda_1^2\Big) \nn \\
 && + \Big(c+ \ln(\frac{3T}{\Lambda_{3d}})\Big) \Big(\frac{39}{16}g_{2,3}^4 -\frac{5}{16}g_{1,3}^4 -\frac{9}{8}g_{2,3}^2g_{1,3}^2+ 12 g_{2,3}^2 h_3 - 6h_3^2 -2 h_3'^2-3h_3''^2+3\lambda_{1,3}(3g_{2,3}^2+g_{1,3}^2)-12\lambda_{1,3}^2\Big)\nn\\
 && -y_t^2\Big(\frac{3}{16}g_2^2 + \frac{11}{48}g_1^2+2 g_s^2\Big)+(\frac{1}{12}g_2^4+\frac{5}{108}g_1^4)N_f + L_f \Big(y_t^2\Big(\frac{9}{16}g_2^2 + \frac{17}{48}g_1^2+2g_s^2-3\lambda_1\Big)+\frac{3}{8}y_t^4-(\frac{1}{4}g_2^4+\frac{5}{36}g_1^4)N_f\Big)\nn \\
 && + \ln(2)\Big(y_t^2\Big(-\frac{21}{8}g_2^2-\frac{47}{72}g_1^2+\frac{8}{3}g_s^2+9\lambda_1\Big)-\frac{3}{2}y_t^4 +(\frac{3}{2}g_2^4+\frac{5}{6}g_1^4)N_f\Big)\Big)\Big],
 \eea
 and
 \bea
 m_{T,3}^2 & = & -m_T^2 + T^2\Big(\frac{1}{6}\lambda_{ht}(\Lambda)+\frac{5}{12}\lambda_t(\Lambda)+\frac{1}{2}g_2^2(\Lambda)\Big)\frac{1}{16\pi^2}\Big[-(6g_2^2-5\lambda_t)m_T^2L_b + 2 \mu^2\lambda_{ht}L_b + T^2 \Big(\big(\frac{71}{18}+\frac{2}{9}N_f\big)g_2^4+\frac{5}{3}\lambda_tg_2^2 \nn \\
 && + \frac{1}{4}\lambda_{ht}g_2^2  + \frac{1}{12}\lambda_{ht}g_1^2 +L_b\Big(\frac{5}{12}g_2^4-\frac{3}{4}\lambda_{ht}^2 -\frac{55}{12}\lambda_t^2 + \frac{11}{8}\lambda_{ht}g_2^2+\frac{1}{8}\lambda_{ht}g_1^2+5 \lambda_t g_2^2 -\frac{5}{6}\lambda_{ht}\lambda_t-\lambda_{ht}\lambda_1\Big)\nn \\
 && +\Big(c+ \ln(\frac{3T}{\Lambda_{3d}})\Big)\Big(-2 \lambda_{ht,3}^2 -10 \lambda_{t,3}^2 + \lambda_{ht,3}(3g_{2,3}^2+ g_{1,3}^2)+20\lambda_{t,3}g_{2,3}^2-3g_{2,3}^4+24 g_{2,3}^2\delta_3 -24 \delta_3^2 + 8 g_{2,3}^2\delta_3'-16 \delta_3 \delta_3'\nn\\
 &&-16\delta_3'^2\Big)-L_f\Big(\frac{1}{2}\lambda_{ht}y_t^2 + \frac{2}{3}g_2^4 N_f\Big)+\ln(2)\Big(3\lambda_{ht}y_t^2 + 4 g_2^4 N_f\Big)\Big)\Big].
 \eea
Below are the expressions for the quantities which are used above:
\bea
g_{2,3}^2 & = & g_2^2(\Lambda)T\Big[1+\frac{g_2^2}{(4\pi)^2}\Big(\frac{44-N_d-2N_t}{6}L_b +\frac{2}{3}-\frac{4N_f}{3}L_f\Big)\Big],\\
g_{1,3}^2 & = & g_1^2(\Lambda)T\Big[1+\frac{g_1^2}{(4\pi^2)}\Big(-\frac{N_d}{6}L_b-\frac{20 N_f}{9}L_f\Big)\Big], \\
h_3 & = & \frac{g_2^2(\Lambda)T}{3}\Big(1+\frac{1}{(4\pi)^2}\Big[\Big(\frac{44-N_d-2N_t}{6}L_b+\frac{53}{6}-\frac{N_d}{3}-\frac{2N_t}{3}-\frac{4N_f}{3}(L_f-1)\Big)g_2^2+\frac{g_1^2}{2}-6y_t^2+12\lambda_1+8\lambda_{ht}\Big]\Big), \nn \\
h_3' & = & \frac{g_1^2(\Lambda)T}{4}\Big(1+ \frac{1}{(4\pi)^2}\Big[\frac{3g_2^2}{2}+\Big(\frac{1}{2}-\frac{N_d}{6}(2+L_b)-\frac{20 N_f}{9}(L_f-1)\Big)g_1^2-\frac{34}{3}y_t^2+12\lambda_1\Big]\Big), \\
h_3'' & = & \frac{g_2(\Lambda)g_1(\Lambda)T}{2}\Big(1+\frac{1}{(4\pi)^2}\Big[-\frac{5+N_d}{6}g_2^2 + \frac{3-N_d}{6}g_1^2 + L_b (\frac{44-N_d}{12}g_2^2-\frac{N_d}{12}g_1^2) \nn\\
&& -N_f(L_f-1)\Big(\frac{2}{3}g_2^2 + \frac{10}{9}g_1^2\Big)+2y_t^2 + 4\lambda_1\Big]\Big),\\
\delta_3 & = & \frac{1}{2}g_2^2(\Lambda)T\Big(1+\frac{1}{(4\pi)^2}\Big[\lambda_{ht}+8\lambda_t+g_2^2\Big(\frac{16-N_d-2N_t}{3}-\frac{4}{3}N_f(L_f-1)+L_b\frac{44-N_d-2N_t}{6}\Big)\Big]\Big),\\
\delta_3' & = & -\frac{1}{2}g_2^2(\Lambda)T\Big(1+\frac{1}{(4\pi)^2}\Big[4\lambda_t + g_2^2\Big(-\frac{20+N_d+2N_t}{3}-\frac{4}{3}N_f(L_f-1)+L_b\frac{44-N_d-2N_t}{6}\Big)\Big]\Big).
\eea
where, $N_d=1$, $N_t= 1$ and $N_f=3$ to identify the contributions from the SM Higgs doublet, the real triplet and the fermions, respectively. \\
In case of triplet, there is significant change after the two-loop corrections to the thermal masses are added. The upper mass bound which was previously 310 GeV is now constrained more and reduced to 259 GeV.

\subsection{Constraints from DM relic}

Both complex singlet and the inert triplet scenarios considered here offer a dark matter candidate being odd under $Z_2$. In order to fulfil the criteria of  only dark matter candidate, the neutral component in both scenarios independently should satisfy the observed dark matter relic by the  
Planck experiments \cite{Planck:2013pxb}
\be\label{dm}
\Omega_{\rm DM} h^2 =0.1199 \pm 0.0027. 
\ee

The interactions of the $Z_2$ odd particles with the particles in the thermal bath are the gauge couplings and the quartic couplings and the values of these couplings are quite large. Hence, the DM particle is considered to be in equilibrium with the thermal bath initially. As the Universe expands, the interaction rates of the DM falls short to maintain the equilibrium number density and freezes out. After freeze out, the number density of the DM remains constant in the comoving frame which gives the DM relic abundance in the current epoch. So, we constraint our parameter space to satisfy the thermal relic abundance as given in \autoref{dm}. In the case of singlet the main annihilation comes via s-channel Higgs boson on- or off-shell. 
It is noticed that for maximum region of parameter space the singlet dark matter matter can satisfy the required observed dark matter relic \cite{sneha, acpb,Robens:2015gla}, thus seems phenomenologically much more viable. Contrastingly, the inert triplet scenario, the neutral  part $T^0$ annihilates mainly $W^+W^-$ and co-annihilates via $T^\pm T^0 \to W^\pm Z$ and thus  
 demands $m_{T^0}\geq 1176$ GeV \cite{Jangid:2020qgo} to satisfy the required dark matter relic in \autoref{dm}. This is incompatible with the demand of first order phase transition at $m_h=125.5$ GeV that we just observed in the previous section which states $m_T  < 193.00$ GeV and $m_{T^0}< 310.24$ GeV. For the first-order phase transition occuring at temperatures after the freeze-out of species, the entropy injection during the first-order phase transition can lead to dilution of the relic species that has decoupled from the thermal bath in the early universe. This dilution factor can only reach on order of 10, in case of purely bosonic models which still does not make inert triplet model relic mass bound of TeV order viable \cite{Wainwright:2009mq}. Certainly, the inert triplet scenario can not satisfy both demands: of obtaining the first order phase transition consistent with current experimental Higgs boson mass bound and satisfying the dark matter relic. A simple gateway would be one more contributor viz. singlet, in order to satisfy the dark matter relic which would also  enhance the possibility of the first order phase transition even further \cite{Paul:2019pgt,tripletsinglet2021,Shajiee,Hitoshi,Garcia-Pepin:2016hvs}.

\section{Calculating frequency detectable by LISA, LIGO and BBO }\label{GW}
The phase transition from symmetric phase to broken phase proceeds via bubble nucleation when bubbles of the false vacua nucleate in the sea of symmetric phase and then keep on expanding. These expanding bubbles collide and gives rise to Gravitational waves (GW) which is described below. The frequencies of such gravitational waves can be estimated via thermal parameters which are described in the next subsections. Before we move on to the calculation  of the frequencies of the gravitational waves, let us revisit the effective potential in order to implement in the  {\tt CosmoTransition } \cite{Wainwright}. The effective potential at finite temperature which can be written as;
\bea\label{vefft}
\rm V_{eff} = V_{tree} + V_1(\phi, 0) + V_1(\phi, T),
\eea 
 where $\rm V_{tree}$ is the tree-level potential, $V_1(\phi, 0) $ is the quantum correction at the zero temperature and $ V_1(\phi, T)= \Delta V_1(\phi, T) +  \Delta V_{\rm daisy/ring}(\phi, T) $  
  as shown in \autoref{vpoteq}.  The one-loop quantum correction  at zero-temperature is estimated via Coleman-Weinberg  method \cite{Coleman} working in the Feynman gauge and also implemented in  {\tt CosmoTransition} \cite{Wainwright};
 \bea
 V_{1}(\phi,0) = \pm \frac{1}{64 \pi^2} \sum_{i} n_i m_i^4 \Big[log\frac{m_i^2}{Q^2}-c_i\Big],
 \eea
 where $n_i$ and $m_i$ are the degrees of freedom and field-dependent masses as described in  \autoref{eq:2.2}, \autoref{eq:2.3} and \autoref{eq:2.9}, respectively.  Here $+(-)$  signs come for bosonic(fermionic) degrees of freedom. The expression for the  potential coming from non-zero temperature including the daisy/ring resummation (also in the Feynman gauge) are expressed as   \cite{Wainwright}:
 \bea
 V_{1}(\phi,T) = \frac{T^2}{2 \pi^2}\sum_{i}n_iJ_{\pm}\Big[\frac{m_i^2}{T^2}\Big],
 \eea
 where $J_{\pm}$ are spline functions with $+(-)$ for bosons(fermions), respectively and are defined as;
 \bea
 J_{\pm}(\frac{m_i^2}{T^2})= \pm \int_{0}^{\infty} dy y^2 log\Big(1 \mp e^{-\sqrt{y^2+\frac{m_i^2}{T^2}}}\Big). 
 \eea
 Next we discuss the relevant parameters needed to calculate the frequencies of the Gravitational Waves(GW) using  {\tt CosmoTransition} \cite{Wainwright} and {\tt Bubbleprofiler}\cite{Athron:2019nbd}.
\subsection{Thermal parameters}
The Gravitational Waves(GW) are created when bubble collision occurs and thus depends on the bubble nucleation rate as given below \cite{LINDE1983421}
\bea
\Gamma(t)=A(t)e^{-S_3(t)} ,
\eea
where $S_3$ is the Euclidean action of the background field $\phi$ written in spherical polar coordinate, of the critical bubble as  follows\cite{LINDE1983421}:
\bea
S_3= 4 \pi \int dr r^2 \Big[\frac{1}{2}(\partial_r \vec{\phi})^2+V_{\rm eff}\Big]. 
\eea
Here, $V_{\rm eff}$ is the total potential as given in \autoref{vefft}.

The temperature of the thermal bath at time $t_*$ is defined as $T_*$ and without significant reheating effect, $T_* \sim T_n$, the nucleation temperature. At the nucleation temperature $T_n$, the bubble nucleation starts and the bubble nucleation rate, $\Gamma$, should be large enough that a bubble is nucleated per horizon volume with probability of order 1\cite{LINDE1983421}. In terms of bubble nucleation rate,  inverse time duration of the phase transition, $\beta$ is given as
\bea
\beta =-\frac{dS}{dt}\Big|_{t=t_*} \simeq \frac{\dot{\Gamma}}{\Gamma}.
\eea
$t_*$ being the instant of time where the first order phase transition completes. The parameter $\beta$ defines the time variation of the bubble nucleation rate and therefore describe the length of the time in which the phase transition occurs. There are two relevant parameters which control the Gravitational Wave (GW) signal, one of them is the fraction $\frac{\beta}{H_*}$, where $H_*$ is the Hubble parameter at temperature $T_*$. To achieve large Gravitational wave (GW) signal, relatively slow phase transition is required and hence the fraction, $\frac{\beta}{H_*}$ should be small for stronger signals. This ratio $\frac{\beta}{H_*}$ instrumental for this is defined as 
\bea
\frac{\beta}{H_*} = T_* \frac{dS}{dt}\Big|_{T_*},
\eea
where $T_*$ is the temperature at time $t_*$, i.e. $T_*=T|_{t_*}$ and it becomes $T_* \simeq T_n$ with negligible reheating effect. The ratio $\frac{\beta}{H_*}$ required for the visible signal in LISA is $\frac{\beta}{H_*}\lsim 10^3$ \cite{Randall:2006py}. This is a dimensionless quantity and it mainly depends on the effective potential size at the nucleation temperature. The another essential parameter is $\alpha$, defined as the ratio of the vacuum energy density which is released during the phase transition to that of radiation bath and it is defined as below; 
\bea
\alpha= \frac{\rho_{vac}}{\rho_{rad}^*},
\eea
where $\rho^*_{rad}=g_*\pi^2T_*^4/30$, and $g_*$ is the number of relativistic degrees of freedom at temperature $T_*$ in plasma. Other relevant  parameters for the appraisal of the GW frequencies are 
\bea\label{alp}
\kappa_v=\frac{\rho_v}{\rho_{vac}}, \, \qquad  \kappa_\phi=\frac{\rho_\phi}{\rho_{vac}},
\eea 
where $\kappa_v$ is the fraction of vacuum energy that is converted into bulk motion of the fluid and $\kappa_\phi$ is the fraction of vacuum energy converted into gradient energy of the Higgs-like field. And $v_w$ is defined as the fluid bubble wall velocity.
\subsection{Production of the Gravitational Wave signal}
The first order phase transition happen via bubble nucleation and  because of the pressure difference between the false and true vacua these bubbles start expanding. The collision of these bubbles then break the spherical symmetry of each bubble and Gravitational waves(GW) are produced while for uncollided bubbles, the spherical symmetry remains preserved and no Gravitational waves(GW) are produced. The Gravitational wave background spectrum arising from cosmological phase transition depends on various sources. The sources which are most relevant for the GW,  depend on the dynamics of bubble expansion an the plasma as we discuss below.
\subsection{Relevant contributions to the Gravitational Wave spectrum}
The following processes are involved in first-order phase transition for the production of Gravitational Waves:
\begin{itemize}
	\item Bubble wall collision \cite{Kosowsky,Turner,Huber_2008,Watkins,Marc,Caprini_2008} and shocks in the plasma. The technique referred as 'envelope approximation' is used in this scenario. In this approximation, the contribution of scalar field, $\phi$, is considered in computing the GW spectrum.
	\item Sound waves in the plasma: when a part of energy released in the transition is dissipated as kinetic energy resulting in the bulk motion of fluid in plasma. \cite{Hindmarsh,Leitao:2012tx,Giblin:2013kea,Giblin:2014qia,Hindmarsh_2015}.
	\item Bubble collision leads to the formation of Magnetohydrodynamic turbulence in the plasma \cite{Chiara,Kahniashvili,Kahniashvili:2008pe,Kahniashvili:2009mf,Caprini:2009yp}.
\end{itemize}
These three processes generally coexist and linearly combine to give the contribution to the GW background as follows \cite{Caprini:2015zlo};
\bea
h^2 \Omega_{GW}\simeq h^2 \Omega_{\phi}+h^2 \Omega_{sw}+h^2 \Omega_{turb}.
\eea
The detailed forms  of each contributions are discussed successively.

\underline{Bubble Collision:}  The scalar field contribution to the Gravitational Wave(GW), involved in the phase transition can be treated by envelope approximation \cite{Turner,Watkins}. In "envelope approximation", the expanding bubbles are configured with the overlapping of corresponding set of infinitely thin shells. Once the phase transition is completed, the envelope disappears and the production of Gravitational waves(GW) stops. It has been found that the peak frequency for the Gravitational wave (GW) signal is determined by the average size of the bubble at collision. The GW contribution to the spectrum using the envelope approximation via numerical simulations can be written as,
\bea
h^2 \Omega_{env}(f) = 1.67 \times 10^{-5} \Big(\frac{\beta}{H}\Big)^{-2} \Big(\frac{\kappa_{\phi} \alpha}{1 + \alpha}\Big)^2 \Big(\frac{100}{g_*}\Big)^{1/3} \Big(\frac{0.11 v_w^3}{0.42 + v_w^2}\Big)\frac{3.8 (f/f_{env})^{2.8}}{1+2.8(f/f_{env})^{3.8}},
\eea
with 
\bea
\beta=\Big[HT\frac{d}{dT}\Big(\frac{S_3}{T}\Big)\Big]\Big|_{T_n},
\eea
where $T_n$ is defined as the nucleation temperature and $H_n$ is the Hubble parameter at temperature $T_n$. The estimation of the bubble wall velocity $v_w$ used in the above equation is given as \cite{Marc,Chao:2017vrq,Dev:2019njv,Paul};
\bea
v_w=\frac{1/\sqrt{3}+\sqrt{\alpha^2+2\alpha/3}}{1+\alpha}.
\eea
The $\kappa_{\phi}$ parameter used in the calculation is defined as the fraction of latent heat deposited in a thin shell and is expressed as,
\bea
\kappa_{\phi}= 1-\frac{\alpha_\infty}{\alpha},
\eea
with \cite{Shajiee:2018jdq,Caprini:2015zlo}
\bea
\alpha_{\infty}=\frac{30}{24 \pi^2 g_*}\Big(\frac{v_n}{T_n}\Big)^2\Big[6\Big(\frac{m_W}{v}\Big)^2+3\Big(\frac{m_Z}{v}\Big)^2+6\Big(\frac{m_t}{v}\Big)^2\Big].
\eea
where $v,\,v_n$ are the vacuum expectation values of Higgs filed at the electroweak scale and at the nucleation temperature $T_n$, respectively. $M_W$, $M_Z$ and $M_t$ are the W boson, Z boson and top quark masses, respectively. $\alpha$ is defined in \autoref{alp} at the nucleation temperature, where 
\bea
\rho_{vac}=\Big[\Big(V^{high}_{eff}-T \frac{dV^{high}_{eff}}{dT}\Big)-\Big(V^{low}_{eff}-T\frac{dV^{low}_{eff}}{dT}\Big)\Big]
\eea
and 
\bea
\rho^*_{rad}=\frac{g_*\pi^2T^4_n}{30}.
\eea
Finally we receive the expression of the  peak frequency $f_{\rm env}$, produced by bubble collisions, which contribute to the GW spectrum as
\bea
f_{\rm env}=16.5 \times 10^{-6} Hz \Big(\frac{0.62}{v^2_w-0.1 v_w+1.8}\Big)\Big(\frac{\beta}{H}\Big)\Big(\frac{T_n}{100 \rm GeV}\Big)\Big(\frac{g_*}{100}\Big)^{\frac{1}{6}}.
\eea

\underline{Sound wave:} The latent heat is released at the phase boundary during bubble expansion. This released energy in transition grows with the volume of the bubble as $\sim R^3$ and the energy that is transferred to the scalar bubble wall grows with the surface of bubble $\sim R^2$, where R is the radius of the bubble. This energy which is released into the fluid mostly contributes in reheating the plasma. A small fraction of this energy goes into the bulk motion of fluid which can give rise to Gravitational waves(GW). Therefore, the contribution to the Gravitational wave from sound wave (SW) can be estimated as follows
\bea
h^2 \Omega_{\rm SW}=2.65\times 10^{-6}\Big(\frac{\beta}{H}\Big)^{-1}v_w \Big(\frac{\kappa_v \alpha}{1+\alpha}\Big)^2 \Big(\frac{g_*}{100}\Big)^{-\frac{1}{3}}\Big(\frac{f}{f_{\rm SW}}\Big)^3\Big[\frac{7}{4+3\Big(\frac{f}{f_{\rm SW}}\Big)^2}\Big]^2,
\eea
where the parameter $\kappa_v$, earlier defined in \autoref{alp} as the fraction of latent heat which is transferred to the bulk motion of the fluid, can be rewritten as
\bea \label{eq:4.22}
\kappa_v=\frac{\alpha_{\infty}}{\alpha}\Big[\frac{\alpha_{\infty}}{0.73+0.083 \sqrt{\alpha_{\infty}}+\alpha_{\infty}}\Big].
\eea
The peak frequency contribution $f_{\rm SW}$ to the GW spectrum produced by sound wave mechanisms is 
\bea
f_{\rm SW}=1.9\times 10^{-5}Hz \Big(\frac{1}{v_w}\Big)\Big(\frac{\beta}{H}\Big)\Big(\frac{T_n}{100 \rm GeV}\Big)\Big(\frac{g_*}{100}\Big)^{\frac{1}{6}}.
\eea

\underline{Turbulence: }  The collision of bubbles can also induce turbulent motion of fluid \cite{Kamionkowski:1993fg}. This can give rise to Gravitational waves(GW) even after the transition is finished. Lastly the contribution to GW from the Magnetohydrodynamic turbulence can be evaluated as 
\bea
h^2 \Omega_{\rm turb}=3.35 \times 10^{-4}\Big(\frac{\beta}{H}\Big)^{-1}v_w \Big(\frac{\epsilon \kappa_v \alpha}{1+\alpha}\Big)^{\frac{3}{2}}\Big(\frac{g_*}{100}\Big)^{-\frac{1}{3}}\frac{(\frac{f}{f_{turb}})^3\Big(1+\frac{f}{f_{\rm turb}}\Big)^{-\frac{11}{3}}}{\Big(1+\frac{8\pi f}{h_*}\Big)},
\eea
where $\epsilon =0.1$ and $f_{\rm turb}$ is again the peak frequency contribution to the GW spectrum produced by the turbulence mechanism 
\bea
f_{\rm turb} = 2.7 \times 10^{-5} Hz \Big(\frac{1}{v_w}\Big)\Big(\frac{\beta}{H}\Big)\Big(\frac{T_n}{100 \rm GeV}\Big)\Big(\frac{g_*}{100}\Big)^{\frac{1}{6}}.
\eea
where
\bea
h_*=16.5 \times 10^{-6} Hz \Big(\frac{T_n}{100 \rm GeV}\Big)\Big(\frac{g_*}{100}\Big)^{\frac{1}{6}}.
\eea
The updated expression for $\kappa_v$ given in Eq: \eqref{eq:4.22} which is used in this analysis is as follows\cite{Ellis:2018mja,Espinosa:2010hh}:
\bea
\kappa_v \simeq \Big[\frac{\alpha_{\infty}}{0.73+0.083\sqrt{\alpha_{\infty}}+\alpha_{\infty}}\Big]
\eea

\subsection{Benchmark points}
In this  section we compare the triplet and the  singlet  scenarios with their gravitational wave frequencies  detectable  detectable by LISA, LIGO and BBO experiments \cite{ LISA:2017pwj,Yagi:2011wg,KAGRA:2013rdx}. For this purpose we choose the benchmark points in the  singlet  and the triplet cases as given in Table \ref{tab:table3}. 
\begin{table}[h]
	\begin{center}
		\begin{tabular}{|c|c|c|c|}\hline
			& $m_{S}/m_T$ &$\lambda_{s}/\lambda_t$ & $\lambda_{hs}/\lambda_{ht}$  \\ \hline
			BP1 & 150.23 & 0.10 & 0.10  \\ 
			BP2 & 120.23 & 0.01 & 0.01  \\ 
			\hline 
		\end{tabular}
	\end{center}
	\caption{BPs for frequency analysis for singlet and triplet scenario.}	\label{tab:table3}
\end{table}

The thermal parameters required for the calculation of GW spectrum are mainly the nucleation temperature $T_n$, the strength of phase transition $\alpha$, length of the time of phase transition $\beta$, Higgs vev at the nucleation temperature $v_n$ and the bubble wall velocity $v_w$. The calculation of the Gravitational Wave(GW) intensity requires the phase transition temperature. Hence, the finite temperature effective potential is computed for the calculation of  transition temperature. These calculations are performed using a publicly available package {\tt CosmoTransition}\cite{Wainwright}. The tree-level potential is given as an input to this package and it provides the thermal parameters required for the calculation of Gravitational Wave(GW) intensity. These thermal parameters corresponding  to the  benchmark points  in \autoref{tab:table3}, predicting strongly first order phase transition and  allowed by 125.5 GeV Higgs boson are  shown in  \autoref{tab:table4}-\autoref{tab:table5} for the singlet and the triplet scenarios, respectively.
\begin{table}[h!]
	\begin{center}
		\begin{tabular}{|c|c|c|}\hline
			& BP1 & BP2  \\ \hline
			$T_n$[GeV] & 121.03 & 119.25   \\ 
			$\alpha$ & 0.17 & 0.18   \\ 
			$\beta/H$ &332.83 & 327.94  \\ 
			$v_n/T_n$ & 1.10 & 1.16  \\ \hline
		\end{tabular}
	\end{center}
	\caption{Thermal parameters required for frequency analysis of the singlet for the chosen benchmark points, where $T_n$ is the nucleation temperature, $\alpha$ is the strength of transition, $\beta$ is the length of the time of phase transition and $v_n$ is the Higgs vev at the nucleation temperature. }	\label{tab:table4}
\end{table}

\begin{table}[h!]
	\begin{center}
		\begin{tabular}{|c|c|c|}\hline
			& BP1 & BP2   \\ \hline
			$T_n$[GeV] & 115.07 & 113.55   \\ 
			$\alpha$ & 0.86& 0.89 \\ 
			$\beta/H$ &284.22 & 278.87  \\ 
			$v_n/T_n$ & 1.16 & 1.22   \\ \hline 
		\end{tabular}
	\end{center}
	\caption{Thermal parameters required for frequency analysis of the  inert triplet for the chosen benchmark points where $T_n$ is the nucleation temperature, $\alpha$ is the strength of transition, $\beta$ is the length of the time of phase transition and $v_n$ is the Higgs vev at the nucleation temperature.}	\label{tab:table5}
\end{table}

\begin{figure}
	\begin{center}
		\mbox{\subfigure[Singlet]{\includegraphics[width=0.5\linewidth,angle=-0]{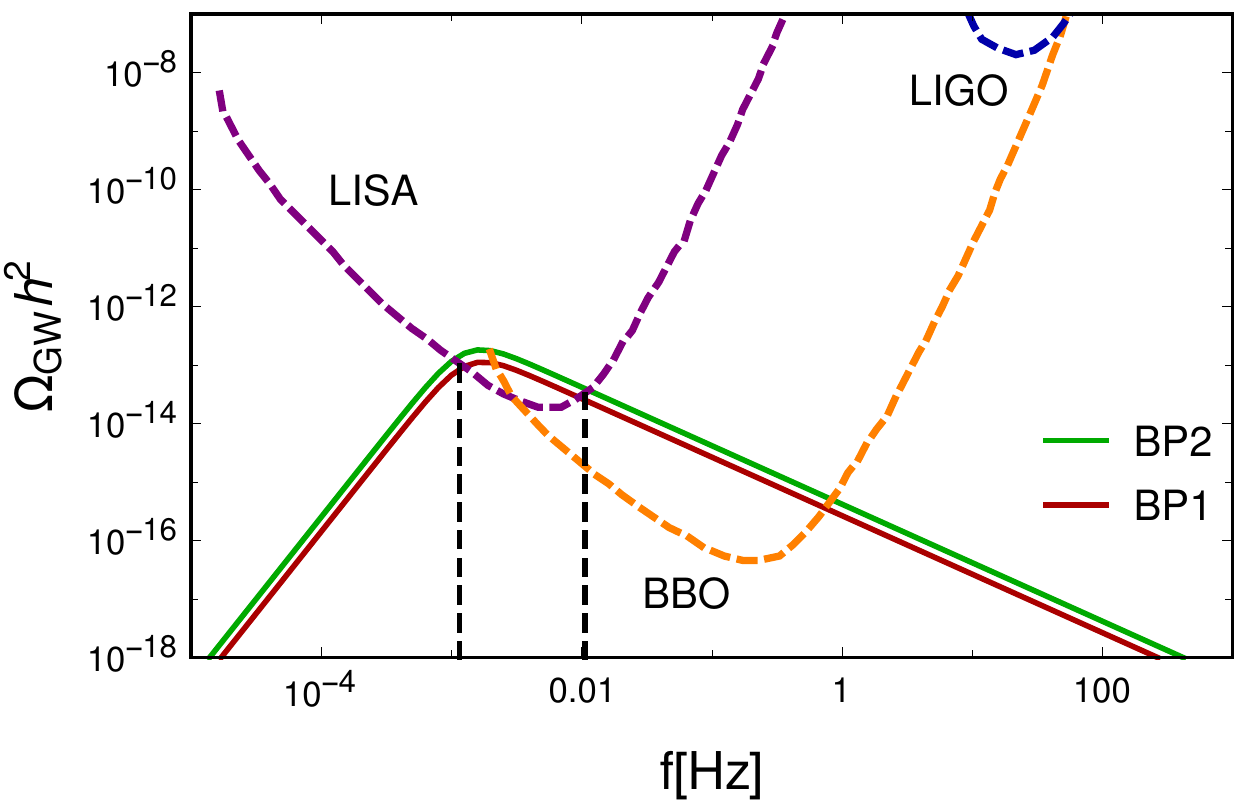}\label{f31a}}
			\subfigure[Triplet]{\includegraphics[width=0.5\linewidth,angle=-0]{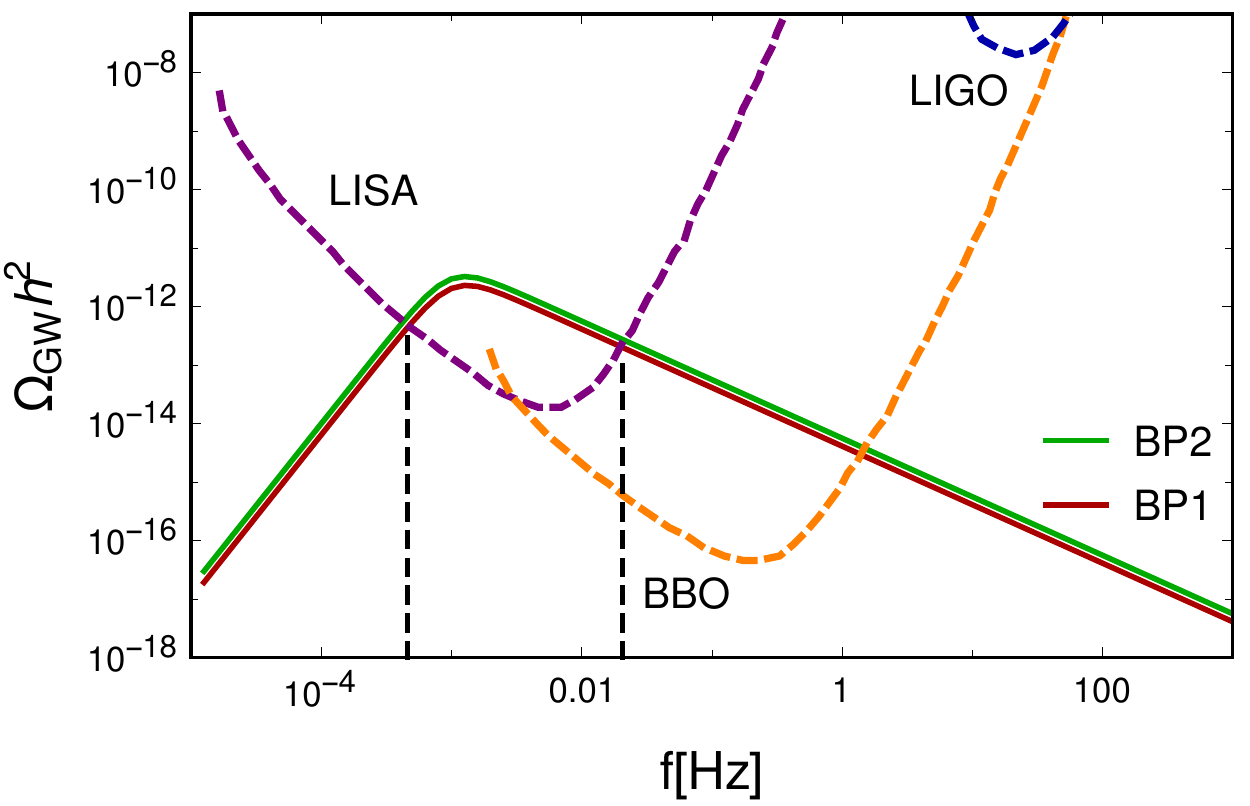}\label{f32a}}}
		\caption{Gravitational Wave(GW) spectrum for the BPs allowed by strongly first-order phase transition in comparison with the sensitivity curves based on noise curves of experiments i.e. LISA, LIGO and BBO. The value of nucleation temperature ($T_n$),  the strength of transition ( $\alpha$)  and the length of the time of phase transition ($\beta$ )  are computed and used for the Gravitational Wave(GW) intensity calculation.}\label{fig:GWI}
	\end{center}
\end{figure}

The Gravitational wave(GW) spectrum arising from the first-order phase transition for the benchmark points  are given in \autoref{fig:GWI}. The constraints for different experiments are drawn by the respective sensitivity curves for the different GW  detectors viz.  LISA, LIGO and BBO. The higher value of $\alpha$ and lower value of $\beta$ actually provides stronger GW signals. It is clear from Table \ref{tab:table4} and Table \ref{tab:table5} that the nucleation temperature $T_n$ is lower than the critical temperature $T_c$ for all benchmark points in singlet and triplet and the value of ratio $v_n/T_n$ is $\gsim$ 1, giving strongly first-order phase transition. The values of nucleation temperature for inert triplet, $T_n=115.07(113.55)$ GeV are lower compared to singlet ones, $T_n=121.03(119.25)$ GeV, that  ensure stronger signals detectable by various experiments. For both the benchmark points, the GW intensity lie within the sensitivity curves of LISA and BBO in the singlet and the triplet scenarios, respectively. The detectable frequencies  for singlet lie between $\sim 1.15 \times 10^{-3}-1.06\times 10^{-2}$ Hz, while for the triplet, the allowed ranges enhance to range $\sim 4.18 \times 10^{-4}-1.99\times 10^{-2}$ Hz, for the LISA experiment as can be seen from \autoref{fig:GWI}. It is also inferred from \autoref{fig:GWI} that the Gravitational Wave(GW) intensity mainly depends on the parameter $\beta$. The smallest value of parameter $\beta$ is attained for BP2 of the inert triplet scenario, which leads to highest Gravitational wave(GW) intensity. For LIGO, the Gravitational Wave(GW) intensities lie outside the detectable region in both the singlet and the triplet scenarios.  In comparison BBO has more region of parameter space that can be detected for both, with triplet having larger spectrum with slight larger frequency range  compared to the singlet case. 
Also the signal to noise ratio(SNR) for a particular detector, which is given as 
	\begin{eqnarray*}
		\rm SNR =\sqrt{2t_{obs}\int_{f_{min}}^{f_{max}}df\Big[\frac{\rm \Omega_{GW}(f)h^2}{\Omega_{noise}(f)h^2}\Big]^2},
	\end{eqnarray*}
where $t_{obs}$ is defined as the duration of the observation in unit of seconds and $\rm\Omega_{noise}(f)h^2$ is defined as the effective strain noise power spectral density for the considered detector. $\rm\Omega_{GW}(f)$ is detectable for signal to noise ratio(SNR) SNR>1, which is possible for $\rm\Omega_{noise}(f)< \Omega_{GW}$. Therefore, there is a finite chance that the frequency range in \autoref{fig:GWI} covered by BBO experiment is detectable. \cite{Barish:2020vmy,Aoki:2019mlt,Yagi:2013du, Moore:2014lga,Yagi:2011yu,Thrane:2013oya}

However, the future advanced Gravitational wave(GW) detectors such as eLISA and BBO are expected to explore millihertz to decihertz of frequency ranges in future.  Similarly the ground based detector like aLIGO  can explore the lower frequency range with much higher sensitivity. There can be two to three orders-of-magnitude theoretical uncertainity in the peak GW amplitude using daisy-resummation approach due to renormalization scale dependence. Using higher order terms in the perturbative calculations i.e. dimensional reduction approach, the scale dependence can be reduced and the theoretical unceratinity can be reduced to $\mathcal{O}(10^0-10^1)$ \cite{Croon:2020cgk,Gould:2021oba}.

In order to ensure that the physical quantities in any field theory are independent of the particular renormalization scheme (RS), if the true result is exactly RS independent then the best approximation should be least sensitive to the small changes in RS. This is known as principle of minimal sensitivity \cite{PhysRevD.23.2916}. This principle states that for unphysical parameters, the exact result is a constant. Hence, the calculated result cannot be a successful approximation where it is varying rapidly. If the variation is considered with the renomalization scale then the extrapolation from $10^2$ GeV can be judged by observing how flat the result is at higher scales. The variation will not be flat everywhere, so one can always choose the scale to lie in the middle of the flat portion of the variation. The variation of all the quartic couplings become almost constant after $10^6$ GeV, and further higher scales. Therefore, we consider the variation of the quartic couplings using the two-loop $\beta$-functions in the Daisy resummation approach including the two-loop potential from \autoref{eq:5.6}-\autoref{eq:5.7} at three different scales i.e. $10^2$ GeV, $10^3$ GeV and $10^6$ GeV, respectively. The benchmark points and the thermal parameters for singlet and triplet scenario are given in Table \ref{tab:table6} to \ref{tab:table8}.

\textcolor{blue}{\begin{table}[h]
		\begin{center}
			\begin{tabular}{|c|c|c|c|c|}\hline
				$\mu$(GeV)& $m_{S}/m_T$ (GeV) &$\lambda_{s}/\lambda_t$ & $\lambda_{hs}/\lambda_{ht}$ & $\lambda_h$ \\ \hline
				$10^2$ & 190.23 & 0.10 & 0.33 & 0.1264\\ 
				$10^3$ & 190.23 & 0.11/0.11 & 0.36/0.36 & 0.1040/0.1040 \\ 
				$10^6$ & 190.23 & 0.19/0.22 & 0.48/0.53 & 0.1198/0.1253 \\ 
				\hline 
			\end{tabular}
		\end{center}
		\caption{Allowed benchmark point for the frequency analysis in singlet and triplet scenario at three different renormalization scales i.e $10^2$ GeV, $10^3$ GeV and $10^6$ GeV using Daisy resummation method with two-loop potential and two-loop $\beta$-functions. The quartic couplings are given for three different renormalization scale variation followed by the running of the two-loop $\beta$-functions.}	\label{tab:table6}
\end{table}}
\begin{table}[h!]
	\begin{center}
		\begin{tabular}{|c|c|c|c|}\hline
			& $10^2$ & $10^3$ &$10^6$ \\ \hline
			$T_n$[GeV] & 130.73 & 131.69 & 185.91 \\ 
			$\alpha$ & 0.15 & 0.15  &  0.10\\ 
			$\beta/H$ &292.83 & 295.94  & 327.68 \\ 
			$v_n/T_n$ & 0.97 & 0.97 & 0.83\\ \hline
		\end{tabular}
	\end{center}
	\caption{Thermal parameters required for the frequency analysis in case of singlet for chosen benchmark points where $T_n$ is the nucleation temperature, $\alpha$ is the strength of transition, $\beta$ is the length of the time of phase transition and $v_n$ is the Higgs vev at the nucleation temperature.}	\label{tab:table7}
\end{table}

\begin{table}[h!]
	\begin{center}
		\begin{tabular}{|c|c|c|c|}\hline
			& $10^2$ & $10^3$ &$10^6$ \\ \hline
			$T_n$[GeV] & 128.50 & 129.60 & 181.07 \\ 
			$\alpha$ & 0.16 & 0.16  &  0.11\\ 
			$\beta/H$ &291.56 & 294.23  & 320.68 \\ 
			$v_n/T_n$ & 0.98 & 0.98 & 0.84\\ \hline
		\end{tabular}
	\end{center}
	\caption{Thermal parameters required for the frequency analysis in case of  the inert  triplet for chosen benchmark points where $T_n$ is the nucleation temperature, $\alpha$ is the strength of transition, $\beta$ is the length of the time of phase transition and $v_n$ is the Higgs vev at the nucleation temperature.}	\label{tab:table8}
\end{table}

\begin{figure}[hbt]
	\begin{center}
		\mbox{\subfigure[$10^2$ GeV]{\includegraphics[width=0.5\linewidth,angle=-0]{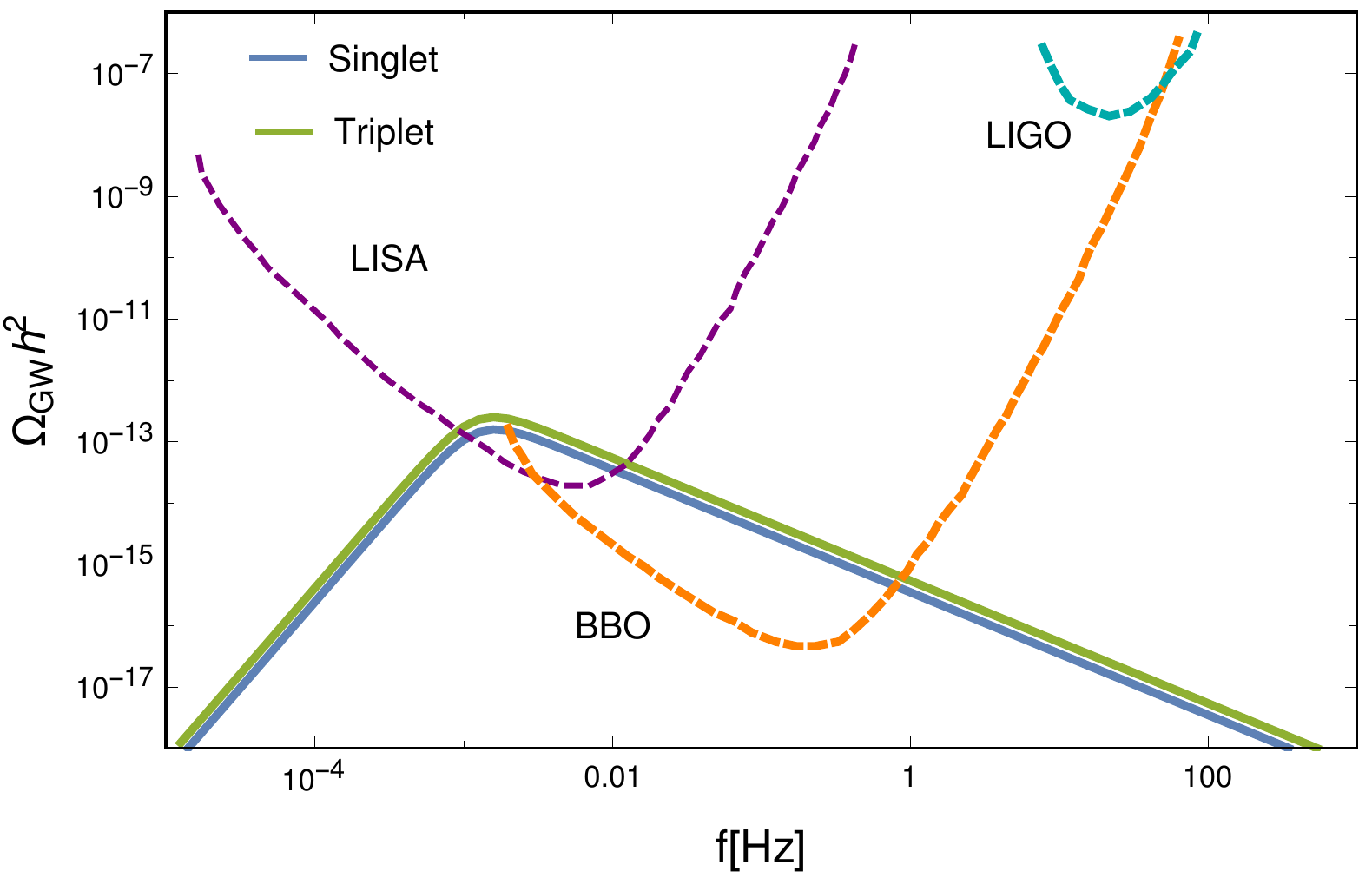}\label{f41a}}
			/			\subfigure[$10^3$ GeV]{\includegraphics[width=0.5\linewidth,angle=-0]{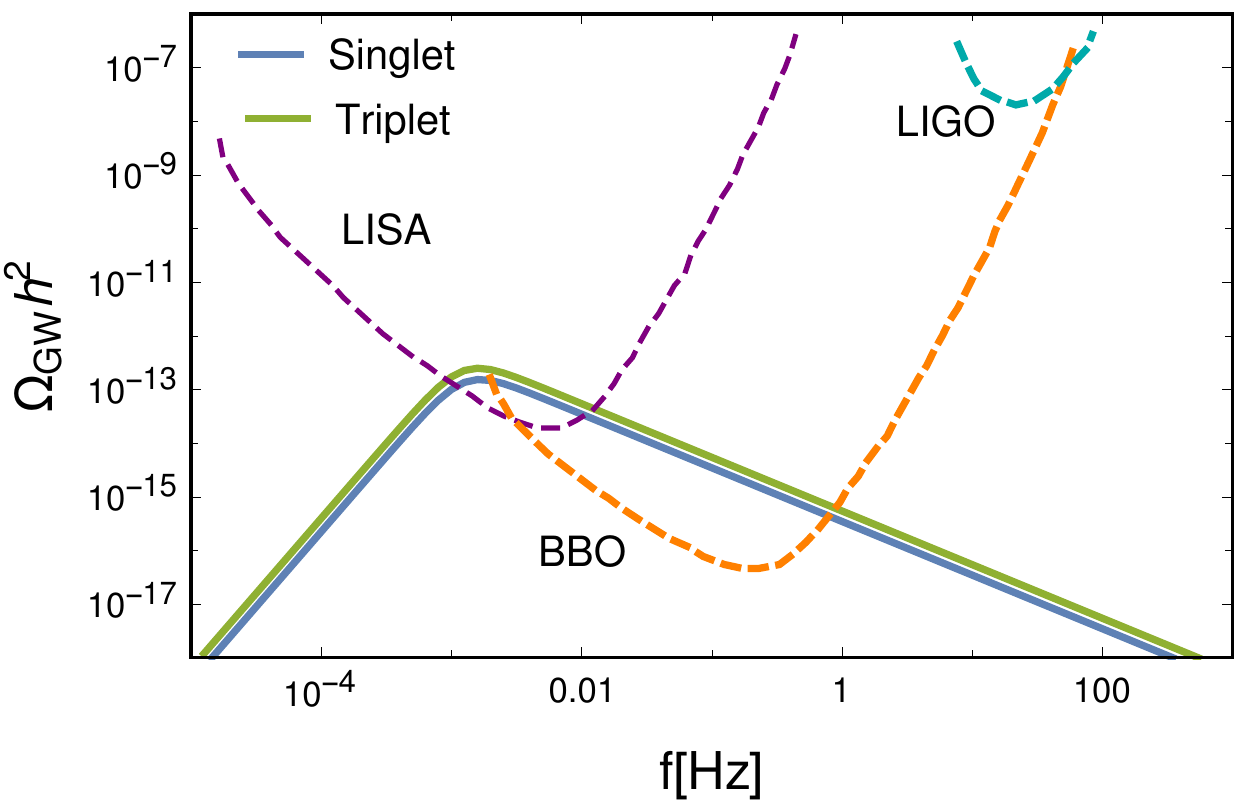}\label{f41b}}}
		\mbox{\subfigure[$10^6$ GeV]{\includegraphics[width=0.5\linewidth,angle=-0]{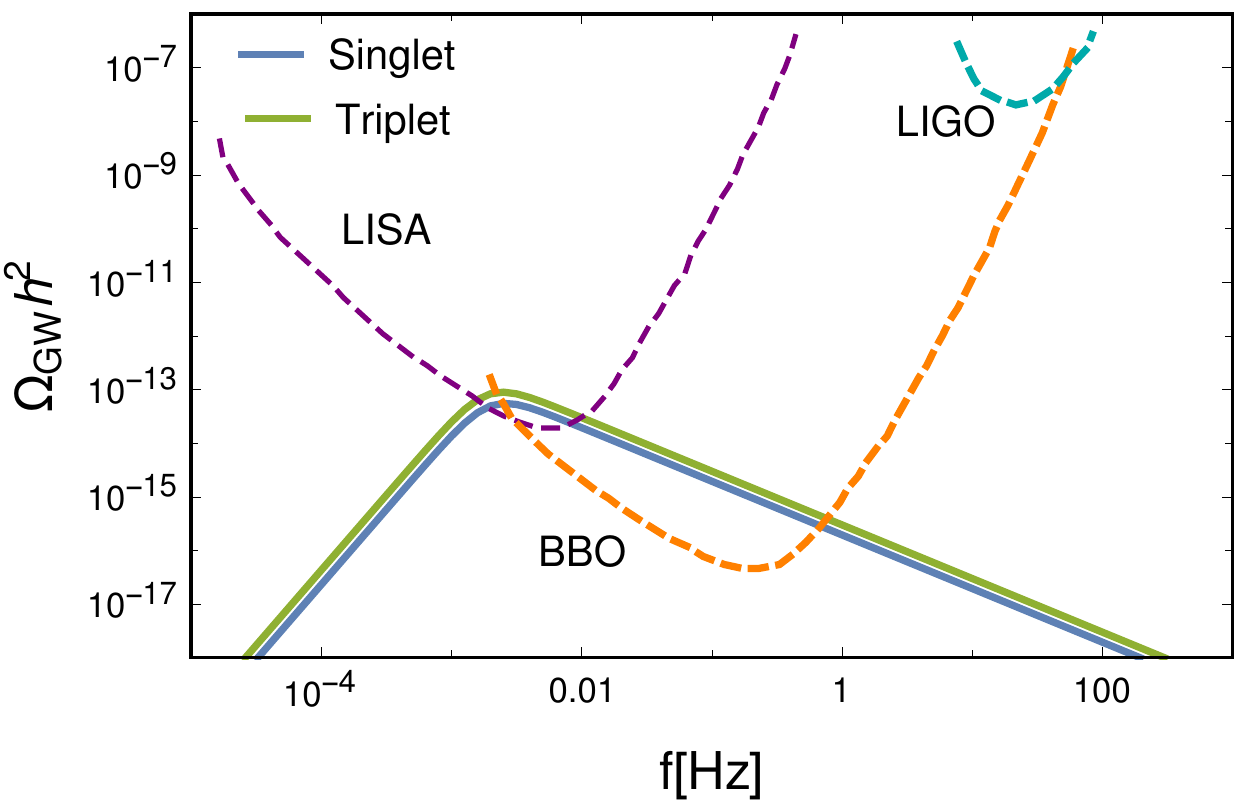}\label{f1a}}
		}
		\caption{Gravitational Wave(GW) spectrum for the BP allowed by strongly first-order phase transition and perturbative unitarity in comparison with the sensitivity curves based on noise curves of experiments i.e. LISA, LIGO and BBO for three different renormalization scales i.e. $10^2$ GeV, $10^3$ GeV and $10^6$ GeV using Daisy resummation method including the two-loop potential and two-loop $\beta$-functions. The blue and green curve corresponds to the Singlet and the triplet scenarios. The dotted purple, orange and cyan curves denotes the sensitivity curves LISA, BBO and LIGO experiments.}\label{fig:freqs}
	\end{center}
\end{figure}

\begin{figure}
	\begin{center}
		\mbox{\subfigure[Singlet]{\includegraphics[width=0.5\linewidth,angle=-0]{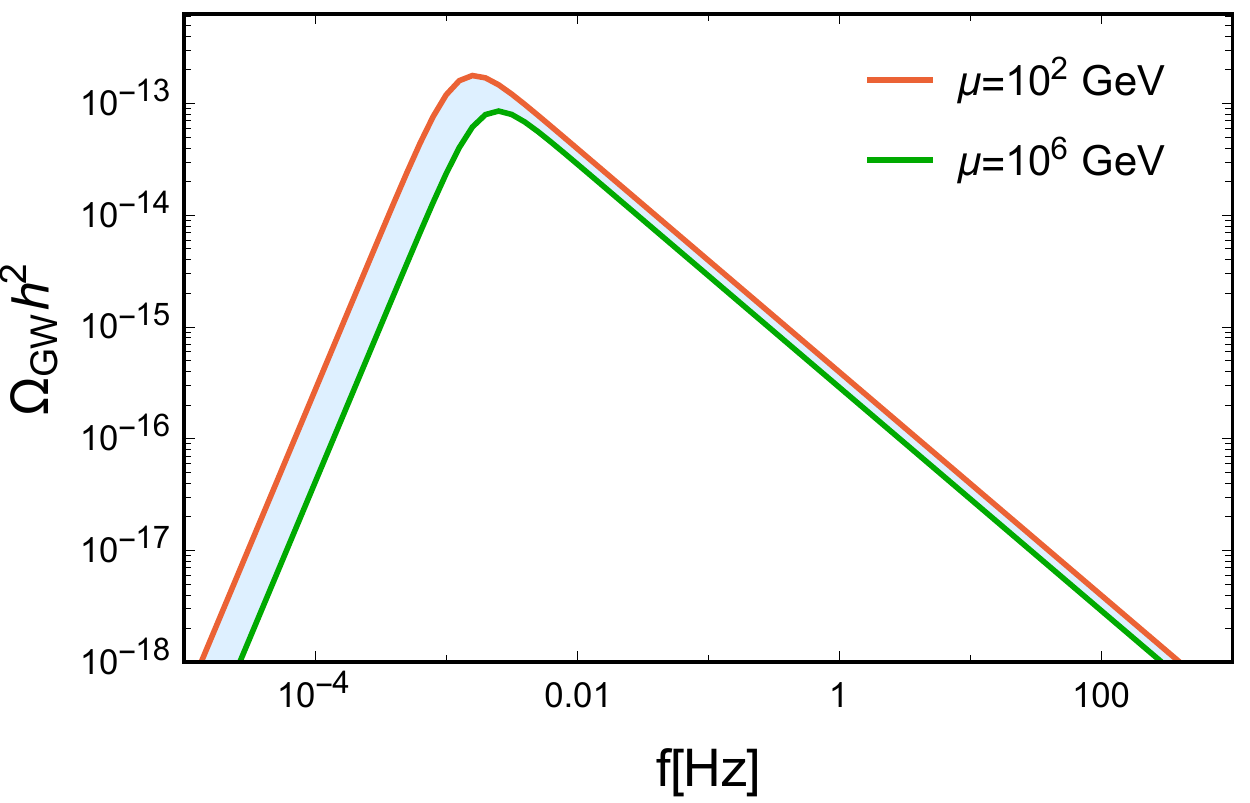}\label{f43a}}
			\subfigure[Triplet]{\includegraphics[width=0.5\linewidth,angle=-0]{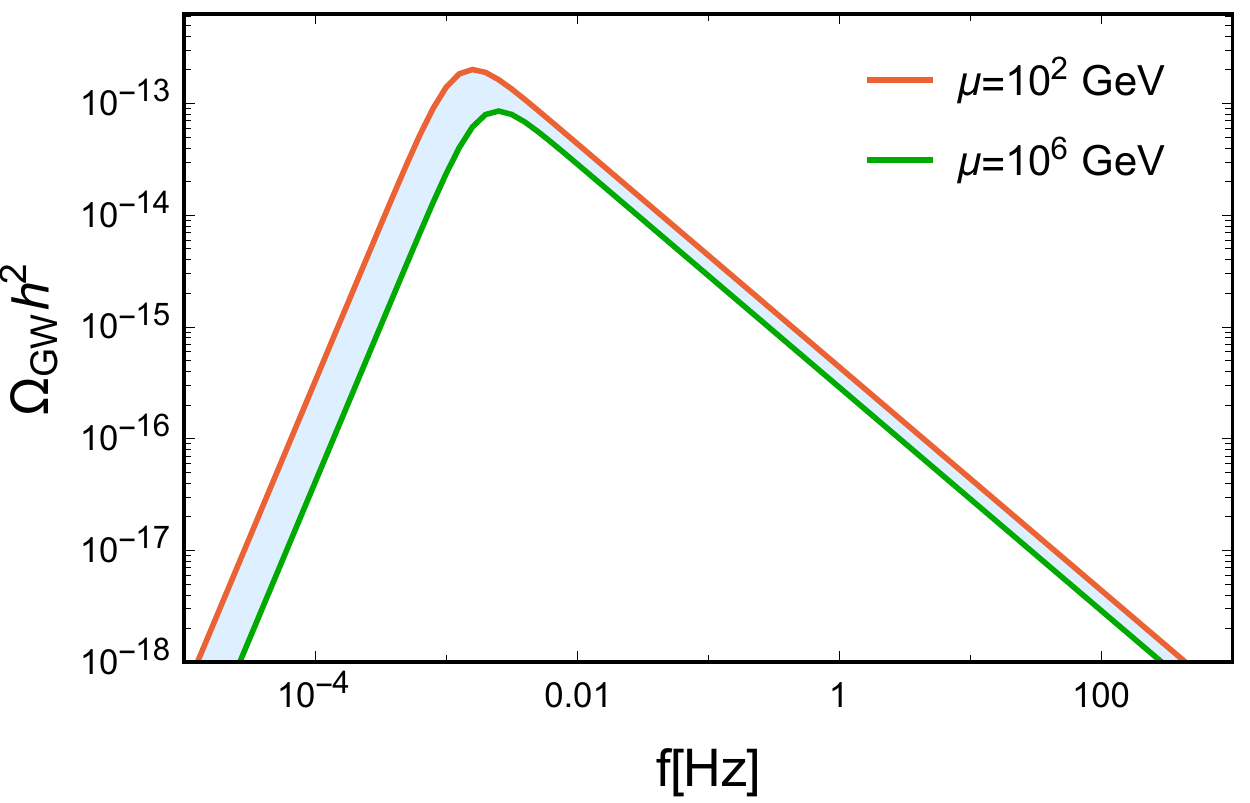}\label{f44a}}}
		\caption{Uncertainity band for the scale dependence of the Gravitational wave intensity in both singlet and the triplet using Daisy resummation method including the two-loop potential and two-loop $\beta$-functions. The uncertainity band depicts the variation as a result of the change in the renormalization scale. The red and the green colors correspond to the Gravitational wave intensity for $10^2$ GeV and $10^6$ GeV renormalization scale, respectively.}\label{fig:GW2}
	\end{center}
\end{figure}

 The Gravitational Wave(GW) spectrum for the singlet and the triplet scenarios are given in \autoref{fig:freqs}. The blue and the green intensity curves correspond to the singlet and the triplet scenarios and the dotted purple, orange and cyan curves denote the intensity spectrum for LISA, BBO and LIGO experiments, respectively. The Gravitational wave spectrum is then considered for three different renormalization scales($\mu$) i.e. $10^2$ GeV, $10^3$GeV and $10^6$ GeV in \autoref{fig:freqs}(a), (b), (c), respectively. The nucleation temperature for the triplet is lower than the singlet scenarios for the three renormalization scales as can be read from \autoref{tab:table7} and \autoref{tab:table8}. It can be noted that the nucleation temperature increases with the renormalization scale, nevertheless remains lower for the triplet compared to the singlet.  The increase in the nucleation temperature with the renormalization scale, actually reduces the Gravitational Wave(GW) intensity, and thus the detectable frequency range by LISA, LIGO and BBO experiments for both singlet and triplet scenarios. Using \autoref{tab:table7} and \autoref{tab:table8}, the uncertainity in the computation of the Gravitational wave intensity due to the renormalization scale dependency is given in \autoref{fig:GW2} for both the singlet and the triplet scenarios. The uncertainity band depicts the variation of the results with the renormalization scale using Daisy resummation method including the two-loop potential and two-loop $\beta$-functions. It can be seen from \autoref{fig:freqs} that the changes in the Gravitational wave intensity from $10^2$ GeV to $10^3$ GeV scale are minuscule. Hence, in order to show a significant amount of scale dependency, we choose  $10^2$ GeV and $10^6$ GeV scales, respectively in \autoref{tab:table7} and \autoref{tab:table8}.

 In this article we showed how the parameters satisfying the first order phase transition and the Gravitational Wave are different for both scenarios. The effect of two-loop corrections to the potentials are considered, in which only the triplet mass bound gets  effected slightly.  Another interesting fact is that the upper mass bound for singlet is larger i.e. $\sim$ 1 TeV compared to the triplet $\lesssim$ 320 GeV. The nucleation temperature is lower for triplet in comparison to singlet and the detectable frequency range by LISA is more for the triplet i.e. $\sim 4.18 \times 10^{-4}-1.99\times 10^{-2}$ Hz, in comparison to the singlet i.e. $\sim 1.15 \times 10^{-3}-1.06\times 10^{-2}$ Hz. 
 
 There are other  aspects of the collider and dark matter phenomenology  in which the scenarios can be easily discerned.  Singlet does not have  a charged Higgs boson  like the triplet one,  which can give rise  to displaced pion at  the collider   due to the compressed spectrum \cite{Jangid:2020qgo,sneha, Bandyopadhyay:2020otm,SabanciKeceli:2018fsd}. $Y=0$, triplet  unlike  the  SM doublet does not couple to fermions which alters  the bounds on rare $B$-decays \cite{Bandyopadhyay:2013lca,Bandyopadhyay:2014tha} and also difficult  to produced at the collider.  However, vector boson fusion to  charged Higgs and other  associate production can be analysed in the $ZW$ decay mode of the charged Higgs via multi-lepton final states, where the triplet takes vev \cite{Bandyopadhyay:2014vma,Bandyopadhyay:2017klv,Bandyopadhyay:2015ifm}.   On the contrary, the singlet does not have any charged Higgs bosons and being gauge singlet, it cannot  be produced via gauge bosons. The productions are mainly come via  the  mixing  with  the SM Higgs bosons  or SM Higgs boson decay to  singlet pair \cite{Profumo:2007wc, Kosowsky} and for inert singlet it  is bounded by  Higgs to invisible  decay width \cite{Barger:2008jx,Turner}.  Lastly, inert singlet model satisfies the DM relic density bound even with the very small singlet mass along with the first-order phase transition \cite{Wainwright,Athron:2019nbd,Randall:2006py}, but  for  the triplet it shows under abundance demanding  such  low triplet mass required  for the first-order phase transition \cite{sneha}.
\section{Conclusion}\label{concl}
In this article we study the  $Y=0$  $SU(2)$ inert triplet which successfully stabilises the electroweak vacuum at the zero temperature and  also provide the DM candidate\cite{Jangid:2020qgo}, at the finite temperature. The regions of parameter space suitable for the first order phase transitions are designated considering  perturbative unitarity at one- and two-loop level along with the demand of a SM-like Higgs boson around $125.5$ GeV. It has been noticed that no consistent solutions have been found at one-loop perturbativity till Planck scale consistent with first order phase transition, and current Higgs boson and top quark masses. Considering the two-loop beta functions with the one-loop resummed potential, one can find the maximum mass values for the singlet and the triplet field as $909,\, 310$ GeV, respectively predicting the first order phase transition which are also  consistent with the currently measured Higgs boson and top quark masses. Including the two-loop contributions coming from the effective potential as well as the thermal masses, the mass bound for the singlet remains the same, while satisfying the current Higgs mass within the uncertainity of 1$\sigma$. On the other hand, the mass bound for the inert triplet is further constrained to $\lesssim$ 259 GeV with these corrections. However, these maximum allowed values of mass correspond to relatively larger values of $\lambda_{hs}(\lambda_{ht})=4.00(1.95)$, respectively. For lower values of these masses correspond to the regions with higher $\frac{\phi_+(T_c)}{T_c}$ i.e., more strongly first order phase transition. The self couplings of the singlet and the triplet
 are considered to be zero to maximize the $\frac{\phi_+(T_c)}{T_c}$. 

It is interesting to note here that for the singlet mass $\lesssim 909$ GeV one not only realises first order phase transition along with a Higgs boson mass around $125.5$ GeV, but also find the parameter space consistent with DM relic \cite{acpb, sneha, Robens:2015gla}. On the contrary, the situation looks grim for the triplet scenario as the correct DM relic abundance demands the  triplet scalar mass $\lesssim 1.2-1.8$ TeV. Thus with only triplet extension of  the SM, we cannot  have the first order phase transition along with the correct DM relic. Triplet DM mass $< 320$ GeV gives rise to under abundance for the DM and we need additional fields to satisfy the correct relic \cite{sneha}. 

First order phase transition in both cases can give rise to gravitational wave coming from the bubble collision, sound wave of the plasma and  the turbulence. These add up to the frequencies  that can be  observed via the space and earth bases experiments like LISA \cite{LISA:2017pwj}, BBO \cite{Yagi:2011wg} and LIGO \cite{KAGRA:2013rdx}. To observe and distinguish the singlet and the triplet scenarios we  benchmark both singlet and  triplet scenarios and predict their frequencies observed by various different detectors. The detectable frequency range by LISA is more for the triplet i.e. $\sim 4.18 \times 10^{-4}-1.99\times 10^{-2}$ Hz, in comparison to the singlet i.e. $\sim 1.15 \times 10^{-3}-1.06\times 10^{-2}$ Hz. For all the benchmark points, the Gravitational wave(GW) intensities lie within the detectable range of LISA and BBO in both singlet and triplet scenarios. With the increase of the renormalization scale using the Daisy resummation method with two-loop $\beta$-functions and two-loop potential, the Gravitational Wave (GW) intesity and also the detectable frequency drop. Thus, the singlet model, constrained from perturbative unitarity and DM relic, is in agreement with the sensitivity curves of Gravitational wave(GW) detectors. However, for the  triplet case, the strongly first order phase transition predicts relatively lower mass for the triplet ($\lesssim 320$ GeV), demanding additional multiplets to satisfy the DM relic.

\section*{Acknowledgements} PB wants to thank SERB project (CRG/2018/004971) and MATRICS Grant MTR/2020/000668 for the financial support towards this work. PB also thanks to Rajesh Gupta, Luigi Delle Rose on clearing some doubts. SJ thanks DST/INSPIRES/03/2018/001207 for the financial support towards finishing this work. This research was also supported by an appointment to the YST Program at the APCTP through the Science and Technology Promotion Fund and Lottery Fund of the Korean Government. This was also supported by the Korean Local Governments - Gyeongsangbuk-do Province and Pohang City. SJ thanks Mariano Quiros, Nikita Blinov, Michael Bardsley, Csaba Balazs and Anirban Karan for useful discussions.
\appendix

\section{Two-loop $\beta$-functions for ITM} \label{betaf1}
\subsection{Scalar Quartic Couplings}\label{A1}
\footnotesize{
	\begingroup
	\allowdisplaybreaks
	\begin{align*}
	\beta_{\lambda=\lambda_{1}} \ =  \ &
	\frac{1}{16\pi^2} \Bigg[	\frac{27}{200} g_{1}^{4} +\frac{9}{20} g_{1}^{2} g_{2}^{2} +\frac{9}{8} g_{2}^{4} -\frac{9}{5} g_{1}^{2} \lambda_1 -9 g_{2}^{2} \lambda_1 +24 \lambda_1^{2} +8 \lambda_{ht}^{2} +12 \lambda_1 \mbox{Tr}\Big({Y_d  Y_{d}^{\dagger}}\Big) +4 \lambda_1 \mbox{Tr}\Big({Y_e  Y_{e}^{\dagger}}\Big) \nonumber \\ 
	&\quad \quad +12 \lambda_1 \mbox{Tr}\Big({Y_u  Y_{u}^{\dagger}}\Big) -6 \mbox{Tr}\Big({Y_d  Y_{d}^{\dagger}  Y_d  Y_{d}^{\dagger}}\Big) -2 \mbox{Tr}\Big({Y_e  Y_{e}^{\dagger}  Y_e  Y_{e}^{\dagger}}\Big) -6 \mbox{Tr}\Big({Y_u  Y_{u}^{\dagger}  Y_u  Y_{u}^{\dagger}}\Big) \Bigg] \nonumber \\
	& +\frac{1}{(16\pi^2)^2}\Bigg[-\frac{3411}{2000} g_{1}^{6} -\frac{1677}{400} g_{1}^{4} g_{2}^{2} -\frac{317}{80} g_{1}^{2} g_{2}^{4} +\frac{277}{16} g_{2}^{6} +\frac{1887}{200} g_{1}^{4} \lambda_1 +\frac{117}{20} g_{1}^{2} g_{2}^{2} \lambda_1 -\frac{29}{8} g_{2}^{4} \lambda_1  \nonumber \\ 
	&\quad \quad +\frac{108}{5} g_{1}^{2} \lambda_1^{2} +108 g_{2}^{2} \lambda_1^{2} -312 \lambda_1^{3} +10 g_{2}^{4} \lambda_{ht} +32 g_{2}^{2} \lambda_{ht}^{2} -80 \lambda_1 \lambda_{ht}^{2} -128 \lambda_{ht}^{3} \nonumber \\ 
	&\quad \quad +\frac{1}{20} \Big(-5 \Big(64 \lambda_1 \Big(-5 g_{3}^{2}  + 9 \lambda_1 \Big) -90 g_{2}^{2} \lambda_1  + 9 g_{2}^{4} \Big) + 9 g_{1}^{4}  + g_{1}^{2} \Big(50 \lambda_1  + 54 g_{2}^{2} \Big)\Big)\mbox{Tr}\Big({Y_d  Y_{d}^{\dagger}}\Big) \nonumber \\ 
	&\quad \quad -\frac{3}{20} \Big(15 g_{1}^{4}  -2 g_{1}^{2} \Big(11 g_{2}^{2}  + 25 \lambda_1 \Big) + 5 \Big(-10 g_{2}^{2} \lambda_1  + 64 \lambda_1^{2}  + g_{2}^{4}\Big)\Big)\mbox{Tr}\Big({Y_e  Y_{e}^{\dagger}}\Big) -\frac{171}{100} g_{1}^{4} \mbox{Tr}\Big({Y_u  Y_{u}^{\dagger}}\Big) \nonumber \\ 
	&\quad \quad +\frac{63}{10} g_{1}^{2} g_{2}^{2} \mbox{Tr}\Big({Y_u  Y_{u}^{\dagger}}\Big) -\frac{9}{4} g_{2}^{4} \mbox{Tr}\Big({Y_u  Y_{u}^{\dagger}}\Big) +\frac{17}{2} g_{1}^{2} \lambda_1 \mbox{Tr}\Big({Y_u  Y_{u}^{\dagger}}\Big) +\frac{45}{2} g_{2}^{2} \lambda_1 \mbox{Tr}\Big({Y_u  Y_{u}^{\dagger}}\Big) \nonumber \\ 
	&\quad \quad +80 g_{3}^{2} \lambda_1 \mbox{Tr}\Big({Y_u  Y_{u}^{\dagger}}\Big) -144 \lambda_1^{2} \mbox{Tr}\Big({Y_u  Y_{u}^{\dagger}}\Big) +\frac{4}{5} g_{1}^{2} \mbox{Tr}\Big({Y_d  Y_{d}^{\dagger}  Y_d  Y_{d}^{\dagger}}\Big) -32 g_{3}^{2} \mbox{Tr}\Big({Y_d  Y_{d}^{\dagger}  Y_d  Y_{d}^{\dagger}}\Big) \nonumber \\ 
	&\quad \quad -3 \lambda_1 \mbox{Tr}\Big({Y_d  Y_{d}^{\dagger}  Y_d  Y_{d}^{\dagger}}\Big) -42 \lambda_1 \mbox{Tr}\Big({Y_d  Y_{u}^{\dagger}  Y_u  Y_{d}^{\dagger}}\Big) -\frac{12}{5} g_{1}^{2} \mbox{Tr}\Big({Y_e  Y_{e}^{\dagger}  Y_e  Y_{e}^{\dagger}}\Big) - \lambda_1 \mbox{Tr}\Big({Y_e  Y_{e}^{\dagger}  Y_e  Y_{e}^{\dagger}}\Big) \nonumber \\ 
	&\quad \quad -\frac{8}{5} g_{1}^{2} \mbox{Tr}\Big({Y_u  Y_{u}^{\dagger}  Y_u  Y_{u}^{\dagger}}\Big) -32 g_{3}^{2} \mbox{Tr}\Big({Y_u  Y_{u}^{\dagger}  Y_u  Y_{u}^{\dagger}}\Big) -3 \lambda_1 \mbox{Tr}\Big({Y_u  Y_{u}^{\dagger}  Y_u  Y_{u}^{\dagger}}\Big) +30 \mbox{Tr}\Big({Y_d  Y_{d}^{\dagger}  Y_d  Y_{d}^{\dagger}  Y_d  Y_{d}^{\dagger}}\Big) \nonumber \\ 
	&\quad \quad -12 \mbox{Tr}\Big({Y_d  Y_{d}^{\dagger}  Y_d  Y_{u}^{\dagger}  Y_u  Y_{d}^{\dagger}}\Big) +6 \mbox{Tr}\Big({Y_d  Y_{u}^{\dagger}  Y_u  Y_{d}^{\dagger}  Y_d  Y_{d}^{\dagger}}\Big) -6 \mbox{Tr}\Big({Y_d  Y_{u}^{\dagger}  Y_u  Y_{u}^{\dagger}  Y_u  Y_{d}^{\dagger}}\Big) \nonumber \\ 
	&\quad \quad +10 \mbox{Tr}\Big({Y_e  Y_{e}^{\dagger}  Y_e  Y_{e}^{\dagger}  Y_e  Y_{e}^{\dagger}}\Big) +30 \mbox{Tr}\Big({Y_u  Y_{u}^{\dagger}  Y_u  Y_{u}^{\dagger}  Y_u  Y_{u}^{\dagger}}\Big)\Bigg] \, . \\
	\beta_{\lambda_t} \  = \ &
	\frac{1}{16\pi^2}\Bigg[
	-24 g_{2}^{2} \lambda_t  + 88 \lambda_{t}^{2}  + 8 \lambda_{ht}^{2}  + \frac{3}{2} g_{2}^{4}\Bigg] \nonumber \\
	& +\frac{1}{(16\pi^2)^2}\Bigg[-\frac{68}{3} g_{2}^{6} +10 g_{2}^{4} \lambda_{ht} +\frac{48}{5} g_{1}^{2} \lambda_{ht}^{2} +48 g_{2}^{2} \lambda_{ht}^{2} -128 \lambda_{ht}^{3} +\frac{94}{3} g_{2}^{4} \lambda_t -320 \lambda_{ht}^{2} \lambda_t +640 g_{2}^{2} \lambda_{t}^{2}\nonumber \\ 
	&\quad \quad  -4416 \lambda_{t}^{3} -48 \lambda_{ht}^{2} \mbox{Tr}\Big({Y_d  Y_{d}^{\dagger}}\Big) -16 \lambda_{ht}^{2} \mbox{Tr}\Big({Y_e  Y_{e}^{\dagger}}\Big) -48 \lambda_{ht}^{2} \mbox{Tr}\Big({Y_u  Y_{u}^{\dagger}}\Big)\Bigg] \, .  \\
	\beta_{\lambda_{ht}} \ =  \ &
	\frac{1}{16\pi^2}\Bigg[\frac{3}{4} g_{2}^{4} -\frac{9}{10} g_{1}^{2} \lambda_{ht} -\frac{33}{2} g_{2}^{2} \lambda_{ht} +12 \lambda_1 \lambda_{ht} +16 \lambda_{ht}^{2} +24 \lambda_{ht} \lambda_t +6 \lambda_{ht} \mbox{Tr}\Big({Y_d  Y_{d}^{\dagger}}\Big) +2 \lambda_{ht} \mbox{Tr}\Big({Y_e  Y_{e}^{\dagger}}\Big) \nonumber \\ 
	&\quad \quad +6 \lambda_{ht} \mbox{Tr}\Big({Y_u  Y_{u}^{\dagger}}\Big)\Bigg]\nonumber \\
	&+\frac{1}{(16\pi^2)^2}\Bigg[-\frac{9}{16} g_{1}^{2} g_{2}^{4} +\frac{329}{48} g_{2}^{6} +\frac{15}{2} g_{2}^{4} \lambda_1 +\frac{1671}{400} g_{1}^{4} \lambda_{ht} +\frac{9}{8} g_{1}^{2} g_{2}^{2} \lambda_{ht} -\frac{1087}{48} g_{2}^{4} \lambda_{ht} +\frac{72}{5} g_{1}^{2} \lambda_1 \lambda_{ht}  \nonumber \\ 
	&\quad \quad +72 g_{2}^{2} \lambda_1 \lambda_{ht}-60 \lambda_1^{2} \lambda_{ht} +\frac{12}{5} g_{1}^{2} \lambda_{ht}^{2} +44 g_{2}^{2} \lambda_{ht}^{2} -288 \lambda_1 \lambda_{ht}^{2} -168 \lambda_{ht}^{3} +20 g_{2}^{4} \lambda_t +144 g_{2}^{2} \lambda_{ht} \lambda_t  \nonumber \\ 
	&\quad \quad -576 \lambda_{ht}^{2} \lambda_t-544 \lambda_{ht} \lambda_{t}^{2} -\frac{1}{4} \Big(3 g_{2}^{4}  -45 g_{2}^{2} \lambda_{ht}  + \lambda_{ht} \Big(-160 g_{3}^{2}  + 192 \lambda_{ht}  + 288 \lambda  -5 g_{1}^{2} \Big)\Big)\mbox{Tr}\Big({Y_d  Y_{d}^{\dagger}}\Big) \nonumber \\ 
	&\quad \quad -\frac{1}{4} \Big(-15 g_{2}^{2} \lambda_{ht}  + \lambda_{ht} \Big(-15 g_{1}^{2}  + 64 \lambda_{ht}  + 96 \lambda_1 \Big) + g_{2}^{4}\Big)\mbox{Tr}\Big({Y_e  Y_{e}^{\dagger}}\Big) -\frac{3}{4} g_{2}^{4} \mbox{Tr}\Big({Y_u  Y_{u}^{\dagger}}\Big)  \nonumber \\ 
	&\quad \quad +\frac{17}{4} g_{1}^{2} \lambda_{ht} \mbox{Tr}\Big({Y_u  Y_{u}^{\dagger}}\Big)+\frac{45}{4} g_{2}^{2} \lambda_{ht} \mbox{Tr}\Big({Y_u  Y_{u}^{\dagger}}\Big) +40 g_{3}^{2} \lambda_{ht} \mbox{Tr}\Big({Y_u  Y_{u}^{\dagger}}\Big) -72 \lambda_1 \lambda_{ht} \mbox{Tr}\Big({Y_u  Y_{u}^{\dagger}}\Big) \nonumber \\ 
	&\quad \quad  -48 \lambda_{ht}^{2} \mbox{Tr}\Big({Y_u  Y_{u}^{\dagger}}\Big)-\frac{27}{2} \lambda_{ht} \mbox{Tr}\Big({Y_d  Y_{d}^{\dagger}  Y_d  Y_{d}^{\dagger}}\Big) -21 \lambda_{ht} \mbox{Tr}\Big({Y_d  Y_{u}^{\dagger}  Y_u  Y_{d}^{\dagger}}\Big) -\frac{9}{2} \lambda_{ht} \mbox{Tr}\Big({Y_e  Y_{e}^{\dagger}  Y_e  Y_{e}^{\dagger}}\Big) \nonumber
	\\
	&\quad \quad  -\frac{27}{2} \lambda_{ht} \mbox{Tr}\Big({Y_u  Y_{u}^{\dagger}  Y_u  Y_{u}^{\dagger}}\Big) \Bigg] \, . \\
	\end{align*}
	\endgroup
	\section{Two-loop $\beta$-functions for Singlet} \label{betaf2}
	\subsection{Scalar Quartic Couplings}\label{B1}
	\footnotesize{
		\begingroup
		\allowdisplaybreaks
		\begin{align*}
		\beta_{\lambda}^{(1)} & =  
		+\frac{27}{200} g_{1}^{4} +\frac{9}{20} g_{1}^{2} g_{2}^{2} +\frac{9}{8} g_{2}^{4} -\frac{9}{5} g_{1}^{2} \lambda_1 -9 g_{2}^{2} \lambda_1 +24 \lambda_1^{2} +4 \lambda_{hs}^{2} +12 \lambda_1 \mbox{Tr}\Big({Y_d  Y_{d}^{\dagger}}\Big) +4 \lambda_1 \mbox{Tr}\Big({Y_e  Y_{e}^{\dagger}}\Big) \nonumber \\ 
		&+12 \lambda_1 \mbox{Tr}\Big({Y_u  Y_{u}^{\dagger}}\Big) -6 \mbox{Tr}\Big({Y_d  Y_{d}^{\dagger}  Y_d  Y_{d}^{\dagger}}\Big) -2 \mbox{Tr}\Big({Y_e  Y_{e}^{\dagger}  Y_e  Y_{e}^{\dagger}}\Big) -6 \mbox{Tr}\Big({Y_u  Y_{u}^{\dagger}  Y_u  Y_{u}^{\dagger}}\Big) \Bigg] \nonumber \\
		& +\frac{1}{(16\pi^2)^2}\Bigg[-\frac{3411}{2000} g_{1}^{6} -\frac{1677}{400} g_{1}^{4} g_{2}^{2} -\frac{289}{80} g_{1}^{2} g_{2}^{4} +\frac{305}{16} g_{2}^{6} +\frac{1887}{200} g_{1}^{4} \lambda_1 +\frac{117}{20} g_{1}^{2} g_{2}^{2} \lambda_1  \nonumber \\ 
		&-\frac{73}{8} g_{2}^{4} \lambda_1 +\frac{108}{5} g_{1}^{2} \lambda_1^{2} +108 g_{2}^{2} \lambda_1^{2}-312 \lambda_1^{3} -40 \lambda_1 \lambda_{hs}^{2} -32 \lambda_{hs}^{3} \nonumber \\ 
		&+\frac{1}{20} \Big(-5 \Big(64 \lambda_1 \Big(-5 g_{3}^{2}  + 9 \lambda_1 \Big) -90 g_{2}^{2} \lambda_1  + 9 g_{2}^{4} \Big) + 9 g_{1}^{4}  + g_{1}^{2} \Big(50 \lambda_1  + 54 g_{2}^{2} \Big)\Big)\mbox{Tr}\Big({Y_d  Y_{d}^{\dagger}}\Big) \nonumber \\ 
		&-\frac{3}{20} \Big(15 g_{1}^{4}  -2 g_{1}^{2} \Big(11 g_{2}^{2}  + 25 \lambda_1 \Big) + 5 \Big(-10 g_{2}^{2} \lambda_1  + 64 \lambda_1^{2}  + g_{2}^{4}\Big)\Big)\mbox{Tr}\Big({Y_e  Y_{e}^{\dagger}}\Big) -\frac{171}{100} g_{1}^{4} \mbox{Tr}\Big({Y_u  Y_{u}^{\dagger}}\Big) \nonumber \\ 
		&+\frac{63}{10} g_{1}^{2} g_{2}^{2} \mbox{Tr}\Big({Y_u  Y_{u}^{\dagger}}\Big) -\frac{9}{4} g_{2}^{4} \mbox{Tr}\Big({Y_u  Y_{u}^{\dagger}}\Big) +\frac{17}{2} g_{1}^{2} \lambda_1 \mbox{Tr}\Big({Y_u  Y_{u}^{\dagger}}\Big) +\frac{45}{2} g_{2}^{2} \lambda_1 \mbox{Tr}\Big({Y_u  Y_{u}^{\dagger}}\Big) \nonumber \\ 
		&+80 g_{3}^{2} \lambda_1 \mbox{Tr}\Big({Y_u  Y_{u}^{\dagger}}\Big) -144 \lambda_1^{2} \mbox{Tr}\Big({Y_u  Y_{u}^{\dagger}}\Big) +\frac{4}{5} g_{1}^{2} \mbox{Tr}\Big({Y_d  Y_{d}^{\dagger}  Y_d  Y_{d}^{\dagger}}\Big) -32 g_{3}^{2} \mbox{Tr}\Big({Y_d  Y_{d}^{\dagger}  Y_d  Y_{d}^{\dagger}}\Big) \nonumber \\ 
		&-3 \lambda_1 \mbox{Tr}\Big({Y_d  Y_{d}^{\dagger}  Y_d  Y_{d}^{\dagger}}\Big) -42 \lambda \mbox{Tr}\Big({Y_d  Y_{u}^{\dagger}  Y_u  Y_{d}^{\dagger}}\Big) -\frac{12}{5} g_{1}^{2} \mbox{Tr}\Big({Y_e  Y_{e}^{\dagger}  Y_e  Y_{e}^{\dagger}}\Big) - \lambda_1 \mbox{Tr}\Big({Y_e  Y_{e}^{\dagger}  Y_e  Y_{e}^{\dagger}}\Big) \nonumber \\ 
		&-\frac{8}{5} g_{1}^{2} \mbox{Tr}\Big({Y_u  Y_{u}^{\dagger}  Y_u  Y_{u}^{\dagger}}\Big) -32 g_{3}^{2} \mbox{Tr}\Big({Y_u  Y_{u}^{\dagger}  Y_u  Y_{u}^{\dagger}}\Big) -3 \lambda_1 \mbox{Tr}\Big({Y_u  Y_{u}^{\dagger}  Y_u  Y_{u}^{\dagger}}\Big) +30 \mbox{Tr}\Big({Y_d  Y_{d}^{\dagger}  Y_d  Y_{d}^{\dagger}  Y_d  Y_{d}^{\dagger}}\Big) \nonumber \\ 
		&-12 \mbox{Tr}\Big({Y_d  Y_{d}^{\dagger}  Y_d  Y_{u}^{\dagger}  Y_u  Y_{d}^{\dagger}}\Big) +6 \mbox{Tr}\Big({Y_d  Y_{u}^{\dagger}  Y_u  Y_{d}^{\dagger}  Y_d  Y_{d}^{\dagger}}\Big) -6 \mbox{Tr}\Big({Y_d  Y_{u}^{\dagger}  Y_u  Y_{u}^{\dagger}  Y_u  Y_{d}^{\dagger}}\Big) \nonumber \\ 
		&+10 \mbox{Tr}\Big({Y_e  Y_{e}^{\dagger}  Y_e  Y_{e}^{\dagger}  Y_e  Y_{e}^{\dagger}}\Big) +30 \mbox{Tr}\Big({Y_u  Y_{u}^{\dagger}  Y_u  Y_{u}^{\dagger}  Y_u  Y_{u}^{\dagger}}\Big)\Bigg] \, . \\
		\beta_{\lambda_s} \  = \ &
		\frac{1}{16\pi^2}\Bigg[
		20 \lambda_{t}^{2}  + 8 \lambda_{hs}^{2}\Bigg] \nonumber \\
		& +\frac{1}{(16\pi^2)^2}\Bigg[\frac{16}{5} \Big(3 g_{1}^{2} \lambda_{hs}^{2} +15 g_{2}^{2} \lambda_{hs}^{2} -20 \lambda_{hs}^{3} -25 \lambda_{hs}^{2} \lambda_t -75 \lambda_{t}^{3} -15 \lambda_{hs}^{2} \mbox{Tr}\Big({Y_d  Y_{d}^{\dagger}}\Big) -5 \lambda_{hs}^{2} \mbox{Tr}\Big({Y_e  Y_{e}^{\dagger}}\Big) \nonumber \\ 
		&-15 \lambda_{hs}^{2} \mbox{Tr}\Big({Y_u  Y_{u}^{\dagger}}\Big) \Big)\Bigg] \, .  \\
		\beta_{\lambda_{hs}} \ =  \ &
		\frac{1}{16\pi^2}\Bigg[\frac{1}{10} \lambda_{hs} \Big(120 \lambda_1  + 20 \mbox{Tr}\Big({Y_e  Y_{e}^{\dagger}}\Big)  -45 g_{2}^{2}  + 60 \mbox{Tr}\Big({Y_d  Y_{d}^{\dagger}}\Big)  + 60 \mbox{Tr}\Big({Y_u  Y_{u}^{\dagger}}\Big)  + 80 \lambda_{hs}  + 80 \lambda_t  -9 g_{1}^{2} \Big)\Bigg]\nonumber \\
		&+\frac{1}{(16\pi^2)^2}\Bigg[+\frac{1671}{400} g_{1}^{4} \lambda_{hs} +\frac{9}{8} g_{1}^{2} g_{2}^{2} \lambda_{hs} -\frac{145}{16} g_{2}^{4} \lambda_{hs} +\frac{72}{5} g_{1}^{2} \lambda_1 \lambda_{hs} +72 g_{2}^{2} \lambda_1 \lambda_{hs} -60 \lambda_1^{2} \lambda_{hs} \nonumber \\ 
		& +\frac{6}{5} g_{1}^{2} \lambda_{hs}^{2} +6 g_{2}^{2} \lambda_{hs}^{2}-144 \lambda_1 \lambda_{hs}^{2} -44 \lambda_{hs}^{3} -96 \lambda_{hs}^{2} \lambda_t -40 \lambda_{hs} \lambda_{t}^{2} \nonumber \\
		& +\frac{1}{4} \Big(32 \Big(-3 \lambda_{hs}  + 5 g_{3}^{2}  -9 \lambda_1 \Big) + 45 g_{2}^{2}  + 5 g_{1}^{2} \Big)\lambda_{hs} \mbox{Tr}\Big({Y_d  Y_{d}^{\dagger}}\Big) \nonumber \\ 
		&+\frac{1}{4} \lambda_{hs} \Big(15 g_{1}^{2}  + 15 g_{2}^{2}  -32 \Big(3 \lambda_1  + \lambda_{hs}\Big)\Big)\mbox{Tr}\Big({Y_e  Y_{e}^{\dagger}}\Big) +\frac{17}{4} g_{1}^{2} \lambda_{hs} \mbox{Tr}\Big({Y_u  Y_{u}^{\dagger}}\Big) +\frac{45}{4} g_{2}^{2} \lambda_{hs} \mbox{Tr}\Big({Y_u  Y_{u}^{\dagger}}\Big) \nonumber \\ 
		&+40 g_{3}^{2} \lambda_{hs} \mbox{Tr}\Big({Y_u  Y_{u}^{\dagger}}\Big) -72 \lambda_1 \lambda_{hs} \mbox{Tr}\Big({Y_u  Y_{u}^{\dagger}}\Big) -24 \lambda_{hs}^{2} \mbox{Tr}\Big({Y_u  Y_{u}^{\dagger}}\Big) -\frac{27}{2} \lambda_{hs} \mbox{Tr}\Big({Y_d  Y_{d}^{\dagger}  Y_d  Y_{d}^{\dagger}}\Big) \nonumber \\ 
		&-21 \lambda_{hs} \mbox{Tr}\Big({Y_d  Y_{u}^{\dagger}  Y_u  Y_{d}^{\dagger}}\Big) -\frac{9}{2} \lambda_{hs} \mbox{Tr}\Big({Y_e  Y_{e}^{\dagger}  Y_e  Y_{e}^{\dagger}}\Big) -\frac{27}{2} \lambda_{hs} \mbox{Tr}\Big({Y_u  Y_{u}^{\dagger}  Y_u  Y_{u}^{\dagger}}\Big) \Bigg] \, . \\
		\end{align*}
		\endgroup
\bibliography{References}

\providecommand{\href}[2]{#2}\begingroup\begin{thebibliography}{100}

\bibitem{ATLAS:2012yve}
{\scshape ATLAS} collaboration, G.~Aad et~al., \textit{{Observation of a new
  particle in the search for the Standard Model Higgs boson with the ATLAS
  detector at the LHC}},
  \href{https://doi.org/10.1016/j.physletb.2012.08.020}{\textit{Phys. Lett. B}
  {\bfseries 716} (2012) 1--29},
  [\href{https://arxiv.org/abs/1207.7214}{{\ttfamily 1207.7214}}].

\bibitem{CMS:2012qbp}
{\scshape CMS} collaboration, S.~Chatrchyan et~al., \textit{{Observation of a
  New Boson at a Mass of 125 GeV with the CMS Experiment at the LHC}},
  \href{https://doi.org/10.1016/j.physletb.2012.08.021}{\textit{Phys. Lett. B}
  {\bfseries 716} (2012) 30--61},
  [\href{https://arxiv.org/abs/1207.7235}{{\ttfamily 1207.7235}}].

\bibitem{KAJANTIE1996189}
K.~Kajantie, M.~Laine, K.~Rummukainen and M.~Shaposhnikov, \textit{The
  electroweak phase transition: a non-perturbative analysis},
  \href{https://doi.org/10.1016/0550-3213(96)00052-1}{\textit{Nuclear Physics
  B} {\bfseries 466} (Apr, 1996) 189--258}.

\bibitem{Laine1996}
K.~Kajantie, M.~Laine, K.~Rummukainen and M.~Shaposhnikov, \textit{Is there a
  hot electroweak phase transition at $m_h \gsim m_w$},
  \href{https://doi.org/10.1103/physrevlett.77.2887}{\textit{Physical Review
  Letters} {\bfseries 77} (Sep, 1996) 2887--2890}.

\bibitem{Schiller1997}
M.~Gurtler, E.-M. Ilgenfritz and A.~Schiller, \textit{Where the electroweak
  phase transition ends},
  \href{https://doi.org/10.1103/physrevd.56.3888}{\textit{Physical Review D}
  {\bfseries 56} (Oct, 1997) 3888--3895}.

\bibitem{Csikor1999}
F.~Csikor, Z.~Fodor and J.~Heitger, \textit{End point of the hot electroweak
  phase transition},
  \href{https://doi.org/10.1103/physrevlett.82.21}{\textit{Physical Review
  Letters} {\bfseries 82} (Jan, 1999) 21--24}.

\bibitem{Michela2014}
M.~D'Onofrio, K.~Rummukainen and A.~Tranberg, \textit{Sphaleron rate in the
  minimal standard model},
  \href{https://doi.org/10.1103/physrevlett.113.141602}{\textit{Physical Review
  Letters} {\bfseries 113} (Oct, 2014) }.

\bibitem{Kuzmin:1985mm}
V.~A. Kuzmin, V.~A. Rubakov and M.~E. Shaposhnikov, \textit{{On the Anomalous
  Electroweak Baryon Number Nonconservation in the Early Universe}},
  \href{https://doi.org/10.1016/0370-2693(85)91028-7}{\textit{Phys. Lett. B}
  {\bfseries 155} (1985) 36}.

\bibitem{Riotto:1999yt}
A.~Riotto and M.~Trodden, \textit{{Recent progress in baryogenesis}},
  \href{https://doi.org/10.1146/annurev.nucl.49.1.35}{\textit{Ann. Rev. Nucl.
  Part. Sci.} {\bfseries 49} (1999) 35--75},
  [\href{https://arxiv.org/abs/hep-ph/9901362}{{\ttfamily hep-ph/9901362}}].

\bibitem{Morrissey:2012db}
D.~E. Morrissey and M.~J. Ramsey-Musolf, \textit{{Electroweak baryogenesis}},
  \href{https://doi.org/10.1088/1367-2630/14/12/125003}{\textit{New J. Phys.}
  {\bfseries 14} (2012) 125003},
  [\href{https://arxiv.org/abs/1206.2942}{{\ttfamily 1206.2942}}].

\bibitem{Jangid:2020dqh}
S.~Jangid, P.~Bandyopadhyay, P.~S. Bhupal~Dev and A.~Kumar, \textit{{Vacuum
  stability in inert higgs doublet model with right-handed neutrinos}},
  \href{https://doi.org/10.1007/JHEP08(2020)154}{\textit{JHEP} {\bfseries 08}
  (2020) 154}, [\href{https://arxiv.org/abs/2001.01764}{{\ttfamily
  2001.01764}}].

\bibitem{Jangid:2020qgo}
S.~Jangid and P.~Bandyopadhyay, \textit{{Distinguishing Inert Higgs Doublet and
  Inert Triplet Scenarios}},
  \href{https://doi.org/10.1140/epjc/s10052-020-8271-5}{\textit{Eur. Phys. J.
  C} {\bfseries 80} (2020) 715},
  [\href{https://arxiv.org/abs/2003.11821}{{\ttfamily 2003.11821}}].

\bibitem{Bandyopadhyay:2020djh}
P.~Bandyopadhyay, S.~Jangid and M.~Mitra, \textit{{Scrutinizing Vacuum
  Stability in IDM with Type-III Inverse seesaw}},
  \href{https://doi.org/10.1007/JHEP02(2021)075}{\textit{JHEP} {\bfseries 02}
  (2021) 075}, [\href{https://arxiv.org/abs/2008.11956}{{\ttfamily
  2008.11956}}].

\bibitem{sjak}
P.~Bandyopadhyay, S.~Jangid and A.~Karan, \textit{{Constraining scalar Doublet
  and Triplet Leptoquarks with Vacuum stability and perturbativity }},
  \href{https://arxiv.org/abs/To be appear soon}{{\ttfamily To be appear
  soon}}.

\bibitem{Carena1996}
M.~Carena, M.~Quiros and C.~Wagner, \textit{Opening the window for electroweak
  baryogenesis},
  \href{https://doi.org/10.1016/0370-2693(96)00475-3}{\textit{Physics Letters
  B} {\bfseries 380} (Jul, 1996) 81--91}.

\bibitem{Quiros:1999jp}
M.~Quiros, \textit{{Finite temperature field theory and phase transitions}},
  in \textit{{ICTP Summer School in High-Energy Physics and Cosmology}},
  pp.~187--259, 1, 1999, \href{https://arxiv.org/abs/hep-ph/9901312}{{\ttfamily
  hep-ph/9901312}}.

\bibitem{Delepine1996}
D.~Delepine, J.-M. Gerard, R.~Gonzalez~Felipe and J.~Weyers, \textit{A light
  stop and electroweak baryogenesis},
  \href{https://doi.org/10.1016/0370-2693(96)00921-5}{\textit{Physics Letters
  B} {\bfseries 386} (Oct, 1996) 183--188}.

\bibitem{Laine1998}
M.~Laine and K.~Rummukainen, \textit{The mssm electroweak phase transition on
  the lattice},
  \href{https://doi.org/10.1016/s0550-3213(98)00530-6}{\textit{Nuclear Physics
  B} {\bfseries 535} (Dec, 1998) 423--457}.

\bibitem{Grojean2005}
C.~Grojean, G.~Servant and J.~D. Wells, \textit{First-order electroweak phase
  transition in the standard model with a low cutoff},
  \href{https://doi.org/10.1103/physrevd.71.036001}{\textit{Physical Review D}
  {\bfseries 71} (Feb, 2005) }.

\bibitem{Huber2001}
S.~Huber and M.~Schmidt, \textit{Electroweak baryogenesis: concrete in a susy
  model with a gauge singlet},
  \href{https://doi.org/10.1016/s0550-3213(01)00250-4}{\textit{Nuclear Physics
  B} {\bfseries 606} (Jul, 2001) 183--230}.

\bibitem{Huber2006}
S.~J. Huber, T.~Konstandin, T.~Prokopec and M.~G. Schmidt, \textit{Electroweak
  phase transition and baryogenesis in the nmssm},
  \href{https://doi.org/10.1016/j.nuclphysb.2006.09.003}{\textit{Nuclear
  Physics B} {\bfseries 757} (Nov, 2006) 172–196}.

\bibitem{Kanemura:2012hr}
S.~Kanemura, E.~Senaha, T.~Shindou and T.~Yamada, \textit{{Electroweak phase
  transition and Higgs boson couplings in the model based on supersymmetric
  strong dynamics}},
  \href{https://doi.org/10.1007/JHEP05(2013)066}{\textit{JHEP} {\bfseries 05}
  (2013) 066}, [\href{https://arxiv.org/abs/1211.5883}{{\ttfamily 1211.5883}}].

\bibitem{Cheung:2012pg}
K.~Cheung, T.-J. Hou, J.~S. Lee and E.~Senaha, \textit{{Singlino-driven
  Electroweak Baryogenesis in the Next-to-MSSM}},
  \href{https://doi.org/10.1016/j.physletb.2012.02.070}{\textit{Phys. Lett. B}
  {\bfseries 710} (2012) 188--191},
  [\href{https://arxiv.org/abs/1201.3781}{{\ttfamily 1201.3781}}].

\bibitem{Kanemura:2011fy}
S.~Kanemura, E.~Senaha and T.~Shindou, \textit{{First-order electroweak phase
  transition powered by additional F-term loop effects in an extended
  supersymmetric Higgs sector}},
  \href{https://doi.org/10.1016/j.physletb.2011.10.046}{\textit{Phys. Lett. B}
  {\bfseries 706} (2011) 40--45},
  [\href{https://arxiv.org/abs/1109.5226}{{\ttfamily 1109.5226}}].

\bibitem{Chiang:2009fs}
C.-W. Chiang and E.~Senaha, \textit{{Electroweak phase transitions in the
  secluded U(1)-prime-extended MSSM}},
  \href{https://doi.org/10.1007/JHEP06(2010)030}{\textit{JHEP} {\bfseries 06}
  (2010) 030}, [\href{https://arxiv.org/abs/0912.5069}{{\ttfamily 0912.5069}}].

\bibitem{Carena:2012np}
M.~Carena, G.~Nardini, M.~Quiros and C.~E.~M. Wagner, \textit{{MSSM Electroweak
  Baryogenesis and LHC Data}},
  \href{https://doi.org/10.1007/JHEP02(2013)001}{\textit{JHEP} {\bfseries 02}
  (2013) 001}, [\href{https://arxiv.org/abs/1207.6330}{{\ttfamily 1207.6330}}].

\bibitem{Giudice}
G.~F. Giudice, \textit{Electroweak phase transition in supersymmetry},
  \href{https://doi.org/10.1103/PhysRevD.45.3177}{\textit{Phys. Rev. D}
  {\bfseries 45} (May, 1992) 3177--3182}.

\bibitem{Stanley}
S.~Myint, \textit{Baryogenesis constraints on the minimal supersymmetric
  model}, \href{https://doi.org/10.1016/0370-2693(92)90991-c}{\textit{Physics
  Letters B} {\bfseries 287} (Aug, 1992) 325–330}.

\bibitem{Chowdhury2012}
T.~A. Chowdhury, M.~Nemevsek, G.~Senjanovic and Y.~Zhang, \textit{Dark matter
  as the trigger of strong electroweak phase transition},
  \href{https://doi.org/10.1088/1475-7516/2012/02/029}{\textit{Journal of
  Cosmology and Astroparticle Physics} {\bfseries 2012} (Feb, 2012) 029--029}.

\bibitem{Borah2012}
D.~Borah and J.~M. Cline, \textit{Inert doublet dark matter with strong
  electroweak phase transition},
  \href{https://doi.org/10.1103/physrevd.86.055001}{\textit{Physical Review D}
  {\bfseries 86} (Sep, 2012) }.

\bibitem{Gil2012}
G.~Gil, P.~Chankowski and M.~Krawczyk, \textit{Inert dark matter and strong
  electroweak phase transition},
  \href{https://doi.org/10.1016/j.physletb.2012.09.052}{\textit{Physics Letters
  B} {\bfseries 717} (Oct, 2012) 396--402}.

\bibitem{AbdusSalam2014}
S.~S. AbdusSalam and T.~A. Chowdhury, \textit{Scalar representations in the
  light of electroweak phase transition and cold dark matter phenomenology},
  \href{https://doi.org/10.1088/1475-7516/2014/05/026}{\textit{Journal of
  Cosmology and Astroparticle Physics} {\bfseries 2014} (May, 2014) 026–026}.

\bibitem{Cline2013}
J.~M. Cline and K.~Kainulainen, \textit{Improved electroweak phase transition
  with subdominant inert doublet dark matter},
  \href{https://doi.org/10.1103/physrevd.87.071701}{\textit{Physical Review D}
  {\bfseries 87} (Apr, 2013) }.

\bibitem{Vaskonen:2016yiu}
V.~Vaskonen, \textit{{Electroweak baryogenesis and gravitational waves from a
  real scalar singlet}},
  \href{https://doi.org/10.1103/PhysRevD.95.123515}{\textit{Phys. Rev. D}
  {\bfseries 95} (2017) 123515},
  [\href{https://arxiv.org/abs/1611.02073}{{\ttfamily 1611.02073}}].

\bibitem{Profumo:2007wc}
S.~Profumo, M.~J. Ramsey-Musolf and G.~Shaughnessy, \textit{{Singlet Higgs
  phenomenology and the electroweak phase transition}},
  \href{https://doi.org/10.1088/1126-6708/2007/08/010}{\textit{JHEP} {\bfseries
  08} (2007) 010}, [\href{https://arxiv.org/abs/0705.2425}{{\ttfamily
  0705.2425}}].

\bibitem{Ahriche:2007jp}
A.~Ahriche, \textit{{What is the criterion for a strong first order electroweak
  phase transition in singlet models?}},
  \href{https://doi.org/10.1103/PhysRevD.75.083522}{\textit{Phys. Rev. D}
  {\bfseries 75} (2007) 083522},
  [\href{https://arxiv.org/abs/hep-ph/0701192}{{\ttfamily hep-ph/0701192}}].

\bibitem{Espinosa:2011ax}
J.~R. Espinosa, T.~Konstandin and F.~Riva, \textit{{Strong Electroweak Phase
  Transitions in the Standard Model with a Singlet}},
  \href{https://doi.org/10.1016/j.nuclphysb.2011.09.010}{\textit{Nucl. Phys. B}
  {\bfseries 854} (2012) 592--630},
  [\href{https://arxiv.org/abs/1107.5441}{{\ttfamily 1107.5441}}].

\bibitem{Cline:2012hg}
J.~M. Cline and K.~Kainulainen, \textit{{Electroweak baryogenesis and dark
  matter from a singlet Higgs}},
  \href{https://doi.org/10.1088/1475-7516/2013/01/012}{\textit{JCAP} {\bfseries
  01} (2013) 012}, [\href{https://arxiv.org/abs/1210.4196}{{\ttfamily
  1210.4196}}].

\bibitem{Cline:2013gha}
J.~M. Cline, K.~Kainulainen, P.~Scott and C.~Weniger, \textit{{Update on scalar
  singlet dark matter}},
  \href{https://doi.org/10.1103/PhysRevD.88.055025}{\textit{Phys. Rev. D}
  {\bfseries 88} (2013) 055025},
  [\href{https://arxiv.org/abs/1306.4710}{{\ttfamily 1306.4710}}]. [Erratum:
  Phys.Rev.D 92, 039906 (2015)].

\bibitem{Barger:2008jx}
V.~Barger, P.~Langacker, M.~McCaskey, M.~Ramsey-Musolf and G.~Shaughnessy,
  \textit{{Complex Singlet Extension of the Standard Model}},
  \href{https://doi.org/10.1103/PhysRevD.79.015018}{\textit{Phys. Rev. D}
  {\bfseries 79} (2009) 015018},
  [\href{https://arxiv.org/abs/0811.0393}{{\ttfamily 0811.0393}}].

\bibitem{Gonderinger:2012rd}
M.~Gonderinger, H.~Lim and M.~J. Ramsey-Musolf, \textit{{Complex Scalar Singlet
  Dark Matter: Vacuum Stability and Phenomenology}},
  \href{https://doi.org/10.1103/PhysRevD.86.043511}{\textit{Phys. Rev. D}
  {\bfseries 86} (2012) 043511},
  [\href{https://arxiv.org/abs/1202.1316}{{\ttfamily 1202.1316}}].

\bibitem{Ahriche:2012ei}
A.~Ahriche and S.~Nasri, \textit{{Light Dark Matter, Light Higgs and the
  Electroweak Phase Transition}},
  \href{https://doi.org/10.1103/PhysRevD.85.093007}{\textit{Phys. Rev. D}
  {\bfseries 85} (2012) 093007},
  [\href{https://arxiv.org/abs/1201.4614}{{\ttfamily 1201.4614}}].

\bibitem{Brauner:2016fla}
T.~Brauner, T.~V.~I. Tenkanen, A.~Tranberg, A.~Vuorinen and D.~J. Weir,
  \textit{{Dimensional reduction of the Standard Model coupled to a new singlet
  scalar field}}, \href{https://doi.org/10.1007/JHEP03(2017)007}{\textit{JHEP}
  {\bfseries 03} (2017) 007},
  [\href{https://arxiv.org/abs/1609.06230}{{\ttfamily 1609.06230}}].

\bibitem{Carena:2019une}
M.~Carena, Z.~Liu and Y.~Wang, \textit{{Electroweak phase transition with
  spontaneous Z$_{2}$-breaking}},
  \href{https://doi.org/10.1007/JHEP08(2020)107}{\textit{JHEP} {\bfseries 08}
  (2020) 107}, [\href{https://arxiv.org/abs/1911.10206}{{\ttfamily
  1911.10206}}].

\bibitem{Carena:2018vpt}
M.~Carena, Z.~Liu and M.~Riembau, \textit{{Probing the electroweak phase
  transition via enhanced di-Higgs boson production}},
  \href{https://doi.org/10.1103/PhysRevD.97.095032}{\textit{Phys. Rev. D}
  {\bfseries 97} (2018) 095032},
  [\href{https://arxiv.org/abs/1801.00794}{{\ttfamily 1801.00794}}].

\bibitem{Parsa}
P.~Ghorbani, \textit{Vacuum structure and electroweak phase transition in
  singlet scalar dark matter},
  \href{https://doi.org/10.1016/j.dark.2021.100861}{\textit{Physics of the Dark
  Universe} {\bfseries 33} (Sep, 2021) 100861}.

\bibitem{ghorbani2019strongly}
K.~Ghorbani and P.~H. Ghorbani, \textit{Strongly first-order phase transition
  in real singlet scalar dark matter model},  2019.

\bibitem{Tenkanen}
P.~M. Schicho, T.~V.~I. Tenkanen and J.~Österman, \textit{Robust approach to
  thermal resummation: Standard model meets a singlet},
  \href{https://doi.org/10.1007/jhep06(2021)130}{\textit{Journal of High Energy
  Physics} {\bfseries 2021} (Jun, 2021) }.

\bibitem{Schicho}
L.~Niemi, P.~Schicho and T.~V. Tenkanen, \textit{Singlet-assisted electroweak
  phase transition at two loops},
  \href{https://doi.org/10.1103/physrevd.103.115035}{\textit{Physical Review D}
  {\bfseries 103} (Jun, 2021) }.

\bibitem{Fuyuto:2014yia}
K.~Fuyuto and E.~Senaha, \textit{{Improved sphaleron decoupling condition and
  the Higgs coupling constants in the real singlet-extended standard model}},
  \href{https://doi.org/10.1103/PhysRevD.90.015015}{\textit{Phys. Rev. D}
  {\bfseries 90} (2014) 015015},
  [\href{https://arxiv.org/abs/1406.0433}{{\ttfamily 1406.0433}}].

\bibitem{Huber2013}
G.~C. Dorsch, S.~J. Huber and J.~M. No, \textit{A strong electroweak phase
  transition in the 2hdm after lhc8},
  \href{https://doi.org/10.1007/jhep10(2013)029}{\textit{Journal of High Energy
  Physics} {\bfseries 2013} (Oct, 2013) }.

\bibitem{Kanemura:2005cj}
S.~Kanemura, Y.~Okada and E.~Senaha, \textit{{Electroweak baryogenesis and the
  triple Higgs boson coupling}}, {\textit{eConf} {\bfseries C050318} (2005)
  0704}, [\href{https://arxiv.org/abs/hep-ph/0507259}{{\ttfamily
  hep-ph/0507259}}].

\bibitem{Andersen:2017ika}
J.~O. Andersen, T.~Gorda, A.~Helset, L.~Niemi, T.~V.~I. Tenkanen, A.~Tranberg
  et~al., \textit{{Nonperturbative Analysis of the Electroweak Phase Transition
  in the Two Higgs Doublet Model}},
  \href{https://doi.org/10.1103/PhysRevLett.121.191802}{\textit{Phys. Rev.
  Lett.} {\bfseries 121} (2018) 191802},
  [\href{https://arxiv.org/abs/1711.09849}{{\ttfamily 1711.09849}}].

\bibitem{Barman:2019oda}
B.~Barman, A.~Dutta~Banik and A.~Paul, \textit{{Singlet-doublet fermionic dark
  matter and gravitational waves in a two-Higgs-doublet extension of the
  Standard Model}},
  \href{https://doi.org/10.1103/PhysRevD.101.055028}{\textit{Phys. Rev. D}
  {\bfseries 101} (2020) 055028},
  [\href{https://arxiv.org/abs/1912.12899}{{\ttfamily 1912.12899}}].

\bibitem{Patel:2012pi}
H.~H. Patel and M.~J. Ramsey-Musolf, \textit{{Stepping Into Electroweak
  Symmetry Breaking: Phase Transitions and Higgs Phenomenology}},
  \href{https://doi.org/10.1103/PhysRevD.88.035013}{\textit{Phys. Rev. D}
  {\bfseries 88} (2013) 035013},
  [\href{https://arxiv.org/abs/1212.5652}{{\ttfamily 1212.5652}}].

\bibitem{Kazemi:2021bzj}
M.~J. Kazemi and S.~S. Abdussalam, \textit{{Electroweak Phase Transition in an
  Inert Complex Triplet Model}},
  \href{https://doi.org/10.1103/PhysRevD.103.075012}{\textit{Phys. Rev. D}
  {\bfseries 103} (2021) 075012},
  [\href{https://arxiv.org/abs/2103.00212}{{\ttfamily 2103.00212}}].

\bibitem{niemi}
L.~Niemi, H.~H. Patel, M.~J. Ramsey-Musolf, T.~V.~I. Tenkanen and D.~J. Weir,
  \textit{Electroweak phase transition in the real triplet extension of the sm:
  Dimensional reduction},
  \href{https://doi.org/10.1103/PhysRevD.100.035002}{\textit{Phys. Rev. D}
  {\bfseries 100} (Aug, 2019) 035002}.

\bibitem{Cho:2021itv}
G.-C. Cho, C.~Idegawa and E.~Senaha, \textit{{Electroweak phase transition in a
  complex singlet extension of the Standard Model with degenerate scalars}},
  \href{https://arxiv.org/abs/2105.11830}{{\ttfamily 2105.11830}}.

\bibitem{Chiang:2018gsn}
C.-W. Chiang, Y.-T. Li and E.~Senaha, \textit{{Revisiting electroweak phase
  transition in the standard model with a real singlet scalar}},
  \href{https://doi.org/10.1016/j.physletb.2018.12.017}{\textit{Phys. Lett. B}
  {\bfseries 789} (2019) 154--159},
  [\href{https://arxiv.org/abs/1808.01098}{{\ttfamily 1808.01098}}].

\bibitem{Paul:2019pgt}
A.~Paul, B.~Banerjee and D.~Majumdar, \textit{{Gravitational wave signatures
  from an extended inert doublet dark matter model}},
  \href{https://doi.org/10.1088/1475-7516/2019/10/062}{\textit{JCAP} {\bfseries
  10} (2019) 062}, [\href{https://arxiv.org/abs/1908.00829}{{\ttfamily
  1908.00829}}].

\bibitem{tripletsinglet2021}
N.~F. Bell, M.~J. Dolan, L.~S. Friedrich, M.~J. Ramsey-Musolf and R.~R. Volkas,
  \textit{A real triplet-singlet extended standard model: dark matter and
  collider phenomenology},
  \href{https://doi.org/10.1007/jhep04(2021)098}{\textit{Journal of High Energy
  Physics} {\bfseries 2021} (Apr, 2021) }.

\bibitem{Shajiee}
V.~R. Shajiee and A.~Tofighi, \textit{Electroweak phase transition,
  gravitational waves and dark matter in two scalar singlet extension of the
  standard model},
  \href{https://doi.org/10.1140/epjc/s10052-019-6881-6}{\textit{The European
  Physical Journal C} {\bfseries 79} (Apr, 2019) }.

\bibitem{Hitoshi}
E.~Hall, T.~Konstandin, R.~McGehee, H.~Murayama and G.~Servant,
  \textit{Baryogenesis from a dark first-order phase transition},
  \href{https://doi.org/10.1007/jhep04(2020)042}{\textit{Journal of High Energy
  Physics} {\bfseries 2020} (Apr, 2020) }.

\bibitem{Garcia-Pepin:2016hvs}
M.~Garcia-Pepin and M.~Quiros, \textit{{Strong electroweak phase transition
  from Supersymmetric Custodial Triplets}},
  \href{https://doi.org/10.1007/JHEP05(2016)177}{\textit{JHEP} {\bfseries 05}
  (2016) 177}, [\href{https://arxiv.org/abs/1602.01351}{{\ttfamily
  1602.01351}}].

\bibitem{LINDE1983421}
A.~Linde, \textit{Decay of the false vacuum at finite temperature},
  \href{https://doi.org/https://doi.org/10.1016/0550-3213(83)90293-6}{\textit{Nuclear
  Physics B} {\bfseries 216} (1983) 421--445}.

\bibitem{Witten:1984rs}
E.~Witten, \textit{{Cosmic Separation of Phases}},
  \href{https://doi.org/10.1103/PhysRevD.30.272}{\textit{Phys. Rev. D}
  {\bfseries 30} (1984) 272--285}.

\bibitem{Hogan:1986qda}
C.~J. Hogan, \textit{{Gravitational radiation from cosmological phase
  transitions}}, {\textit{Mon. Not. Roy. Astron. Soc.} {\bfseries 218} (1986)
  629--636}.

\bibitem{Caprini:2019egz}
C.~Caprini et~al., \textit{{Detecting gravitational waves from cosmological
  phase transitions with LISA: an update}},
  \href{https://doi.org/10.1088/1475-7516/2020/03/024}{\textit{JCAP} {\bfseries
  03} (2020) 024}, [\href{https://arxiv.org/abs/1910.13125}{{\ttfamily
  1910.13125}}].

\bibitem{Gould:2019qek}
O.~Gould, J.~Kozaczuk, L.~Niemi, M.~J. Ramsey-Musolf, T.~V.~I. Tenkanen and
  D.~J. Weir, \textit{{Nonperturbative analysis of the gravitational waves from
  a first-order electroweak phase transition}},
  \href{https://doi.org/10.1103/PhysRevD.100.115024}{\textit{Phys. Rev. D}
  {\bfseries 100} (2019) 115024},
  [\href{https://arxiv.org/abs/1903.11604}{{\ttfamily 1903.11604}}].

\bibitem{Weir:2017wfa}
D.~J. Weir, \textit{{Gravitational waves from a first order electroweak phase
  transition: a brief review}},
  \href{https://doi.org/10.1098/rsta.2017.0126}{\textit{Phil. Trans. Roy. Soc.
  Lond. A} {\bfseries 376} (2018) 20170126},
  [\href{https://arxiv.org/abs/1705.01783}{{\ttfamily 1705.01783}}].

\bibitem{Hindmarsh:2017gnf}
M.~Hindmarsh, S.~J. Huber, K.~Rummukainen and D.~J. Weir, \textit{{Shape of the
  acoustic gravitational wave power spectrum from a first order phase
  transition}}, \href{https://doi.org/10.1103/PhysRevD.96.103520}{\textit{Phys.
  Rev. D} {\bfseries 96} (2017) 103520},
  [\href{https://arxiv.org/abs/1704.05871}{{\ttfamily 1704.05871}}]. [Erratum:
  Phys.Rev.D 101, 089902 (2020)].

\bibitem{Guo:2020grp}
H.-K. Guo, K.~Sinha, D.~Vagie and G.~White, \textit{{Phase Transitions in an
  Expanding Universe: Stochastic Gravitational Waves in Standard and
  Non-Standard Histories}},
  \href{https://doi.org/10.1088/1475-7516/2021/01/001}{\textit{JCAP} {\bfseries
  01} (2021) 001}, [\href{https://arxiv.org/abs/2007.08537}{{\ttfamily
  2007.08537}}].

\bibitem{Hindmarsh:2015qjv}
M.~Hindmarsh, S.~Huber, K.~Rummukainen and D.~Weir, \textit{{Gravitational
  waves from cosmological first order phase transitions}},
  \href{https://doi.org/10.22323/1.251.0233}{\textit{PoS} {\bfseries
  LATTICE2015} (2016) 233}, [\href{https://arxiv.org/abs/1511.04527}{{\ttfamily
  1511.04527}}].

\bibitem{Bandyopadhyay:2020otm}
P.~Bandyopadhyay and A.~Costantini, \textit{{Obscure Higgs boson at
  Colliders}}, \href{https://doi.org/10.1103/PhysRevD.103.015025}{\textit{Phys.
  Rev. D} {\bfseries 103} (2021) 015025},
  [\href{https://arxiv.org/abs/2010.02597}{{\ttfamily 2010.02597}}].

\bibitem{sneha}
P.~Bandyopadhyay, S.~Jangid, A.~KT and S.~Parashar, \textit{{Discerning the
  Triplet charged Higgs bosons in BSM scenarios at the LHC and MATHUSLA}},
  \href{https://arxiv.org/abs/To be appear soon}{{\ttfamily To be appear
  soon}}.

\bibitem{Bandyopadhyay:2013lca}
P.~Bandyopadhyay, K.~Huitu and A.~Sabanci, \textit{{Status of $Y=0$ Triplet
  Higgs with supersymmetry in the light of $\sim 125$ GeV Higgs discovery}},
  \href{https://doi.org/10.1007/JHEP10(2013)091}{\textit{JHEP} {\bfseries 10}
  (2013) 091}, [\href{https://arxiv.org/abs/1306.4530}{{\ttfamily 1306.4530}}].

\bibitem{Bandyopadhyay:2015tva}
P.~Bandyopadhyay, C.~Coriano and A.~Costantini, \textit{{Probing the hidden
  Higgs bosons of the $Y = 0$ triplet- and singlet-extended Supersymmetric
  Standard Model at the LHC}},
  \href{https://doi.org/10.1007/JHEP12(2015)127}{\textit{JHEP} {\bfseries 12}
  (2015) 127}, [\href{https://arxiv.org/abs/1510.06309}{{\ttfamily
  1510.06309}}].

\bibitem{Bandyopadhyay:2015oga}
P.~Bandyopadhyay, C.~Coriano and A.~Costantini, \textit{{Perspectives on a
  supersymmetric extension of the standard model with a Y = 0 Higgs triplet and
  a singlet at the LHC}},
  \href{https://doi.org/10.1007/JHEP09(2015)045}{\textit{JHEP} {\bfseries 09}
  (2015) 045}, [\href{https://arxiv.org/abs/1506.03634}{{\ttfamily
  1506.03634}}].

\bibitem{Bandyopadhyay:2014tha}
P.~Bandyopadhyay, S.~Di~Chiara, K.~Huitu and A.~S. Ke\c{c}eli,
  \textit{{Naturality vs perturbativity, B$_{s}$ physics, and LHC data in
  triplet extension of MSSM}},
  \href{https://doi.org/10.1007/JHEP11(2014)062}{\textit{JHEP} {\bfseries 11}
  (2014) 062}, [\href{https://arxiv.org/abs/1407.4836}{{\ttfamily 1407.4836}}].

\bibitem{Bandyopadhyay:2014vma}
P.~Bandyopadhyay, K.~Huitu and A.~Sabanci~Keceli, \textit{{Multi-Lepton
  Signatures of the Triplet Like Charged Higgs at the LHC}},
  \href{https://doi.org/10.1007/JHEP05(2015)026}{\textit{JHEP} {\bfseries 05}
  (2015) 026}, [\href{https://arxiv.org/abs/1412.7359}{{\ttfamily 1412.7359}}].

\bibitem{Bandyopadhyay:2017klv}
P.~Bandyopadhyay and A.~Costantini, \textit{{Distinguishing charged Higgs
  bosons from different representations at the LHC}},
  \href{https://doi.org/10.1007/JHEP01(2018)067}{\textit{JHEP} {\bfseries 01}
  (2018) 067}, [\href{https://arxiv.org/abs/1710.03110}{{\ttfamily
  1710.03110}}].

\bibitem{Bandyopadhyay:2015ifm}
P.~Bandyopadhyay, C.~Coriano and A.~Costantini, \textit{{General analysis of
  the charged Higgs sector of the $Y=0$ triplet-singlet extension of the MSSM
  at the LHC}}, \href{https://doi.org/10.1103/PhysRevD.94.055030}{\textit{Phys.
  Rev. D} {\bfseries 94} (2016) 055030},
  [\href{https://arxiv.org/abs/1512.08651}{{\ttfamily 1512.08651}}].

\bibitem{SabanciKeceli:2018fsd}
A.~Sabanci~Keceli, P.~Bandyopadhyay and K.~Huitu, \textit{{Long-lived triplinos
  and displaced lepton signals at the LHC}},
  \href{https://doi.org/10.1140/epjc/s10052-019-6818-0}{\textit{Eur. Phys. J.
  C} {\bfseries 79} (2019) 345},
  [\href{https://arxiv.org/abs/1810.09172}{{\ttfamily 1810.09172}}].

\bibitem{quiros}
J.~R. Espinosa and M.~Quiros, \textit{{The Electroweak phase transition with a
  singlet}}, \href{https://doi.org/10.1016/0370-2693(93)91111-Y}{\textit{Phys.
  Lett. B} {\bfseries 305} (1993) 98--105},
  [\href{https://arxiv.org/abs/hep-ph/9301285}{{\ttfamily hep-ph/9301285}}].

\bibitem{Yagi:2011wg}
K.~Yagi and N.~Seto, \textit{{Detector configuration of DECIGO/BBO and
  identification of cosmological neutron-star binaries}},
  \href{https://doi.org/10.1103/PhysRevD.83.044011}{\textit{Phys. Rev. D}
  {\bfseries 83} (2011) 044011},
  [\href{https://arxiv.org/abs/1101.3940}{{\ttfamily 1101.3940}}]. [Erratum:
  Phys.Rev.D 95, 109901 (2017)].

\bibitem{LISA:2017pwj}
{\scshape LISA} collaboration, P.~Amaro-Seoane et~al., \textit{{Laser
  Interferometer Space Antenna}},
  \href{https://arxiv.org/abs/1702.00786}{{\ttfamily 1702.00786}}.

\bibitem{KAGRA:2013rdx}
{\scshape KAGRA, LIGO Scientific, Virgo, VIRGO} collaboration, B.~P. Abbott
  et~al., \textit{{Prospects for observing and localizing gravitational-wave
  transients with Advanced LIGO, Advanced Virgo and KAGRA}},
  \href{https://doi.org/10.1007/s41114-020-00026-9}{\textit{Living Rev. Rel.}
  {\bfseries 21} (2018) 3}, [\href{https://arxiv.org/abs/1304.0670}{{\ttfamily
  1304.0670}}].

\bibitem{PhysRevLett.116.241103}
{\scshape LIGO Scientific Collaboration and Virgo Collaboration} collaboration,
  B.~P. Abbott~et al., \textit{Observation of gravitational waves from a
  22-solar-mass binary black hole coalescence},
  \href{https://doi.org/10.1103/PhysRevLett.116.241103}{\textit{Phys. Rev.
  Lett.} {\bfseries 116} (Jun, 2016) 241103}.

\bibitem{PhysRevLett.118.221101}
{\scshape LIGO Scientific and Virgo Collaboration} collaboration, B.~P.
  Abbott~et al., \textit{Observation of a 50-solar-mass binary black hole
  coalescence at redshift 0.2},
  \href{https://doi.org/10.1103/PhysRevLett.118.221101}{\textit{Phys. Rev.
  Lett.} {\bfseries 118} (Jun, 2017) 221101}.

\bibitem{Coleman}
S.~Coleman and E.~Weinberg, \textit{Radiative corrections as the origin of
  spontaneous symmetry breaking},
  \href{https://doi.org/10.1103/PhysRevD.7.1888}{\textit{Phys. Rev. D}
  {\bfseries 7} (Mar, 1973) 1888--1910}.

\bibitem{Weinberg}
S.~Weinberg, \textit{Gauge and global symmetries at high temperature},
  \href{https://doi.org/10.1103/PhysRevD.9.3357}{\textit{Phys. Rev. D}
  {\bfseries 9} (Jun, 1974) 3357--3378}.

\bibitem{Dolan}
L.~Dolan and R.~Jackiw, \textit{Symmetry behavior at finite temperature},
  \href{https://doi.org/10.1103/PhysRevD.9.3320}{\textit{Phys. Rev. D}
  {\bfseries 9} (Jun, 1974) 3320--3341}.

\bibitem{Kirzhnits:1976ts}
D.~A. Kirzhnits and A.~D. Linde, \textit{{Symmetry Behavior in Gauge
  Theories}},
  \href{https://doi.org/10.1016/0003-4916(76)90279-7}{\textit{Annals Phys.}
  {\bfseries 101} (1976) 195--238}.

\bibitem{Gross}
D.~J. Gross, R.~D. Pisarski and L.~G. Yaffe, \textit{Qcd and instantons at
  finite temperature},
  \href{https://doi.org/10.1103/RevModPhys.53.43}{\textit{Rev. Mod. Phys.}
  {\bfseries 53} (Jan, 1981) 43--80}.

\bibitem{Fendley:1987ef}
P.~Fendley, \textit{{The Effective Potential and the Coupling Constant at High
  Temperature}},
  \href{https://doi.org/10.1016/0370-2693(87)90599-5}{\textit{Phys. Lett. B}
  {\bfseries 196} (1987) 175--180}.

\bibitem{Kapusta}
J.~Kapusta, \textit{{Finite temperature Field Theory}}, {\textit{Cambridge
  University Press} (1989) }.

\bibitem{Arnold:1992rz}
P.~B. Arnold and O.~Espinosa, \textit{{The Effective potential and first order
  phase transitions: Beyond leading-order}},
  \href{https://doi.org/10.1103/PhysRevD.47.3546}{\textit{Phys. Rev. D}
  {\bfseries 47} (1993) 3546},
  [\href{https://arxiv.org/abs/hep-ph/9212235}{{\ttfamily hep-ph/9212235}}].

\bibitem{Niemi:2021qvp}
L.~Niemi, P.~Schicho and T.~V.~I. Tenkanen, \textit{{Singlet-assisted
  electroweak phase transition at two loops}},
  \href{https://doi.org/10.1103/PhysRevD.103.115035}{\textit{Phys. Rev. D}
  {\bfseries 103} (2021) 115035},
  [\href{https://arxiv.org/abs/2103.07467}{{\ttfamily 2103.07467}}].

\bibitem{10.1093/ptep/ptaa104}
{Particle Data Group}, P.~A. Zyla et~al., \textit{{Review of Particle
  Physics}}, \href{https://doi.org/10.1093/ptep/ptaa104}{\textit{Prog. Theor.
  Exp. Phys.} {\bfseries 2020} 083C01 (2020)}.

\bibitem{cohen}
A.~Cohen, D.~Kaplan and A.~Nelson{\textit{{Annual Review of Nuclear and
  Particle Science}} {\bfseries 43} (1994) 27}.

\bibitem{Rubakov:1996vz}
V.~A. Rubakov and M.~E. Shaposhnikov, \textit{{Electroweak baryon number
  nonconservation in the early universe and in high-energy collisions}},
  \href{https://doi.org/10.1070/PU1996v039n05ABEH000145}{\textit{Usp. Fiz.
  Nauk} {\bfseries 166} (1996) 493--537},
  [\href{https://arxiv.org/abs/hep-ph/9603208}{{\ttfamily hep-ph/9603208}}].

\bibitem{Staub:2013tta}
F.~Staub, \textit{{SARAH 4 : A tool for (not only SUSY) model builders}},
  \href{https://doi.org/10.1016/j.cpc.2014.02.018}{\textit{Comput. Phys.
  Commun.} {\bfseries 185} (2014) 1773--1790},
  [\href{https://arxiv.org/abs/1309.7223}{{\ttfamily 1309.7223}}].

\bibitem{Carena:2008vj}
M.~Carena, G.~Nardini, M.~Quiros and C.~E.~M. Wagner, \textit{{The Baryogenesis
  Window in the MSSM}},
  \href{https://doi.org/10.1016/j.nuclphysb.2008.12.014}{\textit{Nucl. Phys. B}
  {\bfseries 812} (2009) 243--263},
  [\href{https://arxiv.org/abs/0809.3760}{{\ttfamily 0809.3760}}].

\bibitem{Espinosa:1996qw}
J.~R. Espinosa, \textit{{Dominant two loop corrections to the MSSM finite
  temperature effective potential}},
  \href{https://doi.org/10.1016/0550-3213(96)00297-0}{\textit{Nucl. Phys. B}
  {\bfseries 475} (1996) 273--292},
  [\href{https://arxiv.org/abs/hep-ph/9604320}{{\ttfamily hep-ph/9604320}}].

\bibitem{Carena:1997ki}
M.~Carena, M.~Quiros and C.~E.~M. Wagner, \textit{{Electroweak baryogenesis and
  Higgs and stop searches at LEP and the Tevatron}},
  \href{https://doi.org/10.1016/S0550-3213(98)00187-4}{\textit{Nucl. Phys. B}
  {\bfseries 524} (1998) 3--22},
  [\href{https://arxiv.org/abs/hep-ph/9710401}{{\ttfamily hep-ph/9710401}}].

\bibitem{Laine:2017hdk}
M.~Laine, M.~Meyer and G.~Nardini, \textit{{Thermal phase transition with full
  2-loop effective potential}},
  \href{https://doi.org/10.1016/j.nuclphysb.2017.04.023}{\textit{Nucl. Phys. B}
  {\bfseries 920} (2017) 565--600},
  [\href{https://arxiv.org/abs/1702.07479}{{\ttfamily 1702.07479}}].

\bibitem{Gould:2021oba}
O.~Gould and T.~V.~I. Tenkanen, \textit{{On the perturbative expansion at high
  temperature and implications for cosmological phase transitions}},
  \href{https://doi.org/10.1007/JHEP06(2021)069}{\textit{JHEP} {\bfseries 06}
  (2021) 069}, [\href{https://arxiv.org/abs/2104.04399}{{\ttfamily
  2104.04399}}].

\bibitem{Farakos:1994kx}
K.~Farakos, K.~Kajantie, K.~Rummukainen and M.~E. Shaposhnikov, \textit{{3-D
  physics and the electroweak phase transition: Perturbation theory}},
  \href{https://doi.org/10.1016/0550-3213(94)90173-2}{\textit{Nucl. Phys. B}
  {\bfseries 425} (1994) 67--109},
  [\href{https://arxiv.org/abs/hep-ph/9404201}{{\ttfamily hep-ph/9404201}}].

\bibitem{Schicho:2021gca}
P.~M. Schicho, T.~V.~I. Tenkanen and J.~\"Osterman, \textit{{Robust approach to
  thermal resummation: Standard Model meets a singlet}},
  \href{https://doi.org/10.1007/JHEP06(2021)130}{\textit{JHEP} {\bfseries 06}
  (2021) 130}, [\href{https://arxiv.org/abs/2102.11145}{{\ttfamily
  2102.11145}}].

\bibitem{Niemi:2018asa}
L.~Niemi, H.~H. Patel, M.~J. Ramsey-Musolf, T.~V.~I. Tenkanen and D.~J. Weir,
  \textit{{Electroweak phase transition in the real triplet extension of the
  SM: Dimensional reduction}},
  \href{https://doi.org/10.1103/PhysRevD.100.035002}{\textit{Phys. Rev. D}
  {\bfseries 100} (2019) 035002},
  [\href{https://arxiv.org/abs/1802.10500}{{\ttfamily 1802.10500}}].

\bibitem{Planck:2013pxb}
{\scshape Planck} collaboration, P.~A.~R. Ade et~al., \textit{{Planck 2013
  results. XVI. Cosmological parameters}},
  \href{https://doi.org/10.1051/0004-6361/201321591}{\textit{Astron.
  Astrophys.} {\bfseries 571} (2014) A16},
  [\href{https://arxiv.org/abs/1303.5076}{{\ttfamily 1303.5076}}].

\bibitem{acpb}
P.~Bandyopadhyay and A.~Costantini, \textit{{Exploring the Complex Scalar
  Extensions of the Standard Model}},  \href{https://arxiv.org/abs/To be appear
  soon}{{\ttfamily To be appear soon}}.

\bibitem{Robens:2015gla}
T.~Robens and T.~Stefaniak, \textit{{Status of the Higgs Singlet Extension of
  the Standard Model after LHC Run 1}},
  \href{https://doi.org/10.1140/epjc/s10052-015-3323-y}{\textit{Eur. Phys. J.
  C} {\bfseries 75} (2015) 104},
  [\href{https://arxiv.org/abs/1501.02234}{{\ttfamily 1501.02234}}].

\bibitem{Wainwright:2009mq}
C.~Wainwright and S.~Profumo, \textit{{The Impact of a strongly first-order
  phase transition on the abundance of thermal relics}},
  \href{https://doi.org/10.1103/PhysRevD.80.103517}{\textit{Phys. Rev. D}
  {\bfseries 80} (2009) 103517},
  [\href{https://arxiv.org/abs/0909.1317}{{\ttfamily 0909.1317}}].

\bibitem{Wainwright}
C.~L. Wainwright{\textit{Comput. Phys. Commun.} {\bfseries 183} (2006) }.

\bibitem{Athron:2019nbd}
P.~Athron, C.~Balazs, M.~Bardsley, A.~Fowlie, D.~Harries and G.~White,
  \textit{{BubbleProfiler: finding the field profile and action for
  cosmological phase transitions}},
  \href{https://doi.org/10.1016/j.cpc.2019.05.017}{\textit{Comput. Phys.
  Commun.} {\bfseries 244} (2019) 448--468},
  [\href{https://arxiv.org/abs/1901.03714}{{\ttfamily 1901.03714}}].

\bibitem{Randall:2006py}
L.~Randall and G.~Servant, \textit{{Gravitational waves from warped
  spacetime}},
  \href{https://doi.org/10.1088/1126-6708/2007/05/054}{\textit{JHEP} {\bfseries
  05} (2007) 054}, [\href{https://arxiv.org/abs/hep-ph/0607158}{{\ttfamily
  hep-ph/0607158}}].

\bibitem{Kosowsky}
A.~Kosowsky, M.~S. Turner and R.~Watkins, \textit{Gravitational radiation from
  colliding vacuum bubbles},
  \href{https://doi.org/10.1103/PhysRevD.45.4514}{\textit{Phys. Rev. D}
  {\bfseries 45} (Jun, 1992) 4514--4535}.

\bibitem{Turner}
A.~Kosowsky and M.~S. Turner, \textit{Gravitational radiation from colliding
  vacuum bubbles: Envelope approximation to many-bubble collisions},
  \href{https://doi.org/10.1103/PhysRevD.47.4372}{\textit{Phys. Rev. D}
  {\bfseries 47} (May, 1993) 4372--4391}.

\bibitem{Huber_2008}
S.~J. Huber and T.~Konstandin, \textit{Gravitational wave production by
  collisions: more bubbles},
  \href{https://doi.org/10.1088/1475-7516/2008/09/022}{\textit{Journal of
  Cosmology and Astroparticle Physics} {\bfseries 2008} (Sep, 2008) 022}.

\bibitem{Watkins}
A.~Kosowsky, M.~S. Turner and R.~Watkins, \textit{Gravitational waves from
  first-order cosmological phase transitions},
  \href{https://doi.org/10.1103/PhysRevLett.69.2026}{\textit{Phys. Rev. Lett.}
  {\bfseries 69} (Oct, 1992) 2026--2029}.

\bibitem{Marc}
M.~Kamionkowski, A.~Kosowsky and M.~S. Turner, \textit{Gravitational radiation
  from first-order phase transitions},
  \href{https://doi.org/10.1103/PhysRevD.49.2837}{\textit{Phys. Rev. D}
  {\bfseries 49} (Mar, 1994) 2837--2851}.

\bibitem{Caprini_2008}
C.~Caprini, R.~Durrer and G.~Servant, \textit{Gravitational wave generation
  from bubble collisions in first-order phase transitions: An analytic
  approach}, \href{https://doi.org/10.1103/physrevd.77.124015}{\textit{Physical
  Review D} {\bfseries 77} (Jun, 2008) }.

\bibitem{Hindmarsh}
M.~Hindmarsh, S.~J. Huber, K.~Rummukainen and D.~J. Weir, \textit{Gravitational
  waves from the sound of a first order phase transition},
  \href{https://doi.org/10.1103/PhysRevLett.112.041301}{\textit{Phys. Rev.
  Lett.} {\bfseries 112} (Jan, 2014) 041301}.

\bibitem{Leitao:2012tx}
L.~Leitao, A.~Megevand and A.~D. Sanchez, \textit{{Gravitational waves from the
  electroweak phase transition}},
  \href{https://doi.org/10.1088/1475-7516/2012/10/024}{\textit{JCAP} {\bfseries
  10} (2012) 024}, [\href{https://arxiv.org/abs/1205.3070}{{\ttfamily
  1205.3070}}].

\bibitem{Giblin:2013kea}
J.~T. Giblin, Jr. and J.~B. Mertens, \textit{{Vacuum Bubbles in the Presence of
  a Relativistic Fluid}},
  \href{https://doi.org/10.1007/JHEP12(2013)042}{\textit{JHEP} {\bfseries 12}
  (2013) 042}, [\href{https://arxiv.org/abs/1310.2948}{{\ttfamily 1310.2948}}].

\bibitem{Giblin:2014qia}
J.~T. Giblin and J.~B. Mertens, \textit{{Gravitional radiation from first-order
  phase transitions in the presence of a fluid}},
  \href{https://doi.org/10.1103/PhysRevD.90.023532}{\textit{Phys. Rev. D}
  {\bfseries 90} (2014) 023532},
  [\href{https://arxiv.org/abs/1405.4005}{{\ttfamily 1405.4005}}].

\bibitem{Hindmarsh_2015}
M.~Hindmarsh, S.~J. Huber, K.~Rummukainen and D.~J. Weir, \textit{Numerical
  simulations of acoustically generated gravitational waves at a first order
  phase transition},
  \href{https://doi.org/10.1103/physrevd.92.123009}{\textit{Physical Review D}
  {\bfseries 92} (Dec, 2015) }.

\bibitem{Chiara}
C.~Caprini and R.~Durrer, \textit{Gravitational waves from stochastic
  relativistic sources: Primordial turbulence and magnetic fields},
  \href{https://doi.org/10.1103/PhysRevD.74.063521}{\textit{Phys. Rev. D}
  {\bfseries 74} (Sep, 2006) 063521}.

\bibitem{Kahniashvili}
T.~Kahniashvili, A.~Kosowsky, G.~Gogoberidze and Y.~Maravin,
  \textit{Detectability of gravitational waves from phase transitions},
  \href{https://doi.org/10.1103/PhysRevD.78.043003}{\textit{Phys. Rev. D}
  {\bfseries 78} (Aug, 2008) 043003}.

\bibitem{Kahniashvili:2008pe}
T.~Kahniashvili, L.~Campanelli, G.~Gogoberidze, Y.~Maravin and B.~Ratra,
  \textit{{Gravitational Radiation from Primordial Helical Inverse Cascade MHD
  Turbulence}}, \href{https://doi.org/10.1103/PhysRevD.78.123006}{\textit{Phys.
  Rev. D} {\bfseries 78} (2008) 123006},
  [\href{https://arxiv.org/abs/0809.1899}{{\ttfamily 0809.1899}}]. [Erratum:
  Phys.Rev.D 79, 109901 (2009)].

\bibitem{Kahniashvili:2009mf}
T.~Kahniashvili, L.~Kisslinger and T.~Stevens, \textit{{Gravitational Radiation
  Generated by Magnetic Fields in Cosmological Phase Transitions}},
  \href{https://doi.org/10.1103/PhysRevD.81.023004}{\textit{Phys. Rev. D}
  {\bfseries 81} (2010) 023004},
  [\href{https://arxiv.org/abs/0905.0643}{{\ttfamily 0905.0643}}].

\bibitem{Caprini:2009yp}
C.~Caprini, R.~Durrer and G.~Servant, \textit{{The stochastic gravitational
  wave background from turbulence and magnetic fields generated by a
  first-order phase transition}},
  \href{https://doi.org/10.1088/1475-7516/2009/12/024}{\textit{JCAP} {\bfseries
  12} (2009) 024}, [\href{https://arxiv.org/abs/0909.0622}{{\ttfamily
  0909.0622}}].

\bibitem{Caprini:2015zlo}
C.~Caprini et~al., \textit{{Science with the space-based interferometer eLISA.
  II: Gravitational waves from cosmological phase transitions}},
  \href{https://doi.org/10.1088/1475-7516/2016/04/001}{\textit{JCAP} {\bfseries
  04} (2016) 001}, [\href{https://arxiv.org/abs/1512.06239}{{\ttfamily
  1512.06239}}].

\bibitem{Chao:2017vrq}
W.~Chao, H.-K. Guo and J.~Shu, \textit{{Gravitational Wave Signals of
  Electroweak Phase Transition Triggered by Dark Matter}},
  \href{https://doi.org/10.1088/1475-7516/2017/09/009}{\textit{JCAP} {\bfseries
  09} (2017) 009}, [\href{https://arxiv.org/abs/1702.02698}{{\ttfamily
  1702.02698}}].

\bibitem{Dev:2019njv}
P.~S.~B. Dev, F.~Ferrer, Y.~Zhang and Y.~Zhang, \textit{{Gravitational Waves
  from First-Order Phase Transition in a Simple Axion-Like Particle Model}},
  \href{https://doi.org/10.1088/1475-7516/2019/11/006}{\textit{JCAP} {\bfseries
  11} (2019) 006}, [\href{https://arxiv.org/abs/1905.00891}{{\ttfamily
  1905.00891}}].

\bibitem{Paul}
P.~J. Steinhardt, \textit{Relativistic detonation waves and bubble growth in
  false vacuum decay},
  \href{https://doi.org/10.1103/PhysRevD.25.2074}{\textit{Phys. Rev. D}
  {\bfseries 25} (Apr, 1982) 2074--2085}.

\bibitem{Shajiee:2018jdq}
V.~R. Shajiee and A.~Tofighi, \textit{{Electroweak Phase Transition,
  Gravitational Waves and Dark Matter in Two Scalar Singlet Extension of The
  Standard Model}},
  \href{https://doi.org/10.1140/epjc/s10052-019-6881-6}{\textit{Eur. Phys. J.
  C} {\bfseries 79} (2019) 360},
  [\href{https://arxiv.org/abs/1811.09807}{{\ttfamily 1811.09807}}].

\bibitem{Kamionkowski:1993fg}
M.~Kamionkowski, A.~Kosowsky and M.~S. Turner, \textit{{Gravitational radiation
  from first order phase transitions}},
  \href{https://doi.org/10.1103/PhysRevD.49.2837}{\textit{Phys. Rev. D}
  {\bfseries 49} (1994) 2837--2851},
  [\href{https://arxiv.org/abs/astro-ph/9310044}{{\ttfamily
  astro-ph/9310044}}].

\bibitem{Ellis:2018mja}
J.~Ellis, M.~Lewicki and J.~M. No, \textit{{On the Maximal Strength of a
  First-Order Electroweak Phase Transition and its Gravitational Wave Signal}},
  \href{https://doi.org/10.1088/1475-7516/2019/04/003}{\textit{JCAP} {\bfseries
  04} (2019) 003}, [\href{https://arxiv.org/abs/1809.08242}{{\ttfamily
  1809.08242}}].

\bibitem{Espinosa:2010hh}
J.~R. Espinosa, T.~Konstandin, J.~M. No and G.~Servant, \textit{{Energy Budget
  of Cosmological First-order Phase Transitions}},
  \href{https://doi.org/10.1088/1475-7516/2010/06/028}{\textit{JCAP} {\bfseries
  06} (2010) 028}, [\href{https://arxiv.org/abs/1004.4187}{{\ttfamily
  1004.4187}}].

\bibitem{Barish:2020vmy}
B.~C. Barish, S.~Bird and Y.~Cui, \textit{{Impact of a midband gravitational
  wave experiment on detectability of cosmological stochastic gravitational
  wave backgrounds}},
  \href{https://doi.org/10.1103/PhysRevD.103.123541}{\textit{Phys. Rev. D}
  {\bfseries 103} (2021) 123541},
  [\href{https://arxiv.org/abs/2012.07874}{{\ttfamily 2012.07874}}].

\bibitem{Aoki:2019mlt}
M.~Aoki and J.~Kubo, \textit{{Gravitational waves from chiral phase transition
  in a conformally extended standard model}},
  \href{https://doi.org/10.1088/1475-7516/2020/04/001}{\textit{JCAP} {\bfseries
  04} (2020) 001}, [\href{https://arxiv.org/abs/1910.05025}{{\ttfamily
  1910.05025}}].

\bibitem{Yagi:2013du}
K.~Yagi, \textit{{Scientific Potential of DECIGO Pathfinder and Testing GR with
  Space-Borne Gravitational Wave Interferometers}},
  \href{https://doi.org/10.1142/S0218271813410137}{\textit{Int. J. Mod. Phys.
  D} {\bfseries 22} (2013) 1341013},
  [\href{https://arxiv.org/abs/1302.2388}{{\ttfamily 1302.2388}}].

\bibitem{Moore:2014lga}
C.~J. Moore, R.~H. Cole and C.~P.~L. Berry, \textit{{Gravitational-wave
  sensitivity curves}},
  \href{https://doi.org/10.1088/0264-9381/32/1/015014}{\textit{Class. Quant.
  Grav.} {\bfseries 32} (2015) 015014},
  [\href{https://arxiv.org/abs/1408.0740}{{\ttfamily 1408.0740}}].

\bibitem{Yagi:2011yu}
K.~Yagi, N.~Tanahashi and T.~Tanaka, \textit{{Probing the size of extra
  dimension with gravitational wave astronomy}},
  \href{https://doi.org/10.1103/PhysRevD.83.084036}{\textit{Phys. Rev. D}
  {\bfseries 83} (2011) 084036},
  [\href{https://arxiv.org/abs/1101.4997}{{\ttfamily 1101.4997}}].

\bibitem{Thrane:2013oya}
E.~Thrane and J.~D. Romano, \textit{{Sensitivity curves for searches for
  gravitational-wave backgrounds}},
  \href{https://doi.org/10.1103/PhysRevD.88.124032}{\textit{Phys. Rev. D}
  {\bfseries 88} (2013) 124032},
  [\href{https://arxiv.org/abs/1310.5300}{{\ttfamily 1310.5300}}].

\bibitem{Croon:2020cgk}
D.~Croon, O.~Gould, P.~Schicho, T.~V.~I. Tenkanen and G.~White,
  \textit{{Theoretical uncertainties for cosmological first-order phase
  transitions}}, \href{https://doi.org/10.1007/JHEP04(2021)055}{\textit{JHEP}
  {\bfseries 04} (2021) 055},
  [\href{https://arxiv.org/abs/2009.10080}{{\ttfamily 2009.10080}}].

\bibitem{PhysRevD.23.2916}
P.~M. Stevenson, \textit{Optimized perturbation theory},
  \href{https://doi.org/10.1103/PhysRevD.23.2916}{\textit{Phys. Rev. D}
  {\bfseries 23} (Jun, 1981) 2916--2944}.

\end{thebibliography}\endgroup
\bibliographystyle{Ref}

\end{document}